# Current Status and Prospects of Neutron Detection and Neutron/Gamma Discrimination Technologies in Fusion Applications

Zhuo Zuo[1,2], Bingqi Liu[1,4*], Hao Feng[2], Jie Zhang[2], Xianghe Liu[1], Haoran Liu[1], Peng Li[1], Qibiao Wang[4], Mingzhe Liu[1,3]
1 College of Nuclear Technology and Automation Engineering, Chengdu University of Technology, Chengdu 610059, China
2 Engineering & Technical College of Chengdu University of Technology, Southwestern Institute of Physics, Leshan 614000, China
3 School of Data Science and Artificial Intelligence, Wenzhou University of Technology, Wenzhou 325000, China
4 Artificial Intelligence Key Laboratory of Sichuan Province, Sichuan University of Science and Engineering, Yibin 644000, China
*Corresponding author, liubingqi@cdut.edu.cn (Bingqi Liu)

This work was supported by the National Natural Science Foundation of China under Grant 42574242; Funded by the Sichuan Provincial Regional Innovation Cooperation Project (2026YFHZ0146); Open Fund of the Sichuan Provincial Key Laboratory of Geoscience and Nuclear Technology (GNZDS2026007)

**Abstract:** Fusion research is progressing from physics-oriented experiments toward reactor-oriented engineering applications, placing increasing demands on the accuracy and reliability of neutron measurements. This review summarizes the current status and prospects of neutron detection and neutron/gamma discrimination technologies for fusion applications. The main sources, energy characteristics, and measurement requirements of fusion neutrons are first reviewed, with particular attention to 2.45 MeV D-D and 14.1 MeV D-T neutrons. Four principal neutron detection methods, including nuclear reaction, nuclear recoil, nuclear fission, and neutron activation, are discussed together with the operating characteristics and applicability of scintillation, gas, semiconductor, and other specialized detectors. The roles of neutron/gamma discrimination are then examined in plasma diagnostics, safe operation monitoring, radiation protection monitoring, and fusion reactor design. The review shows that different detection methods and detector types have distinct advantages and application boundaries, and no single detector provides an optimal solution for all fusion measurement tasks. Liquid scintillators and some organic crystals remain important for fast-neutron measurement and neutron/gamma discrimination, whereas gas and semiconductor detectors serve complementary roles in thermal-neutron monitoring, compact detection, radiation tolerance, and fast-neutron spectrometry. Future development is expected to focus on high-performance detectors, material damage assessment, intelligent real-time discrimination, and multi-detector coordination with system-level diagnostic integration.

Nuclear fusion is widely regarded as one of the important candidate pathways for future advanced energy systems because of its high energy density, relatively abundant fuel resources, and low carbon emissions (INTERNATIONAL ATOMIC ENERGY AGENCY, 2025). With the continuous acceleration of the global energy transition and competition in advanced energy technologies, fusion research is gradually shifting from a stage dominated by device-level physics validation toward reactor-oriented engineering implementation. Studies have shown that the focus of international fusion research has clearly expanded from whether fusion reactions can be achieved to whether fusion devices can realize stable, controllable, and long-duration operation, indicating that the importance of diagnostic and measurement capabilities is increasing accordingly (Mohamed et al., 2024). Meanwhile, experimental platforms with deuterium-tritium (D-T) operation capability remain an important foundation for the continuous development and validation of fusion and diagnostic technologies (Solano, 2025). Therefore, fusion research has entered a new stage in which higher requirements are imposed on the characterization of fusion neutron fields, the accuracy of nuclear measurements, and the reliability of diagnostic systems.

Against this background, neutron detection and neutron/gamma discrimination technologies have become increasingly important. For fusion devices, neutrons are not only the most direct products of D-T and deuterium-deuterium (D-D) reactions, but also important information carriers for characterizing key parameters such as fusion power and ion temperature (Du et al., 2022). For future fusion reactors, neutrons are further associated with material irradiation damage and operational safety assessment (Giancarli et al., 2024; Olatujoye et al., 2025). Reviews of fusion diagnostics have further indicated that, as fusion research advances toward high-field, high-power, and reactor-relevant operating conditions, diagnostic systems must simultaneously withstand harsh thermal, electromagnetic, and radiation environments, limited port and penetration resources, and system-integration requirements for both early operation and subsequent upgrades (Reinke et al., 2024). Under these conditions, the acquisition of true neutron signals is no longer merely a matter of detection, but is closely related to gamma-background suppression, signal discrimination, time resolution, and system-level integration in complex mixed radiation fields. Therefore, neutron/gamma discrimination has become an essential technical foundation for reliable neutron measurement in fusion applications.

Existing studies have discussed the future development trends of fusion diagnostic technologies from the perspectives of diagnostic cross-fertilization between magnetic confinement fusion and inertial confinement fusion, as well as methodological transfer and synergistic utilization among different radiation channels (Gatu Johnson et al., 2024). Other studies, starting from the concept of volumetric neutron sources for component testing, have emphasized that the qualification of key components for future fusion power plants will rely on high-flux 14 MeV neutron environments, further highlighting the fundamental role of neutron measurement and diagnostic reliability in the engineering realization of fusion energy (Bachmann et al., 2025). However, current research has mainly focused on fusion development roadmaps, diagnostic requirements of specific devices, individual measurement principles, or specific detector systems. A systematic and integrated discussion along the complete chain of "fusion neutron sources and characteristics-measurement methods-detector types-applications of neutron/gamma discrimination-future development potential for fusion reactors" remains insufficient. In particular, the intrinsic relationship between neutron detection and neutron/gamma discrimination, their functions in different fusion scenarios, and their evolution from localized detection capabilities toward system-level support capabilities in future fusion reactors still require a more focused and systematic review.

Based on the above background, this paper reviews the current status and prospects of neutron detection and neutron/gamma discrimination technologies in fusion applications, aiming to establish a relatively comprehensive analytical framework from the following aspects. First, the main sources, energy characteristics, and measurement requirements of fusion neutrons are summarized in relation to fusion reaction types and typical device operating conditions. Second, neutron detection methods and the applicable ranges of different detector types are reviewed from the perspectives of both measurement principles and device implementation. Third, the application status and functional mechanisms of neutron/gamma discrimination technologies in plasma diagnostics, safe operation monitoring of fusion devices, radiation protection monitoring, and fusion reactor design are analyzed in detail. Furthermore, their development potential is discussed in terms of high-performance detector development, material damage assessment, intelligent real-time discrimination systems, and multi-detector coordination with system-level diagnostic integration. Compared with existing studies, this review does not focus on a single detector type or a specific diagnostic method. Instead, it integrates relevant research distributed across different technical directions and application scenarios, with the aim of providing a more systematic reference for the development and application of diagnostic technologies for future fusion reactors.

# 1. Fusion Neutron Sources and Detection

## 1.1 Characteristics of Neutron Sources and Measurement Technologies

### 1.1.1 Main Neutron Sources in Fusion Reactions

Neutrons in fusion devices are mainly produced by light-nucleus fusion reactions. According to the confinement scheme, current fusion experimental platforms can generally be divided into two categories: magnetic confinement fusion devices and inertial confinement fusion devices (Abdou et al., 2021). Magnetic confinement fusion devices, represented by tokamaks and stellarators, mainly conduct experiments involving deuterium plasmas, D-T plasmas, and fast-ion transport. Inertial confinement fusion devices, represented by laser-driven implosion systems, mainly focus on D-T target compression, ignition, and burning. In these devices, neutron generation can essentially be attributed to D-T, D-D, and a small number of secondary reaction processes, as summarized in Table 1. Among them, the D-T reaction is currently the dominant neutron source in fusion research (IAEA, 2024), and its reaction equation is expressed as follows:

$$^{2}H + ^{3}H \rightarrow ^{4}He(3.5Mev) + n(14.1Mev) \quad (1.1)$$

This reaction releases a total energy of approximately 17.6 MeV, of which about 14.1 MeV is carried by the neutron. It is generally recognized that, under currently achievable fusion temperature ranges and engineering conditions, D-T fuel remains the most practically feasible main route. Therefore, 14.1 MeV neutrons produced by the D-T reaction constitute the most important neutron source in future fusion reactors (Garcia and JET Contributors, 2025).

In magnetic confinement fusion research, tokamaks represent the most mature device type for conducting D-T neutron experiments. The Joint European Torus (JET), as a representative tokamak device, has successively carried out the first Deuterium-Tritium Experiment (DTE1) and the second Deuterium-Tritium Experiment (DTE2). In particular, DTE2 substantially advanced the understanding of D-T plasma physics, nuclear operating conditions, and neutron diagnostic methods, and provided an important experimental basis for neutronics analysis and nuclear technology validation for the International Thermonuclear Experimental Reactor (ITER) (Villari et al., 2025).

In addition to the D-T reaction, the D-D reaction is also an important neutron source commonly encountered in fusion experiments, and its reaction equation is expressed as follows:

$$^{2}H + ^{2}H \rightarrow ^{3}He(0.82Mev) + n(2.45Mev) \quad (1.2)$$

The characteristic neutron energy produced by this reaction is approximately 2.45 MeV. Although the D-D reaction has lower reactivity and a relatively lower neutron yield than the D-T reaction, it does not require the direct introduction of tritium during experimental operation. Therefore, D-D experiments are widely used in current fusion research for studies of fast-ion behavior and validation of neutron diagnostic systems (Ogawa et al., 2021).

In magnetic confinement experiments dominated by the D-D reaction, high-energy neutron components induced by secondary reactions may also appear. Tritium ions produced by D-D reactions can further react with deuterium ions through secondary D-T reactions, thereby generating a small number of 14.1 MeV neutrons. Ogawa et al. performed time-resolved measurements of secondary D-T neutrons in deuterium plasmas of the Experimental Advanced Superconducting Tokamak (EAST) using a scintillating-fiber detector (Kunihiro Ogawa et al., 2025b). Related studies have also experimentally analyzed tritium burnup in EAST using neutron activation methods (Li et al., 2025). These results indicate that, even under non-D-T main-fuel conditions, secondary D-T neutron signals associated with tritium slowing-down and confinement behavior can still be observed in magnetic confinement devices.

In inertial confinement fusion, the D-T reaction is also the most central neutron source.

Laser-driven inertial confinement fusion experiments, represented by the National Ignition Facility (NIF) in the United States, mainly generate a large number of 14.1 MeV primary neutrons during the compression and burn phases through laser-driven implosion of D-T capsules. Recent experimental studies have shown that NIF has achieved the landmark breakthrough of a target gain greater than unity, further confirming the dominant role of the D-T reaction as a high-yield fusion neutron source (Abu-Shawareb et al., 2024).

In addition to the above two reactions, the deuterium-helium-3 (D-$^3$He) reaction is generally regarded as one of the advanced fusion fuel routes. The primary branch of this reaction does not directly produce neutrons, and it is therefore often referred to as a low-neutron or aneutronic fusion mode. However, in practical systems, the deuterium component in the fuel can still undergo D-D reactions and further induce secondary reactions. Therefore, the D-$^3$He system is not completely neutron-free; rather, it is characterized by a total neutron yield significantly lower than that of the D-T system, with secondary neutrons serving as the dominant neutron component (Morandi et al., 2026).

Neutron sources in fusion reactions exhibit a clear hierarchy. The 14.1 MeV neutrons produced by the D-T reaction represent the most important primary neutron source at present, whereas the 2.45 MeV neutrons produced by the D-D reaction are the most commonly encountered neutron source in many current experimental devices. In contrast, neutrons in D-$^3$He systems mainly appear as weak neutron components induced by secondary reactions.

Table 1 Neutron sources and characteristics of different fusion reaction types

| Fusion reaction type | Main reaction equation | Characteristic neutron energy | Characteristics of neutron sources |
|---|---|---|---|
| D-T | $^2H+^3H \rightarrow ^4He+n$ | 14.1 MeV | The dominant high-yield neutron source at present. |
| D-D | $^2H+^2H \rightarrow ^3He+n$ | 2.45 MeV | A common neutron source in current experimental devices; under certain conditions, it can further generate secondary D-T neutrons. |
| D-$^3$He | $^2H+^3He \rightarrow ^4He+^1H$ | The primary reaction itself produces few neutrons. | Low-neutron in nature, but not completely neutron-free. |

### 1.1.2 Neutron Characteristics in Fusion

The characteristics of fusion neutrons are first reflected in the clear correspondence between neutron energy and reaction type. For the currently most widely studied D-T fusion reaction, the characteristic neutron energy is approximately 14.1 MeV, whereas for the D-D fusion reaction, the characteristic neutron energy is approximately 2.45 MeV. Different fuel systems not only determine the neutron energy range but also directly affect fusion reactivity, energy release characteristics, and subsequent engineering loads. Du et al. systematically compared different fusion fuel systems and pointed out that, compared with the D-D system, the D-T reaction generally exhibits higher reaction efficiency and stronger neutron output. In contrast, advanced fuel systems such as D-$^{3}He$ may still produce certain neutron components under practical conditions due to accompanying D-D reactions and secondary reactions (Du et al., 2025).

Fusion neutrons also exhibit pronounced electrical neutrality. Because neutrons are uncharged, they are not directly confined by the magnetic fields used to control plasma ions and electrons. In D-T fusion, each reaction produces a neutron that can travel almost freely through the plasma until reaching the structures surrounding the vacuum chamber, where components such as the breeding blanket are located (Meschini et al., 2023). For this reason, neutrons are among the few particles that can carry reaction information from the plasma core and be directly measured by external detection systems, giving them unique value in fusion diagnostics. From the perspective of plasma diagnostics, Bielecki and Kurowski systematically discussed neutron diagnostic technologies for tokamaks and pointed out that neutron measurement is one of the most important diagnostic methods in fusion devices (Bielecki and Kurowski, 2019). In particular, neutron emission intensity, neutron energy spectra, and spatial distribution can provide key information on fusion power, ion temperature, fuel-ion behavior, plasma position, and fast-ion transport. Therefore, fusion neutrons are not only reaction products, but also important physical quantities reflecting the plasma state.

Another prominent characteristic of fusion neutrons is their strong irradiation effect, particularly the significant impact of 14.1 MeV high-energy neutrons on material service performance. Shengli Chen re-evaluated neutron-induced displacement damage cross sections for EUROFER97, a candidate structural material for fusion reactors, and pointed out that neutron irradiation damage is a critical issue in fusion reactor applications (Chen, 2022). The displacement damage evaluation results for D-T fusion neutrons were significantly higher than earlier estimates, indicating that high-energy fusion neutrons can induce stronger atomic displacement damage in materials and impose more stringent requirements on irradiation damage assessment for candidate structural materials in fusion reactors. Therefore, whether fusion structural materials can be adopted in engineering applications depends not only on irradiation damage itself, but also on the system-level qualification pathway under constraints of lifetime, operating environment, and reliability. Accordingly, high-energy fusion neutrons will directly affect material selection, lifetime assessment, and engineering deployment (Lewis et al., 2026).

The complexity of the fusion neutron environment is another important characteristic of fusion neutrons. Cohen-Tanugi et al., in their review of long-term research strategies for fusion energy materials, pointed out that future fusion reactors will face extreme and complex irradiation environments, with related challenges extending across multiple subsystems, including plasma-facing materials, blankets, magnets, sensors, and maintenance components (Cohen-Tanugi et al., 2024). This means that the effects of fusion neutrons are not limited to energy deposition or localized irradiation damage, but extend to multiple levels, including material selection, component design, lifetime prediction, and system integration in fusion devices. Therefore, in fusion reactor research, neutrons are not only important indicators of reaction performance, but also key environmental factors determining the engineering feasibility of fusion devices.

Neutrons in fusion are characterized by high energy, electrical neutrality, strong irradiation effects, and significant diagnostic value. On the one hand, they directly reflect the fusion reaction type and reaction intensity; on the other hand, they profoundly affect material behavior, structural safety, and engineering design in fusion devices. Therefore, fusion neutrons serve as a core medium linking fusion physics research with the engineering realization of fusion reactors.

### 1.1.3 Detection Methods for Fusion Reaction Neutrons

Because neutrons are electrically neutral, they cannot produce measurable signals in a detection medium through direct ionization in the same way as charged particles. Therefore, neutron detection usually relies on conversion materials that interact with neutrons and transform neutron information into secondary charged particles, fission fragments, or induced radiation signals for subsequent detection (Pietropaolo et al., 2020). In a review of neutron detection technologies, A. Pietropaolo et al. pointed out that the essence of neutron detection lies in utilizing detectable secondary signals generated through neutron-matter interactions. In nature, this is a conversion process from neutrons to charged particles or from neutrons to electromagnetic radiation, which also constitutes the key distinction between neutron detection and other forms of radiation detection. According to the different interaction mechanisms between neutrons and atomic nuclei and the corresponding signal formation processes, neutron detection methods can generally be classified into four categories: the nuclear reaction method (Saengkaew and Ploykrachang, 2026), nuclear recoil method (C. Wang et al., 2025), nuclear fission method, and neutron activation method (Ruddy et al., 2022), as shown in Fig. 1

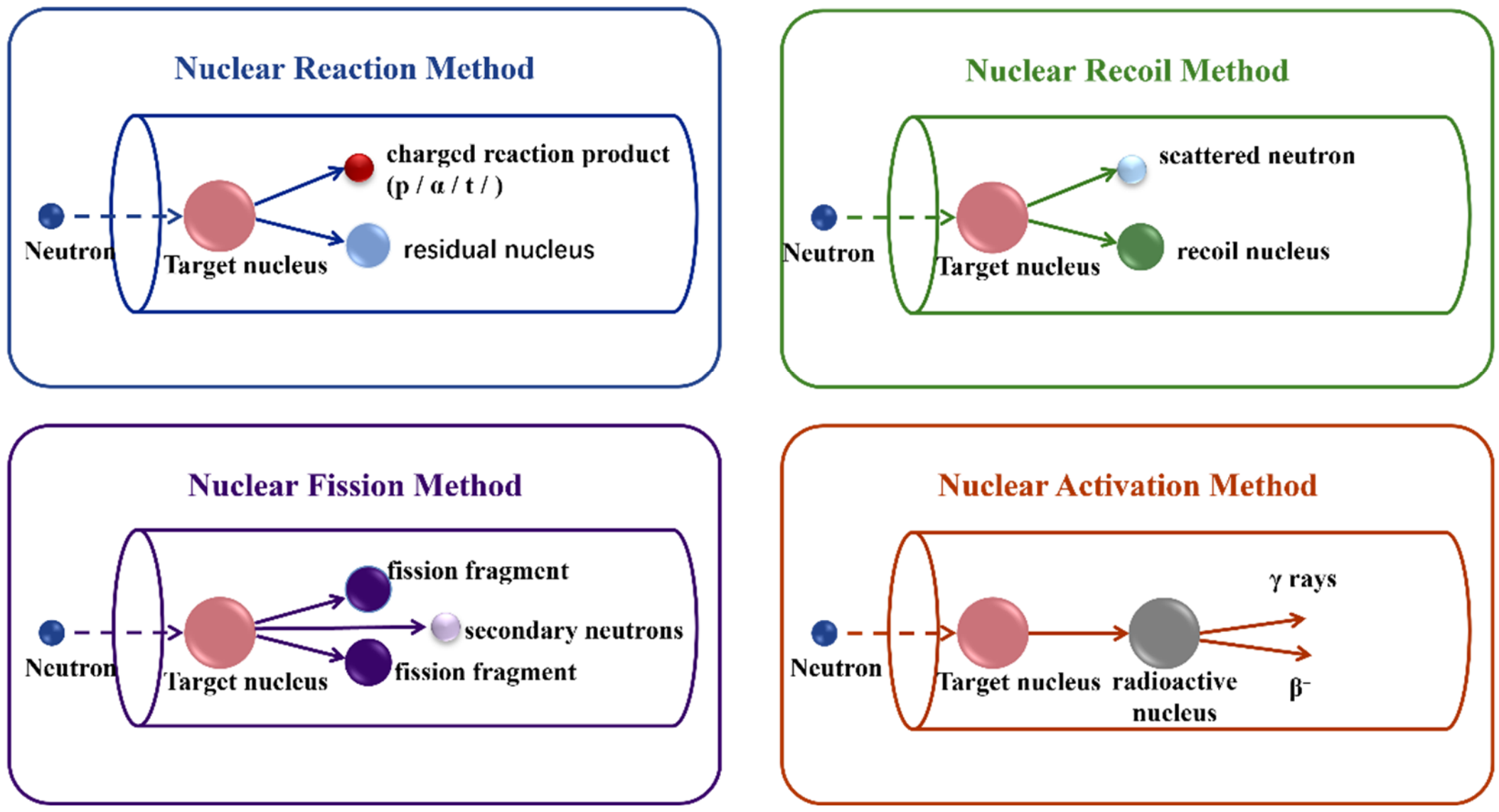

Fig. 1. Principles of the Four Neutron Measurement Methods

The nuclear reaction method mainly utilizes neutron capture or conversion reactions between neutrons and specific nuclides to generate directly detectable charged particles or secondary radiation signals, thereby enabling neutron measurement (Pietropaolo, 2023). This method typically relies on nuclides with relatively large neutron reaction cross sections, such as $^{3}He$, $^{6}Li$, and $^{10}B$. It can be used for thermal neutron detection and can also be extended to higher energy ranges through appropriate converter structures. Pérez et al. demonstrated this by showing that similar detector devices equipped with LiF or $^{10}B$ conversion layers can be applied to thermal neutron detection (Pérez et al., 2024). Berry and Diawara further indicated in a comprehensive review that the nuclear reaction method represents one of the most fundamental and typical principles in neutron detection (Berry and Diawara, 2023).

The nuclear recoil method is based on the detection of recoil-particle signals generated by elastic scattering between neutrons and light nuclei. When neutrons collide with hydrogen nuclei or other light nuclei, part of the neutron energy is transferred to recoil nuclei. Information on the incident neutrons can then be inferred by measuring the energy, count rate, and angular distribution of recoil protons or recoil nuclei (Sangaroon et al., 2024a). Since fast neutrons are more likely to produce recoil particles with sufficiently high energies during scattering, the nuclear recoil method is particularly suitable for fast-neutron flux and spectrum measurements. Ericsson pointed out that the nuclear recoil method provides an important basis for neutron spectrum diagnostics of D-D and D-T fusion neutrons (Ericsson, 2019).

The nuclear fission method relies on neutron-induced fission of heavy nuclei and detects neutrons through fission fragments or related reaction products (Henzlova et al., 2024). Compared with the two methods described above, the nuclear fission method exploits the strong response of heavy nuclei to neutrons. The fission fragments released during fission reactions possess high kinetic energies and can therefore generate large-amplitude signals, thereby improving the resistance of the detection system to gamma background and complex electromagnetic interference. Reviews of fusion-neutron-induced fission diagnostics have shown that the nuclear fission method maintains good applicability in environments characterized by intense radiation fields, strong magnetic fields, and complex electromagnetic interference, making it particularly suitable for monitoring and diagnostics under high-neutron-flux conditions (Kono et al., 2024).

The neutron activation method induces activation reactions in selected materials, converting them into radioactive nuclides. The neutron flux, energy spectrum, or irradiation history can then be inferred by measuring the characteristic radiation emitted during radioactive decay. Compared with real-time detection methods, neutron activation generally does not directly provide instantaneous signals. Instead, it realizes an integral characterization of the neutron field through post-irradiation offline measurements. Therefore, it offers good stability and strong resistance to transient electromagnetic interference. Abdelnour pointed out in a review that neutron activation remains one of the important technical routes for neutron field characterization in complex irradiation environments (Abdelnour et al., 2025).

Overall, these methods correspond to different pathways for converting neutron information into detectable signals. Nuclear reaction methods detect charged particles produced by neutron capture or conversion reactions, while nuclear recoil methods rely on recoil particles generated through neutron scattering. Fission-based detectors use the large ionization signals produced by neutron-induced fission, and activation methods characterize neutron fields through induced radioactivity. In fusion-related neutron diagnostics, methods based on neutron-induced charged-particle production and recoil processes are particularly important for time-resolved or energy-resolved measurements, whereas activation methods are more suitable for integral neutron-field characterization (He et al., 2023; Sykora et al., 2024).

In contrast, the nuclear fission and neutron activation methods are more suitable for monitoring, calibration, or integral measurements in high-flux environments. For the neutron/gamma discrimination issue emphasized in this review, the most directly relevant detection approaches are generally those capable of producing instantaneous pulse signals. This is because neutron/gamma discrimination essentially relies on differences between the two types of radiation in interaction mechanisms, pulse formation processes, deposited energy, or response time (Shen et al., 2024). Therefore, real-time detection systems based on the nuclear reaction and nuclear recoil methods often have greater application value.

### 1.1.4 Types of Neutron Detectors

(1) Scintillation Detectors

Scintillation detectors are among the most widely used devices in neutron detection at present. Their operating principle is based on nuclear reactions or elastic scattering between neutrons and scintillator materials or converter nuclides within the scintillator, producing secondary charged particles such as recoil protons, α particles, and tritons. After depositing their energy in the scintillation medium, these particles excite luminescence centers and generate scintillation light signals, which are then converted into electrical pulses by photomultiplier tubes or other photodetectors for subsequent analysis (Koshimizu, 2025). According to the material type and physical state, scintillation detectors can be classified into crystal scintillators, liquid scintillators, and plastic scintillators. Crystal scintillators can be further divided into organic and inorganic crystal scintillators.

Organic crystal, liquid, and plastic scintillators are usually rich in hydrogen. Therefore, in fast-neutron detection, they mainly rely on elastic scattering between neutrons and hydrogen nuclei, in which recoil protons generate measurable signals. These scintillators are suitable for measuring 2.45 MeV and 14.1 MeV fusion fast neutrons (Ahnouz et al., 2024). Inorganic glass scintillators, especially $Ce^{3+}$-doped lithium silicate glass scintillators, are particularly suitable for thermal-neutron detection. Commercial lithium silicate glass scintillators such as GS20 and NE902 have been developed for thermal-neutron detection, and their relatively high light yield and superior chemical stability compared with hygroscopic $^{6}LiI$ crystals make them reliable candidates for radiation measurements (Sui et al., 2025). Compared with gas detectors and some semiconductor detectors, scintillation detectors offer advantages such as high detection efficiency, fast time response, ease of array configuration, and convenient system integration. Consequently, they show considerable application potential in fusion neutron diagnostics, radiation monitoring, and neutron cameras (An et al., 2025). However, their limitations are also evident, including signal overlap under intense gamma backgrounds, limited radiation tolerance of some materials, insufficient long-term stability, and hygroscopicity in certain inorganic crystals. In fusion environments, strong neutron fluxes, high count rates, and intense electromagnetic interference can further amplify these problems (Richards et al., 2025).

From the perspective of neutron/gamma discrimination capability, scintillation detectors are currently among the most promising approaches, although their performance varies among different scintillator types. Liquid scintillators and some organic crystals can generally achieve effective neutron/gamma discrimination using pulse-shape discrimination (PSD), because neutron- and gamma-ray-induced signals exhibit different pulse decay components. Therefore, they have long been an important technical route for fusion fast-neutron measurement. Baselga and Montbarbon investigated neutron/gamma signal processing in organic scintillators and proposed a PSD method based on robust extraction of decay-shape characteristics. Their results showed that this method can improve the stability and noise resistance of neutron/gamma discrimination in mixed radiation fields, indicating that discrimination performance depends not only on the detector material itself, but also closely on back-end waveform-processing algorithms (Baselga and Montbarbon, 2024). Duan et al. designed a compact detector for tokamak applications, in which the probe employed a BC-501A liquid scintillator coupled with a silicon photomultiplier (SiPM) readout, demonstrating the adaptability of liquid scintillators for compact deployment in fusion devices (Duan et al., 2025). Zhang et al. further applied a liquid scintillator detection system to time-resolved fusion neutron measurements on the HL-3 tokamak, enabling characterization of the evolution of fusion neutron spectra (Zhang et al., 2025). In detector studies for the Advanced Fusion Neutron Source (A-FNS), Ogawa et al. characterized scintillators such as stilbene, EJ-301, and CLYC7, and achieved neutron/gamma discrimination using the charge-comparison method. These results indicate that scintillator-based systems have clear discrimination potential in fusion neutron sources and intense mixed radiation fields (Kunihiro Ogawa et al., 2025a). To illustrate the neutron/gamma discrimination capability of typical scintillation detectors in fusion neutron source-related scenarios, the two-dimensional PSD distributions obtained with stilbene, EJ-301, and CLYC7 detectors during D-D neutron measurements are shown in Fig. 2.

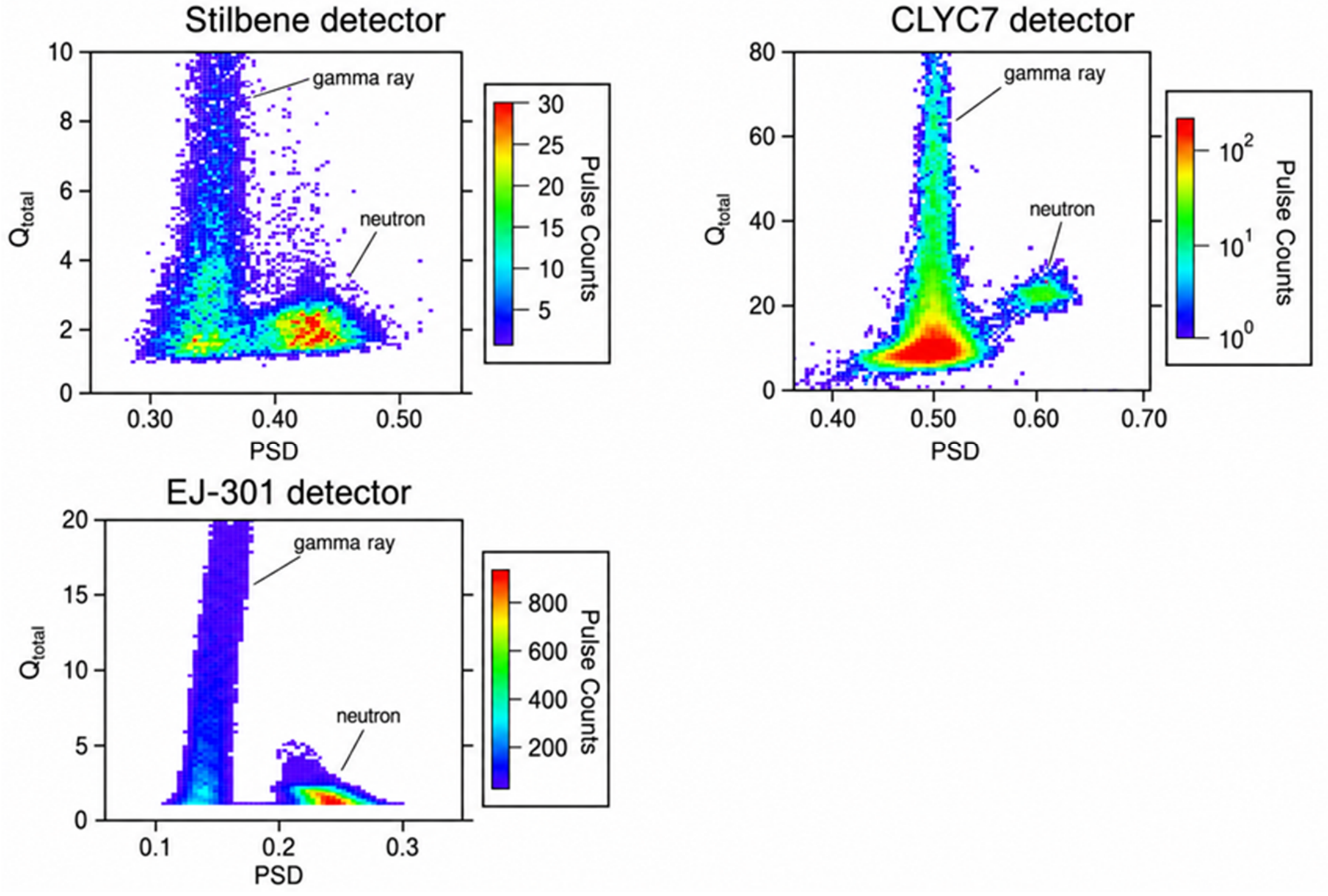

Fig. 2. PSD analysis results of stilbene, EJ-301, and CLYC7 scintillation detectors in D-D neutron measurements (Kunihiro Ogawa et al., 2025a)

As shown in Fig. 2, gamma-ray events are mainly concentrated in the lower PSD region, whereas neutron events are distributed in the higher PSD region, indicating that the charge comparison method based on differences in pulse tail integration can effectively discriminate fast neutron signals from gamma-ray signals.

Among crystal scintillators, organic crystals such as stilbene are more suitable for fast-neutron detection and neutron/gamma discrimination, whereas many Li-containing inorganic scintillators rely on neutron-induced nuclear reactions of specific nuclides, such as $^{6}$Li and $^{35}$Cl, to achieve neutron response. Recent studies have shown that compact detector modules based on CLYC crystals and SiPM readout can achieve thermal neutron/gamma discrimination in a $^{241}$Am-Be mixed radiation field, with a favorable figure of merit (Yücel et al., 2025). A phoswich composite scintillator based on $^{6}LiI:Eu/LaBr_3:Ce$ has also demonstrated effective discrimination between thermal neutrons and gamma rays (Sisodiya et al., 2025). These inorganic crystal schemes indicate that crystal scintillators are not incapable of neutron/gamma discrimination; however, their advantages are mainly reflected in thermal-neutron or mixed-field measurements rather than in high-energy fusion fast-neutron detection. For MeV fast neutrons in the 0.9 – 5.2 MeV range, the direct detection efficiency of $^{6}$Li-enriched CLYC and CLLB scintillators is relatively low, about 0.1% – 0.75% (Song et al., 2023), and a significant enhancement in response typically relies on moderation or composite structures. Therefore, Li-doped scintillators are more suitable for thermal neutron monitoring, mixed-field discrimination, or as part of a composite detector, rather than being simply regarded as a primary detector for fusion fast neutrons.

Compared with liquid and crystal scintillators, plastic scintillators offer advantages such as low cost, good machinability, and ease of fabrication into large-area or fiber-based structures, making them suitable for array configuration and engineering applications. However, their energy resolution and neutron/gamma discrimination capability are generally weaker than those of liquid scintillators and organic crystals (Zaitseva, 2026). Zhang et al. conducted design and simulation studies of a scintillating-fiber detector for HL-3, demonstrating that plastic scintillating-fiber schemes offer structural flexibility, ease of channelization, and good adaptability to complex spatial layouts in D-T fusion neutron measurements (Y. S. Zhang et al., 2026). Inorganic lithium-containing crystals and lithium glass-based schemes are more suitable for enhanced thermal neutron detection, mixed radiation field discrimination, and auxiliary measurement tasks. In a $^{6}$Li-glass composite scintillator, neutron signals can be converted into scintillation light signals through the $^{6}Li(n,\alpha)^{3}H$ reaction, in which the generated charged particles deposit energy in the scintillation medium (Favalli et al., 2025). A typical working principle is illustrated in Fig. 3.

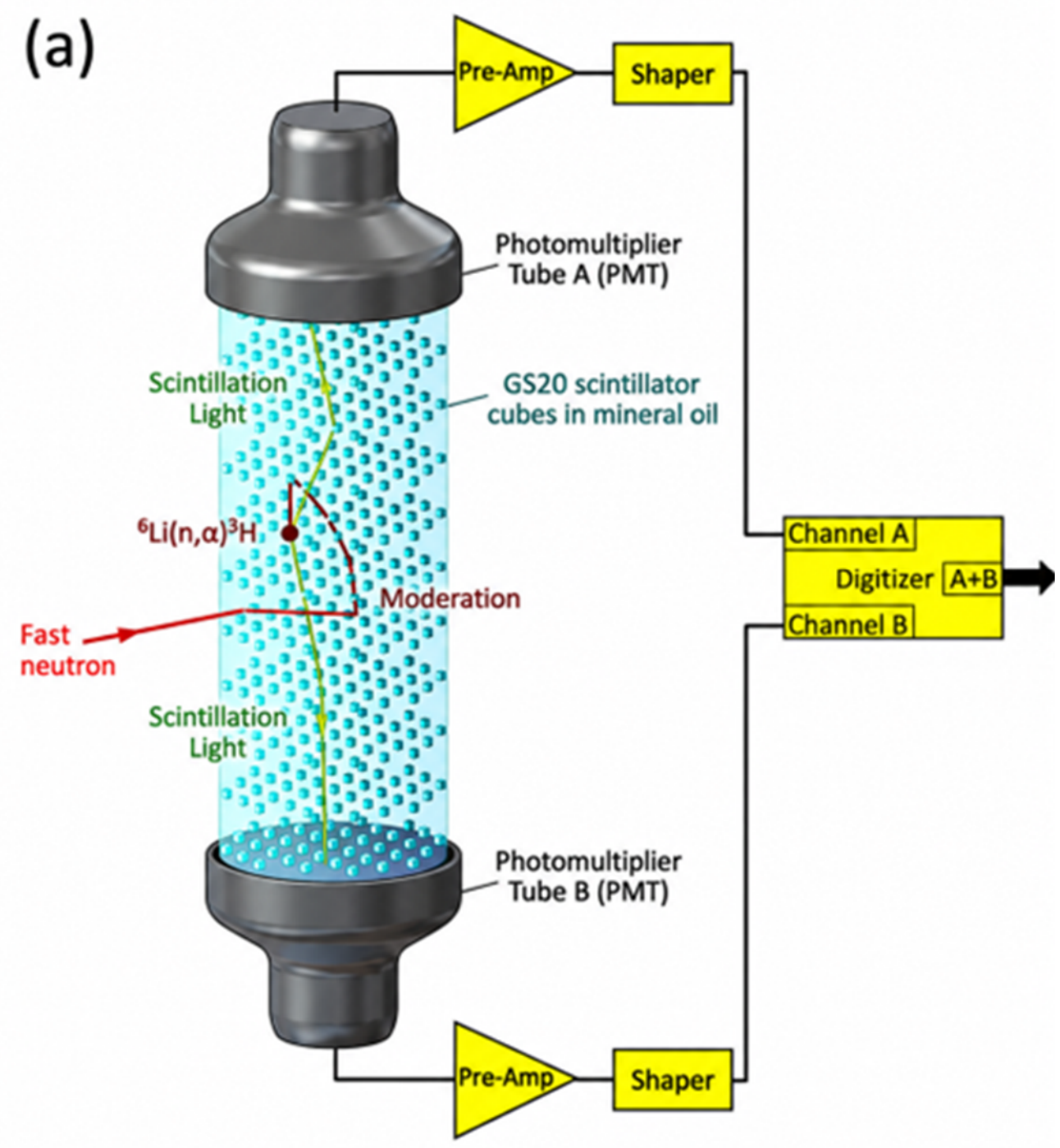

Fig. 3. Schematic diagram of the working principle of a $^{6}$Li-glass composite scintillation neutron detector (Favalli et al., 2025)

(2) Gas Detectors

Gas detectors represent another classical technical route for neutron measurement. Their operating principle is based on nuclear reactions between neutrons and the working gas or converter-layer nuclides, producing charged particles that subsequently induce gas ionization and charge collection under an applied electric field, thereby generating electrical pulse signals (Mauri et al., 2019). Typical gas detectors include $^{3}$He proportional counters, $BF_3$ proportional counters, as well as the increasingly developed $^{10}$B solid converter gas detectors and related multi-wire proportional chambers (multi-wire proportional chamber, MWPC) and gas electron multipliers (gas electron multiplier, GEM). Among them, the signal formation processes of the $^{10}$B thin-film gas detector and the $^{3}$He tube detector are illustrated in Fig. 4. One important reason why $^{3}$He and $BF_3$ gas systems have long been used for thermal-neutron detection is their relatively large thermal-neutron reaction cross sections, which can effectively increase the interaction probability between neutrons and the working medium (Mullen et al., 2024). For recently developed $^{10}$B converter-layer gas detectors, the basic principle is to introduce a boron-containing converter layer, in which neutrons generate charged particles through the $^{10}B(n,\alpha)^{7}Li$ reaction, followed by signal readout through gas multiplication structures. This mechanism also provides the physical basis for the development of boron-lined detectors, GEM-based detectors, and MWPC-based detector systems (Deng et al., 2024).

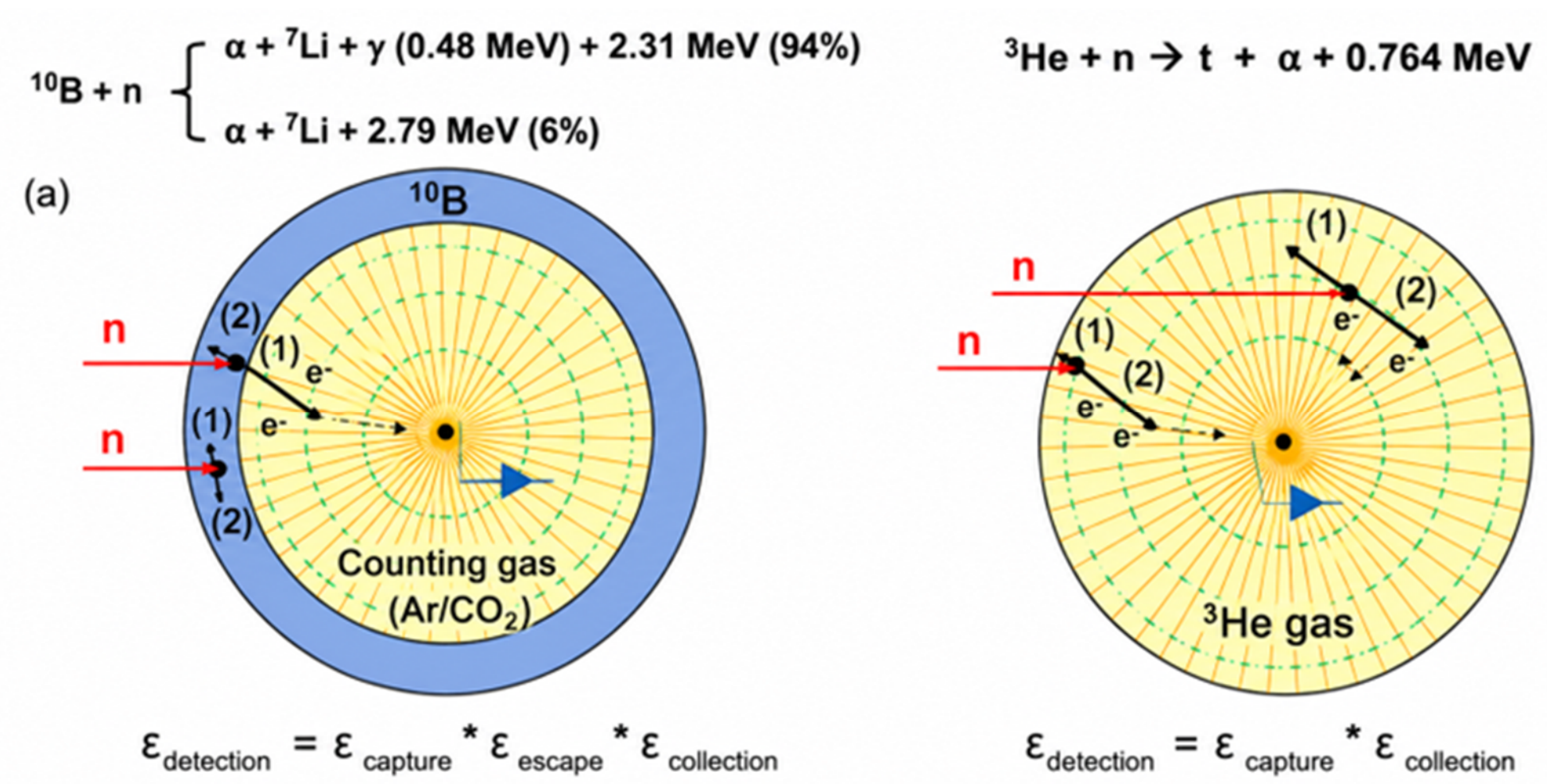

Fig. 4. Schematic diagram of the working principles of the $^{10}B$ thin-film gas detector and the 3He tube detector (Stefanescu et al., 2024)

Because these reactions directly produce charged particles with high linear energy transfer, gas detectors usually exhibit high sensitivity in thermal-neutron measurements. Among them, gas detectors based on $^{10}B$ converter layers generally show low gamma sensitivity, while sealed boron-lined MWPC devices further demonstrate good long-term stability and online monitoring capability. Therefore, they remain of significant application value in thermal-neutron monitoring and in the development of alternatives to $^{3}He$ detectors (Meng et al., 2025). The advantages of gas detectors include a well-established operating mechanism, stable output signals, effective suppression of gamma background, and suitability for long-term online monitoring. Through structural optimization, they can also provide position resolution and relatively high count-rate capability. In recent years, with the limited availability of $^{3}He$, research attention has gradually shifted toward $^{10}B$ converter layers and multilayer gas amplification structures. A review by Kreusch Filho et al. showed that recent studies on gas detectors have mainly focused on efficiency improvement, enhanced spatial resolution, increased high-flux capability, and alternatives to $^{3}He$ detectors (Kreusch Filho et al., 2025). Cancelli et al. developed a multilayer high-efficiency GEM neutron detector by combining gas electron multiplication structures with boron converter layers for thermal and epithermal neutron detection, reflecting the development of gas detectors toward higher flux capability and higher detection efficiency (Cancelli et al., 2024). Meng et al. reported a thermal-neutron detector based on a boron-lined MWPC, indicating that such detector schemes still have strong engineering application potential after structural optimization (Meng et al., 2026).

Gas detectors remain important for thermal-neutron flux monitoring, radiation protection, and beam monitoring. However, the current development of gas detectors is still mainly focused on thermal-neutron detection (Bencivenni et al., 2023). For 2.45 MeV and 14.1 MeV fusion fast neutrons, effective measurement usually requires converter layers, moderators, or specific operating modes, which limit their application to some extent in high-time-resolution and fine diagnostics of fusion fast neutrons.

(3) Semiconductor Detectors

Semiconductor detectors operate by utilizing reactions between neutrons and converter-layer materials or the semiconductor bulk itself. The resulting charged particles generate electron-hole pairs in the sensitive region of the semiconductor, and the charges are then collected under an applied electric field to produce output signals (Meng et al., 2024). Conventional semiconductor neutron detectors commonly employ Si/$^{6}$LiF structures, as shown in Fig. 5, or $^{10}$B converter-layer structures (Takada et al., 2024). In recent years, wide-bandgap semiconductors such as silicon carbide (SiC) and diamond have gradually become research hotspots for fusion neutron detection because of their radiation tolerance, low leakage current, and capability for high-temperature operation (Melbinger et al., 2024). Compared with scintillation detectors, semiconductor detectors offer advantages such as compact size, high spatial resolution, and convenient digital signal readout (Tudisco et al., 2025). However, their limitations include relatively low intrinsic detection efficiency, high requirements for material quality and fabrication processes, and possible displacement damage and performance drift under extreme irradiation environments (Zeng et al., 2026).

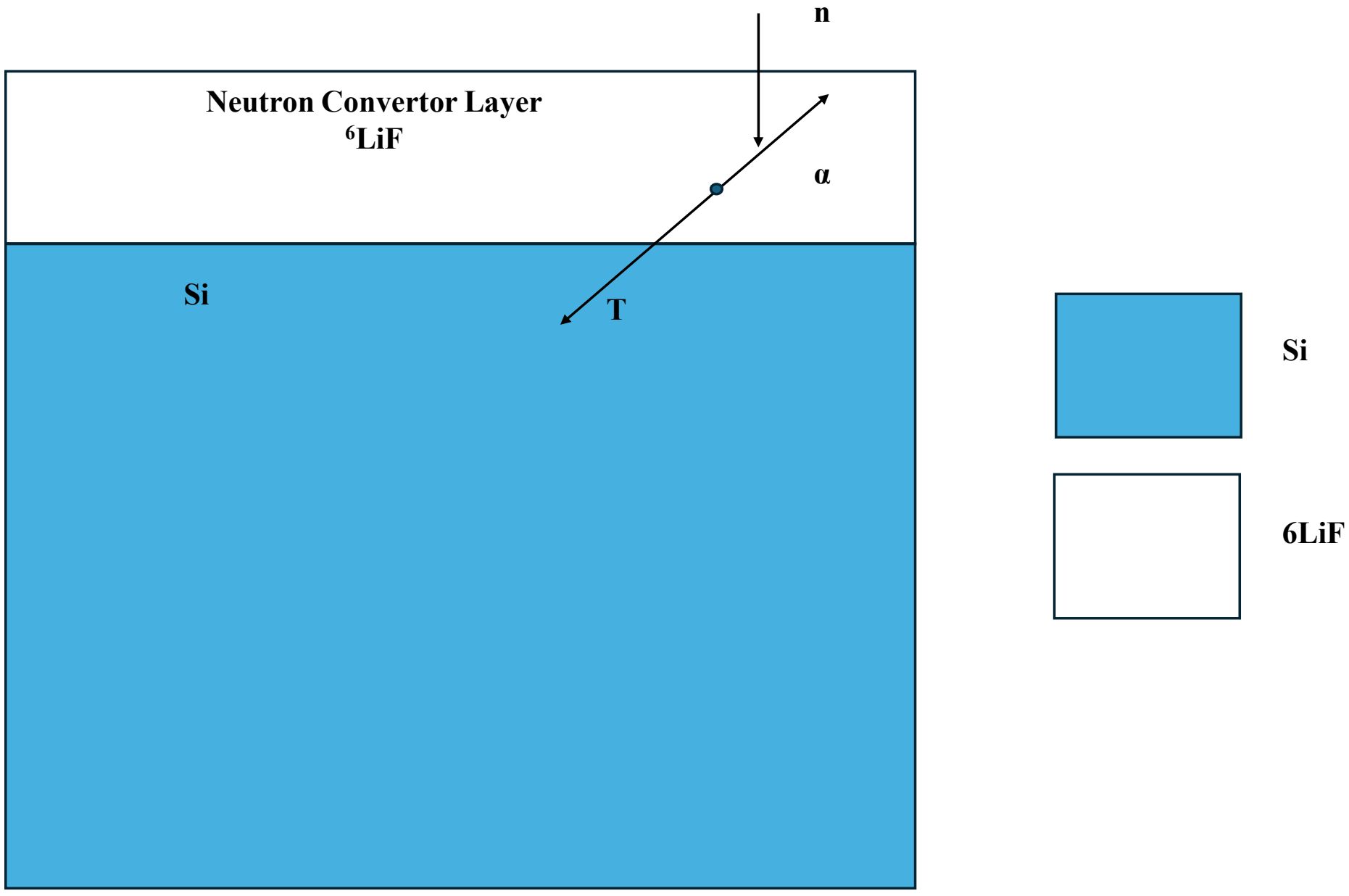

Fig. 5. Schematic diagram of a semiconductor neutron detector (Zhang H. et al., 2026)

In fusion applications, the most prominent potential of semiconductor detectors lies in compact fast-neutron spectrometry and online monitoring under intense radiation environments. Kushoro et al. simulated and experimentally verified the response of SiC detectors under tokamak neutron-spectrum conditions, indicating that SiC is a promising alternative for fast-neutron detection and spectrum measurement in future fusion devices, although its systematic application in tokamaks still requires further validation (Kushoro et al., 2025). Lin et al. directly investigated the use of diamond detectors for fast-neutron spectrum measurement on EAST, demonstrating that diamond detectors have begun to enter real fusion-device application scenarios (Lin et al., 2024).

In terms of neutron/gamma discrimination, semiconductor detectors generally do not rely on classical PSD in the same way as liquid scintillators. However, this does not mean that they lack the potential for neutron/gamma separation. In a study of single-crystal chemical vapor deposition (CVD) diamond detectors, Kobayashi et al. extracted fast-neutron events from a high-dose mixed radiation field using pulse-shape and pulse-width analyses, and further reconstructed the fast-neutron energy spectrum (Kobayashi et al., 2025). This indicates that, for fast-response semiconductor devices such as diamond detectors, although the discrimination mechanism differs from the slow-fast component separation used in scintillators, neutron and gamma backgrounds can still be distinguished to some extent through waveform-feature processing. Therefore, the value of semiconductor detectors in fusion research is more appropriately summarized as having potential for neutron/gamma separation, while their current role is more often reflected in mixed-field signal selection and fast-neutron spectrometry rather than in mature large-scale real-time neutron/gamma discrimination.

(4) Other Detectors

In addition to the three mainstream detector types discussed above, several neutron detectors designed for more specialized scenarios are also used in fusion research, such as self-powered neutron detectors (SPNDs) and solid-state nuclear track detectors, including CR-39 detectors. These detectors are usually not intended primarily for high-time-resolution neutron/gamma discrimination, but instead emphasize long-term online monitoring, integral measurements, or beam-profile analysis.

The basic principle of an SPND is that neutron interactions with the detector emitter produce β particles or induce secondary charges, thereby generating a weak current signal without an external power supply. Its advantages include simple structure, compact size, and good durability, making it suitable for long-term online monitoring (Lanza and Cao, 2026). Álvarez et al. investigated SPNDs for the International Fusion Materials Irradiation Facility-DEMO-Oriented Neutron Source (IFMIF-DONES), showing the feasibility of using SPNDs for online beam monitoring and diagnostics in fusion neutron source environments (Álvarez et al., 2025). However, the output signals of such detectors are usually weak, and their signal formation mechanisms are relatively complex. Therefore, they are more suitable for flux or operational-state monitoring than for real-time neutron/gamma discrimination.

Solid-state nuclear track detectors such as CR-39 achieve integral detection by recording the tracks of charged particles generated after neutron conversion. Their advantages include simple structure, strong resistance to electromagnetic interference, and suitability for irradiation integral measurements and spatial distribution analysis (Akino et al., 2026). However, their main limitation is also evident: they cannot provide a truly real-time readout. Simoni et al. used CR-39 dosimeters to image neutron beam profiles in the energy range of 100 keV – 10 MeV (Simoni et al., 2025). D'Amico et al. further enhanced the detection efficiency of CR-39 for Am-Be, $^{252}$Cf, and D-T neutron sources by using boron nitride coatings (D'Amico et al., 2026). Therefore, these detectors are more suitable for beam distribution measurements, fluence integration, and offline diagnostics in fusion research, rather than serving as primary detectors for neutron/gamma discrimination.

The differences among neutron detectors essentially reflect differences in the neutron-to-signal conversion mechanisms. For neutron/gamma discrimination in fusion applications, liquid scintillators and some organic crystals remain the most practically valuable routes at present (Karmakar et al., 2025). Inorganic Li-containing scintillators are more suitable for thermal-neutron-enhanced detection and mixed-field discrimination (Feng et al., 2026). Gas detectors are more appropriate for thermal-neutron monitoring and low-gamma-background scenarios. Semiconductor detectors show important potential in compact design, radiation tolerance, and fast-neutron spectrometry (Melbinger et al., 2026). Other detectors, such as SPNDs and CR-39 detectors, are more suitable for long-term monitoring and integral measurements. These differences also indicate that, when neutron measurement and neutron/gamma discrimination are carried out in fusion devices, different detector types often need to be selected according to specific diagnostic tasks, or multi-detector collaborative schemes need to be constructed.

## 1.2 Neutron Flux Measurement

Neutron flux measurement is one of the fundamental components of fusion diagnostics, and its results are directly related to the reliability of fusion power evaluation, neutron emissivity reconstruction, and the diagnosis of relevant physical parameters (Esposito et al., 2022). For fusion devices, flux measurement is not merely a counting problem, but rather a problem of extracting true neutron signals under conditions of high count rates, intense mixed radiation fields, and complex electronic environments. Recent reviews of fusion diagnostics and studies on device design have shown that neutron flux measurement systems generally need to simultaneously address a wide dynamic range, resistance to gamma interference, time resolution, and cross-calibration capability with other neutron diagnostic systems (Wang et al., 2024).

### 1.2.1 Total Neutron Flux Measurement Based on Fission Chambers and Flux Monitors

In fusion devices, total neutron flux monitoring is usually centered on fission-chamber-based detectors. The basic principle is that neutrons react with fissile materials deposited on the electrode surface to produce high-energy fission fragments. These fragments deposit a large amount of energy in the working gas and generate electron-ion pairs, which are then collected under an applied electric field to produce pulse or current signals. Compared with the relatively weak electronic response induced by gamma rays in the gas chamber, fission fragments are heavy charged particles with high linear energy transfer and therefore produce signals with significantly larger amplitudes. This means that fission chambers do not rely on typical pulse-shape discrimination, as liquid scintillators do, but mainly achieve interference suppression through the inherent amplitude difference between neutron-induced fission signals and gamma backgrounds.

This is also an important reason why fission chambers have long been used as a main technical route for total neutron flux monitoring in fusion devices (Kovalev et al., 2023).

Recent device design studies have demonstrated that the value of fission chamber-based neutron flux monitors in fusion applications lies in their wide dynamic range and strong adaptability to high-background environments. An overview of the neutron diagnostic system for the compact high-field tokamak SPARC indicates that its flux monitoring subsystem consists of approximately 15 detectors, mainly including ionization chambers and proportional counters, for continuous monitoring of neutron yield (Raj et al., 2024). Corresponding neutronics simulations further compared the design schemes of $^{10}$B gamma-compensated ionization chambers and parallel-plate $^{238}$U fission chambers to simultaneously satisfy the requirements of neutron flux response, gamma background suppression, and engineering feasibility (Wang et al., 2024). To further illustrate the engineering arrangement of total neutron flux monitors in fusion devices, Fig. 6 presents a schematic layout of the SPARC neutron flux monitor array.

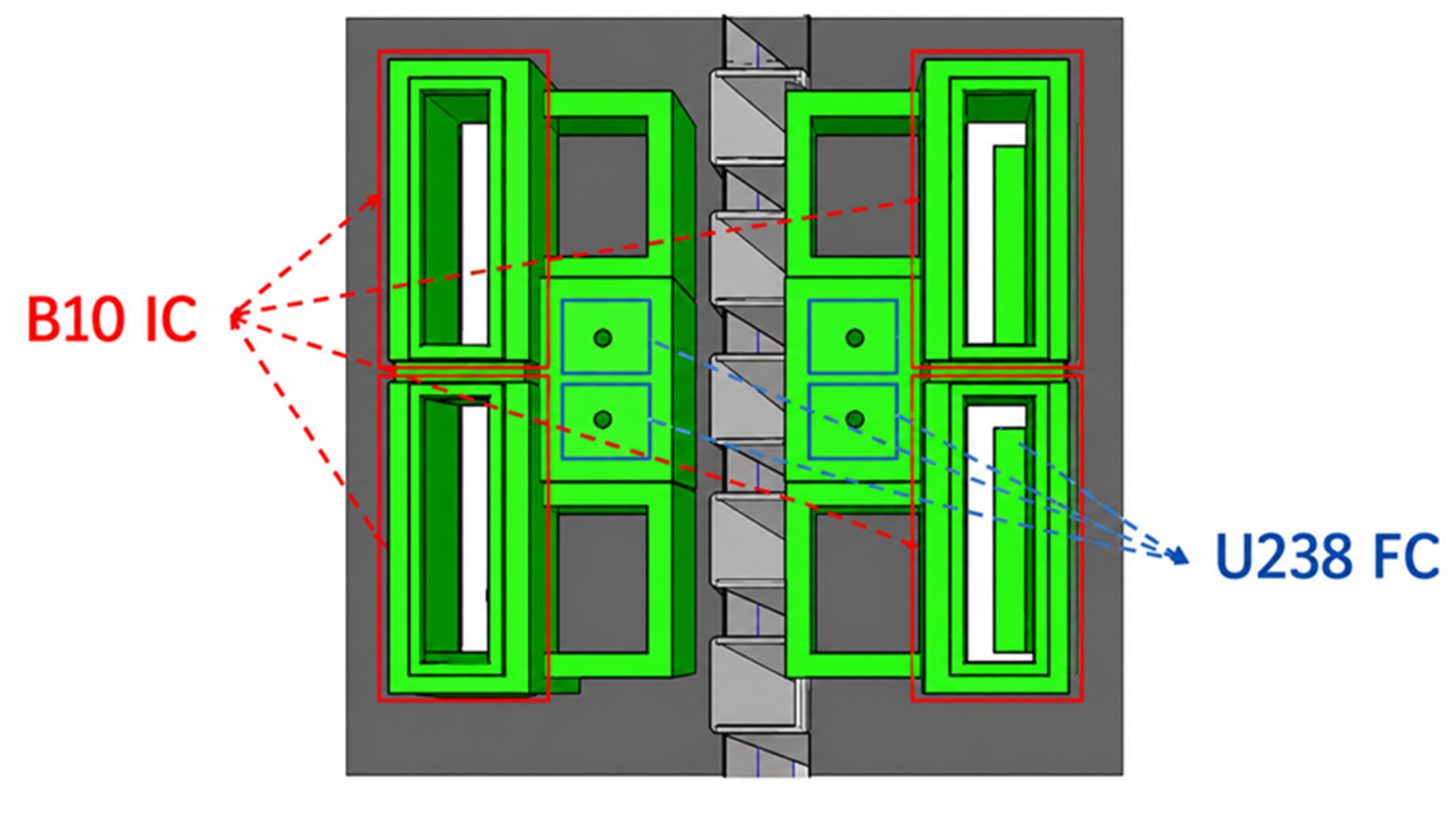

Fig. 6. Schematic layout of the SPARC neutron flux monitor array (Wang et al., 2024)

As shown in Fig. 6, the SPARC neutron flux monitors adopt an array-based arrangement close to the hall wall, while the incident neutron field is regulated through collimation structures and hydrogen-containing moderating/shielding materials. The combination of $^{10}$B gamma-compensated ionization chambers and $^{238}$U fission chambers helps simultaneously satisfy the requirements of neutron response over different energy ranges, wide dynamic range monitoring, and gamma background suppression, reflecting the development trend of total neutron flux monitoring in fusion devices from single-detector configurations toward multi-detector collaborative arrangements. DTT, which is currently under construction, provides a representative example of how neutron monitoring for future fusion devices must combine detector selection with calibration strategy. As discussed above, its Neutron Yield Monitor system adopts multiple detector types to cover different neutron-yield ranges and neutron-energy components (Marocco et al., 2026). Anagnostopoulou et al. further investigated the preliminary calibration procedure for the Phase 1 neutron and gamma-ray diagnostics of DTT, showing that the Neutron Yield Monitors (NYM) and Neutron Activation System (NAS) require absolute calibration using $^{252}$Cf sources and compact D-T neutron generators, together with MCNP-based correction factors and staged D-D/D-T calibration schemes (Anagnostopoulou et al., 2026). This indicates that, for online monitoring in fusion devices, neutron/gamma discrimination must be considered together with detector calibration and system-level response correction, so that neutron-yield measurements remain reliable under complex mixed-radiation and device-geometry conditions. These studies indicate that, in current fusion flux-monitoring designs, fission-chamber-based detectors remain central to total flux monitoring, while their neutron/gamma discrimination capability is mainly reflected in their ability to stably extract neutron signals under strong gamma backgrounds.

Recent studies have also improved neutron flux detector structures using Monte Carlo methods to enhance detection efficiency and response performance (He et al., 2026). Wen et al. conducted neutronics modeling and physical analysis of the neutron yield measurement system for the HL-3 tokamak. Their results showed that detector arrangement, shielding-structure design, and response-function optimization directly affect the accuracy of neutron yield and flux measurements, indicating that flux-monitoring systems in fusion devices must be co-designed with the specific device geometry and radiation environment (Wen et al., 2026). In future fusion reactors, the development of fission-chamber-based flux monitors should focus not only on high-temperature tolerance and miniaturization, but also on maintaining long-term stable operation under strong gamma backgrounds through the combined optimization of detector structure, electronics design, and calibration strategies.

#### 1.2.2 Flux Distribution Measurement Based on Neutron Cameras

Unlike fission chambers, which are mainly used for total flux monitoring, neutron cameras are primarily intended to obtain the spatial distribution of neutron flux. Their operating principle combines a collimation system with position-sensitive detectors. Neutrons from different lines of sight are first angularly restricted by collimators, and the line-integrated signals along each sightline are then recorded by detector arrays. Finally, neutron emission profiles in the plasma are obtained through inversion or tomographic reconstruction (Mazon et al., 2025). This information is directly important for analyzing fast-ion transport, beam-ion confinement, and the spatial distribution of fusion power.

In terms of neutron/gamma discrimination, neutron cameras depend more on system-level background suppression than single-point flux monitors. According to the 2025 prototype characterization results for the ex-port subsystem of the ITER radial neutron camera, the newly developed $^{4}$He scintillation detector exhibited good linear response under different neutron energy and flux conditions, and a clear neutron/gamma separation map was obtained, as shown in Fig. 7. These results indicate that the detector possesses a certain neutron/gamma discrimination capability while still satisfying high-flux measurement requirements (Cesaroni et al., 2025). This suggests that, for neutron cameras, spatial flux distribution measurement and neutron/gamma discrimination must be considered simultaneously in the system design.

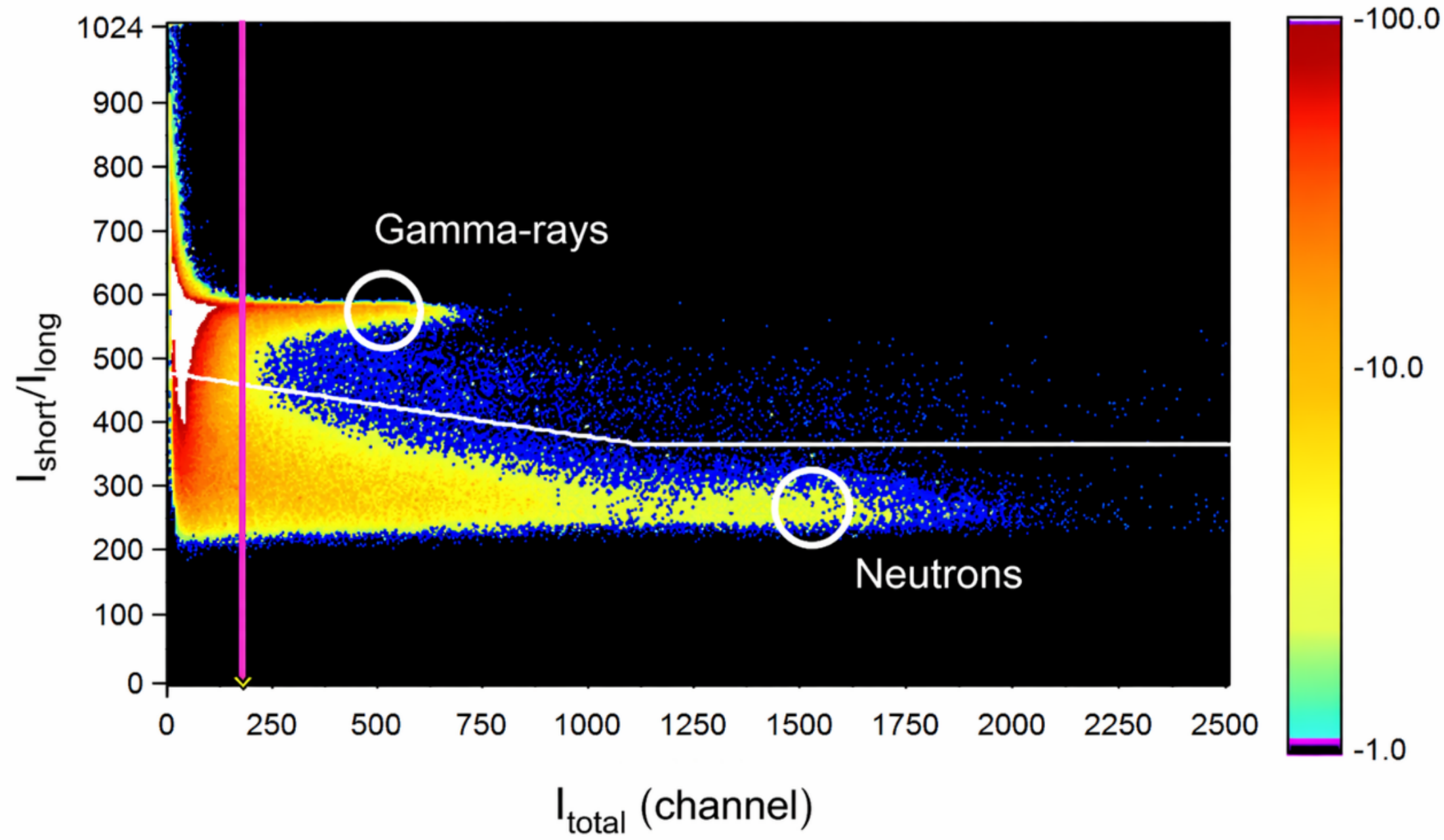

Fig. 7. Neutron/gamma-ray separation map obtained with a $^{4}$He detector (Cesaroni et al., 2025)

Beyond ITER, more direct operational examples have also been reported in other operating magnetic confinement devices. A 2026 study on the Large Helical Device (LHD) showed that three neutron camera systems have been developed for this facility. One system is based on stilbene scintillation detectors, operates in pulse-counting mode, and provides pulse-shape discrimination capability, making it suitable for discharges with medium to high neutron emission rates. The other two systems are based on EJ-410 fast-neutron scintillators and operate in current mode for discharges with low to medium emission rates (Ogawa et al., 2026a). To further demonstrate the structural configuration of neutron cameras in practical fusion devices, Fig. 8 presents a schematic diagram of the collimator, detector array, and readout system of the VNC1 vertical neutron camera on LHD.

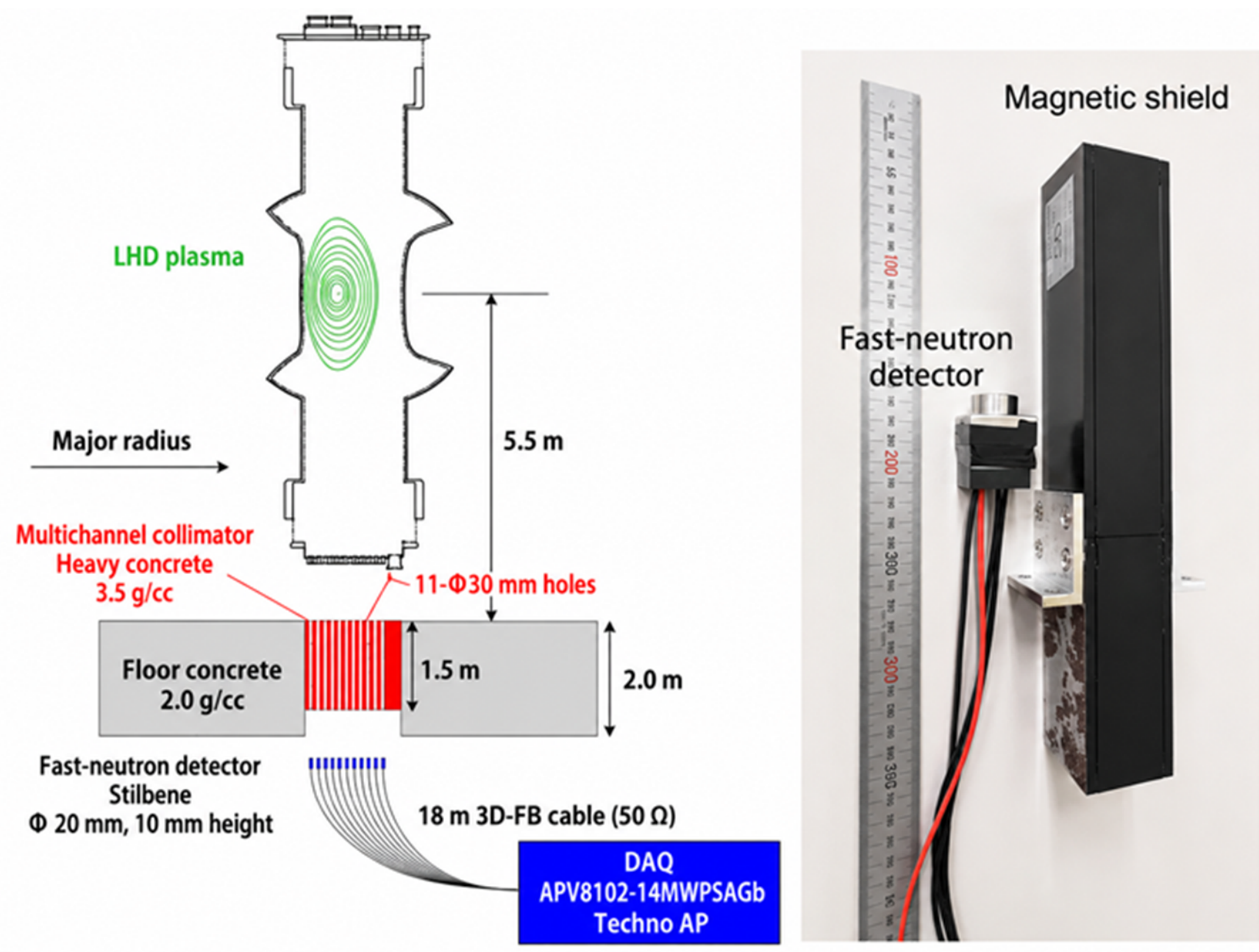

Fig. 8. Schematic diagram of the VNC1 vertical neutron camera structure and stilbene fast neutron detector on LHD (Ogawa et al., 2026a)

The neutron camera employs a multi-channel collimator system to define different lines of sight, while fast neutron detector arrays are arranged behind the collimators to obtain line-integrated neutron signals at different radial positions. Compared with single-point neutron flux monitors, neutron cameras can not only provide information on neutron emission intensity, but also further reconstruct the neutron emission profile of the plasma. Therefore, they are more suitable for diagnostic tasks related to fast-ion confinement, beam-ion transport, and the spatial distribution of fusion power.

These results indicate that relatively mature neutron camera application systems have already been established in existing fusion devices. They also further demonstrate that neutron/gamma discrimination in neutron cameras is not limited to a single detector type, but instead adopts different implementation routes depending on discharge conditions, count-rate levels, and diagnostic tasks.

#### 1.2.3 Long Counters and Their Applicability Boundaries in Fusion

Long counters usually do not serve as primary diagnostic instruments for fusion plasmas, but they remain important for neutron flux calibration and reference measurements. A typical long counter consists of a $BF_3$ or $^{3}He$ proportional counter placed at the center of a moderator, so that fast neutrons are first slowed down in a moderator such as polyethylene and then converted into measurable signals through thermal-neutron capture reactions (Ali et al., 2024; Matsumoto et al., 2024). From the perspective of the operating mechanism, $BF_3$ or $^{3}He$ counters are commonly used in long counters precisely because $^{10}B$ and $^{3}He$ have large thermal-neutron capture cross sections, enabling efficient neutron-to-signal conversion after moderation. For this reason, the overall response of a long counter strongly depends on the moderation process and structural parameters (Matsumoto et al., 2024).

The main advantages of long counters lie in their mature structure, stable operation, low gamma sensitivity, and relatively flat response over a broad energy range after appropriate design. Therefore, they are often used as secondary standards for neutron fluence-rate measurements or as instruments for source-strength calibration (Li et al., 2021). However, these operating characteristics also define their applicability boundaries in fusion diagnostics. Since long counters essentially rely on a signal formation process involving neutron moderation followed by thermal-neutron capture, the initial energy information of incident neutrons is significantly degraded during moderation. As a result, they are more suitable for neutron source-strength calibration, fluence-rate measurement, and auxiliary reference measurements, rather than for directly distinguishing 2.45 MeV neutrons from D-D reactions and 14.1 MeV neutrons from D-T reactions, let alone independently supporting deuterium-tritium ratio diagnostics.

Neutron flux measurement in fusion cannot be accomplished by a single detector alone, but should be hierarchically designed according to specific diagnostic objectives. Fission chambers and compensated flux monitors are suitable for total neutron flux and fusion power monitoring, with their core advantage lying in their strong resistance to gamma background based on reaction mechanisms and signal-amplitude differences. Neutron cameras are suitable for measuring the spatial distribution of neutron flux, and their effectiveness depends on the coordinated optimization of collimation and shielding design, together with detector-level neutron/gamma discrimination capability. Long counters, in contrast, are more appropriate for auxiliary calibration and reference fluence-rate measurements. Therefore, when discussing neutron flux measurement in fusion devices, the key issue is how to integrate total flux monitoring, spatial distribution measurement, and neutron/gamma discrimination into a coordinated diagnostic system for different measurement tasks (Tinguely et al., 2019).

### 1.3 Neutron Spectrum Measurement

The core objective of neutron spectrum measurement is to obtain neutron energy distribution information and, on this basis, infer key parameters such as fusion reaction type, fuel composition, and ion temperature. This is important for fusion research because different fusion reactions correspond to different characteristic neutron energies. Therefore, the relative intensities of different energy components not only reflect the proportions of different reaction channels, but can also be further correlated with fuel-ion ratios and fusion power levels (Pankratenko et al., 2023). Jeet et al. conducted diagnostic studies of upscattered D-T neutrons generated by burning plasmas at NIF, demonstrating that different energy components in the neutron spectrum carry important information on burn conditions and reaction kinetics. This further indicates that neutron spectrum measurement not only serves energy distribution reconstruction, but can also be used to diagnose fusion physics processes (Jeet et al., 2024). Neutron spectrum diagnostics can provide information on ion temperature and fuel-ion ratios, while time-integrated activation measurements provide an important basis for fusion yield, fusion power, and neutronics-related validation (Maggi et al., 2024; X. Wang et al., 2025).

### 1.3.1 Time-of-Flight Method

The time-of-flight (TOF) method is one of the most classical methods for neutron spectrum measurement. Its principle is that when neutrons emitted from a known position are recorded by a detector after traveling along a flight path of length L, the neutron energy can be derived from the neutron velocity determined by the measured flight time t. Therefore, this method is essentially a direct spectrum diagnostic technique based on velocity measurement (Moore et al., 2023). Fig. 9 shows the variation of the DT neutron signal temporal broadening at different detection distances.

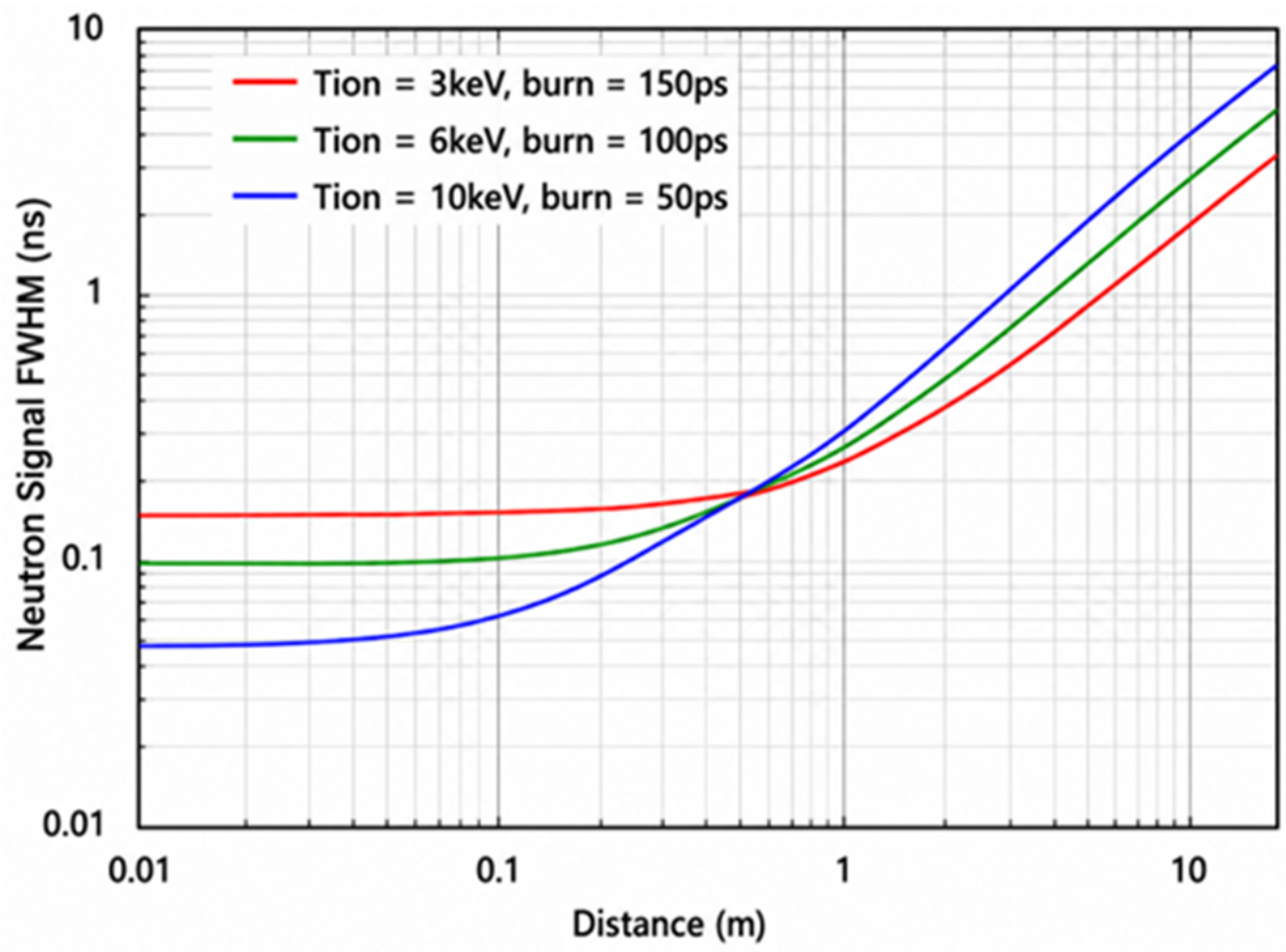

Fig. 9. Schematic diagram of DT neutron signal temporal broadening at different flight distances (Moore et al., 2023)

When the detector is located at a relatively short distance, the measured neutron signal width is mainly determined by the source emission duration and the response function of the detection system. As the flight distance increases, the flight-time differences among neutrons with different energies gradually become larger, and the contribution of neutron energy distribution to temporal signal broadening becomes more pronounced. Therefore, a longer flight path is beneficial for improving neutron energy spectrum resolution, but it also reduces detection efficiency and increases the complexity of system arrangement.

For fusion devices, the main advantage of TOF is its ability to directly resolve the peak position and broadening of fast-neutron spectra. It is sensitive to ion temperature, fast-ion slowing-down behavior, and variations in D-D and D-T reaction components, and can therefore be used for diagnostics related to fuel ratio and fusion power (Schwemmlein et al., 2022). Its limitations are that it generally requires a well-defined time reference, a relatively long flight path, and effective collimation and shielding conditions. Therefore, TOF is more suitable as a high-resolution neutron spectrometer, rather than as the sole online power monitor under continuous long-pulse operating conditions (Su et al., 2025).

### 1.3.2 Recoil Proton Method

The recoil proton method is one of the most representative and engineering-adaptable technical routes for fast-neutron spectrum measurement in fusion applications (Sangaroon et al., 2024b). Its principle is based on elastic scattering between neutrons and hydrogen atoms in hydrogen-containing materials, whereby part of the incident neutron energy is transferred to recoil protons. The neutron energy spectrum can then be unfolded by measuring the energy distribution of the recoil protons (Gerenton et al., 2024). Under ideal n-p elastic scattering conditions, there is a clear kinematic relationship between recoil-proton energy and incident-neutron energy. Therefore, this method is naturally suitable for MeV-level fast-neutron measurements under typical fusion conditions and can cover the main energy ranges relevant to D-D and D-T fusion neutron diagnostics (Liao et al., 2024). Compared with the time-of-flight method, the recoil proton method is less dependent on very long flight paths and strict timing references, making compact system design more feasible (Scholz et al., 2026).

The applications of the recoil proton method can be broadly divided into two categories. The first category uses liquid scintillators or other hydrogen-rich detection media to record recoil-proton pulse-height spectra, followed by neutron spectrum reconstruction using response functions and unfolding algorithms. The second category is the magnetic proton recoil spectrometer (MPR), which usually employs a converter target, such as a polyethylene foil, to convert neutrons into recoil protons and then analyzes the proton momentum using a magnetic field, thereby providing higher-resolution spectroscopic information. At present, thin-foil recoil proton spectrometers have been explicitly proposed as an important option for high-resolution neutron spectrum diagnostics in ITER. Performance studies have shown that such systems can be designed and optimized for tasks including fuel-ion ratio measurement, ion temperature diagnosis, and fine-structure measurement of fusion neutron spectra (Dankowski et al., 2025). Further studies on SPARC indicate that both the ion-optical design of MPR systems and the optimization of hodoscope structures are evolving toward high-resolution and high-flux measurements under compact port constraints. This suggests that the MPR route not only offers strong physics diagnostic capability but also has the potential to be extended toward engineering applications in future fusion devices (Mackie et al., 2024; Rosa et al., 2024).

In addition to high-resolution magnetic-analysis approaches, compact scintillator-based fast-neutron spectrometry schemes have also been developed rapidly. For 2.45 MeV neutron measurements in environments with intense stray magnetic fields, compact detectors based on EJ-276D scintillators and customized readout systems have been developed with particle-discrimination capability and magnetic-field insensitivity. This indicates that miniaturized fast-neutron measurement schemes for near-device deployment in fusion facilities are evolving from laboratory-scale detector studies toward device-adapted system design (Molin et al., 2024). Studies on the energy response of scintillators such as $Cs_2LiYCl_6$:Ce (CLYC) have also provided detector-level foundations and response-function support for D-D fusion fast-neutron spectrometry, suggesting that compact scintillator-based neutron spectrometers offer good deployment flexibility and strong potential for system integration in future fusion devices (Xu et al., 2024). To more intuitively illustrate the response characteristics of CLYC-type scintillators in fast-neutron spectroscopy, Fig. 10 presents the response matrix of a CLYC scintillator to fission-spectrum fast neutrons under proton-discrimination conditions. It can be seen that, after excluding the contributions of reactions involving heavier charged particles, the fast-neutron response matrix of CLYC becomes significantly simplified, thereby providing a clearer response basis for compact scintillator-based fast-neutron spectrometry.

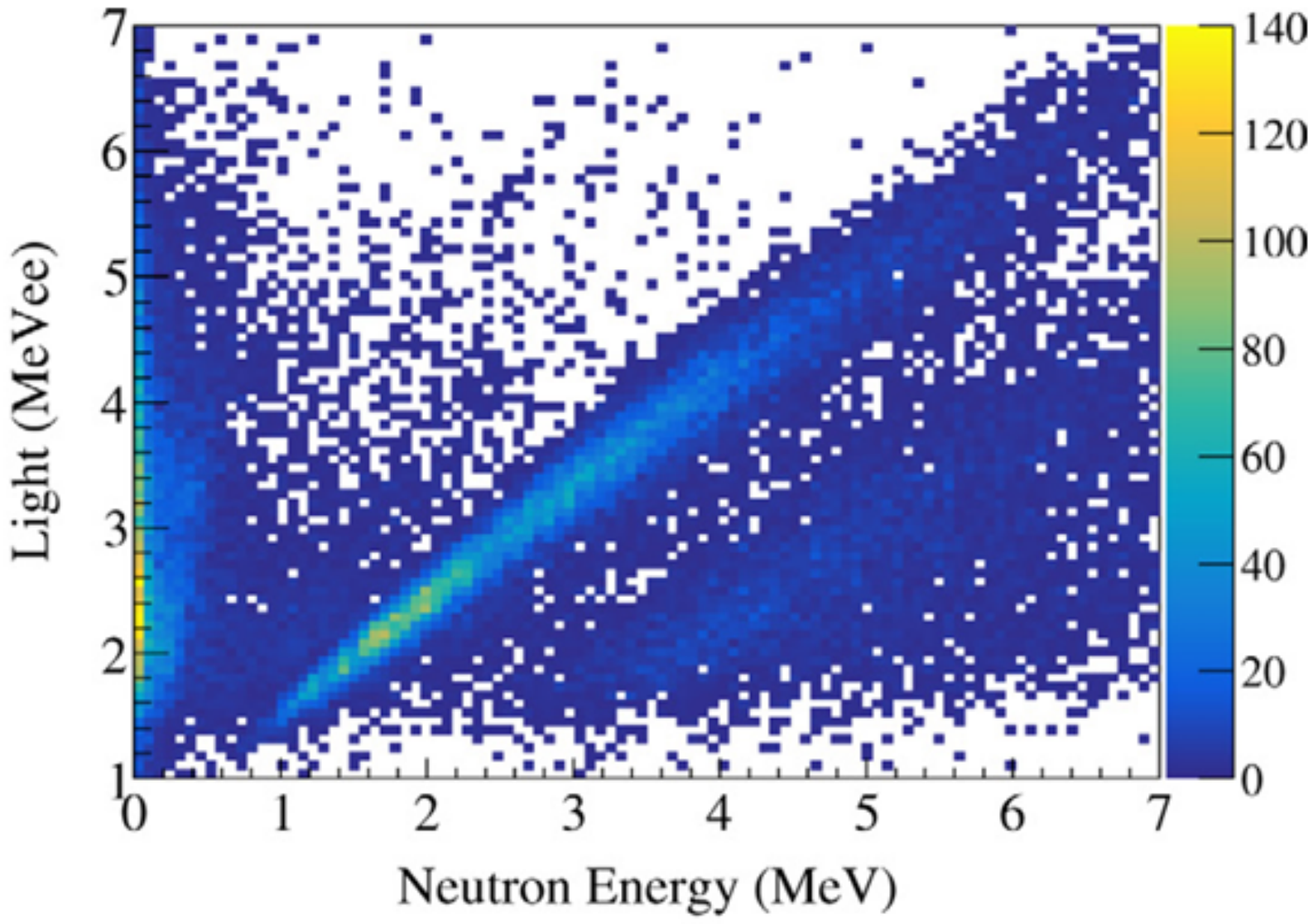

Fig. 10. Response matrix of a CLYC scintillator to fission-spectrum fast neutrons under proton-discrimination conditions (Brown et al., 2024)

The recoil proton method combines a well-defined physical basis with strong engineering adaptability. It can support D-D and D-T neutron spectrum measurements and further contribute to the diagnosis of key fusion parameters, including fuel ratio, ion temperature, and fast-ion behavior.

#### 1.3.3 Multi-sphere Spectrometer Method

The Bonner sphere spectrometer (BSS) is one of the most classical integral methods for neutron spectrum unfolding. Its principle is based on the different moderating capabilities of spheres with different diameters for neutrons of different energies. A set of count results with different energy responses is obtained, and the neutron energy spectrum is then unfolded from the multi-sphere count data using a response matrix and unfolding algorithms. Therefore, this method essentially relies on multiple response functions and mathematical inversion, rather than direct event-by-event energy measurement (Qiao et al., 2025). This feature makes the BSS more suitable for broad-energy-range neutron field characterization than for high-time-resolution core spectrum diagnostics in a single plasma discharge.

From the perspective of fusion applicability, although the BSS can, in principle, cover a broad energy range and can be used for peripheral neutron-field characterization, shielding design validation, and radiation protection assessment (Mayer et al., 2024), it usually relies on sequential measurements with multiple spheres and response-matrix unfolding. As a result, its time resolution is limited, and the results are strongly dependent on the accuracy of the response functions and the stability of the unfolding algorithms. In addition, conventional multi-sphere systems also have certain limitations in terms of mobility and measurement convenience (Tursinah et al., 2025). Hu et al. measured neutron spectra at two locations in the EAST experimental hall using a BSS based on a $^{3}$He thermal-neutron counter, and compared the experimental results with MCNP5 calculations. Their results indicate that this method is more suitable for neutron spectral-field measurement in the device hall, shielding adequacy evaluation, and radiation protection analysis, rather than for high-time-resolution spectral diagnostics of the core plasma (Hu et al., 2018). Fig. 11 presents the response function matrix of the Bonner sphere spectrometer used for neutron field measurements on EAST, illustrating the influence of different moderator sphere sizes on neutron energy response.

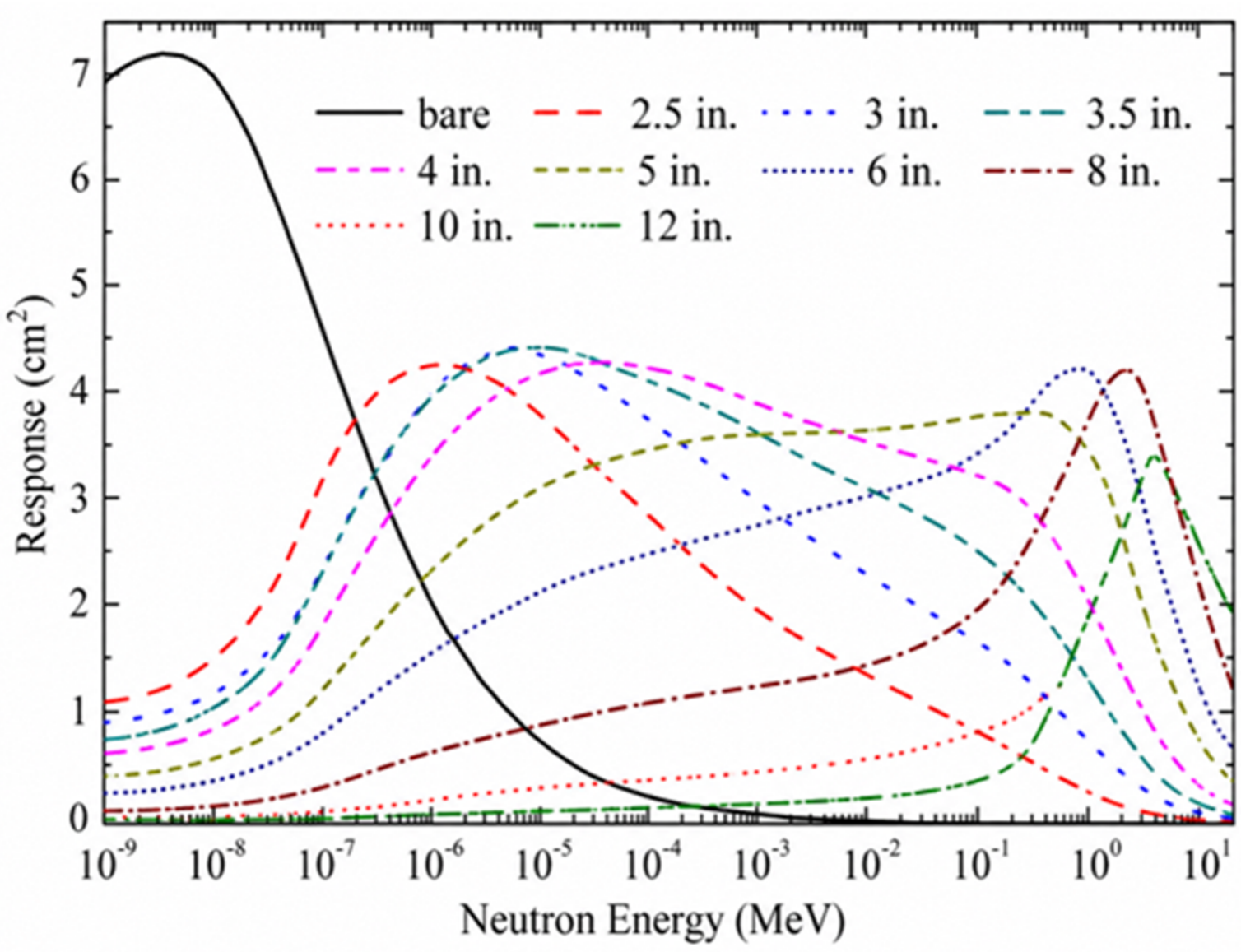

Fig. 11. Response function matrix of the Bonner sphere spectrometer for neutron field measurements on EAST (Hu et al., 2018)

Different moderator spheres exhibit different response peaks and response widths in different energy ranges. Therefore, the Bonner sphere spectrometer can reconstruct neutron energy spectra over a wide energy range through the combined unfolding of multiple counting results and the response function matrix. Although this method has limited energy resolution, it shows good applicability in neutron field measurements, radiation protection dose assessment, and shielding verification in fusion experimental halls. Di Chicco et al. measured high-energy secondary neutron fields produced by the interaction of a 1 GeV/u $^{56}$Fe beam with an aluminum target using an extended-range BSS, and compared the results obtained from MAXED unfolding with those from several Monte Carlo codes. This demonstrates the clear advantages of the method in broad-energy-range spectral-field characterization, while also showing that its results are strongly dependent on response-function calculation and unfolding-algorithm stability (Di Chicco et al., 2024). Vega-Carrillo et al. further pointed out that conventional BSS systems suffer from large volume, heavy weight, and time-consuming measurements, and accordingly designed a more compact, regular parallelepiped alternative spectrometer. This also indirectly indicates that, for core plasma spectral diagnostics requiring discrimination between 2.45 MeV and 14.1 MeV components and further inference of fuel ratio and power evolution, the BSS is generally not the preferred method, but is more suitable as an auxiliary tool for spectral-field characterization and calibration (Vega-Carrillo et al., 2025).

### 1.3.4 Nuclear Fission Method

The nuclear fission method is an important high-flux detection route in fusion neutron measurement, with the prominent advantage of maintaining a high signal-to-noise ratio and strong resistance to gamma interference in intense mixed radiation fields. Its principle is that when neutrons interact with fissile materials such as $^{235}$U and $^{238}$U, high-energy fission fragments are produced. Because the energy deposited by fission fragments in the detection medium is much higher than the electronic response associated with ordinary gamma backgrounds, this method can usually generate large-amplitude and easily identifiable signals in complex neutron/gamma mixed radiation fields (Marocco et al., 2024). For this reason, fission chambers have long been used as one of the core devices for total neutron yield and fusion power monitoring in fusion facilities, especially under operating conditions involving high count rates, wide dynamic ranges, and strong gamma backgrounds. A dual-mode cross-calibration study of the fission-chamber neutron flux monitoring system on EAST by Hong et al. showed that the combination of pulse-counting and Campbell modes can achieve wide-range measurement and improve flux-monitoring accuracy under high-count-rate conditions (Hong et al., 2026). Kovalev et al. further analyzed the radiation conditions of the ITER divertor neutron flux monitoring system and showed that fission-chamber-based monitoring modules can undertake total neutron yield and fusion power measurements in complex neutron/gamma mixed radiation fields, further demonstrating the important application value of fission chambers under strong gamma-background conditions (Kovalev et al., 2022).

In neutron spectrum measurement, the applicability boundary of the nuclear fission method should be defined more carefully. Unlike spectral diagnostic methods based on event-level energy information, such as the time-of-flight method or magnetic proton recoil spectrometers, the output of a fission chamber is usually closer to an integral response dominated by the fission reaction rate. Therefore, it is more suitable for neutron yield, flux, and fusion power monitoring, rather than being directly regarded as a high-resolution neutron spectrum measurement method. Pouradier Duteil et al. modeled and measured the signals of an optical fission chamber in neutron and gamma fields, and pointed out that its output is jointly affected by the fission reaction rate, detection geometry, and background conditions. In essence, it represents a response-function-weighted integral response rather than direct event-level energy information (Duteil et al., 2024). Pu et al. performed neutron monitoring simulations using $^{235}$U micro-fission chambers in a CiADS-like lead-cooled fast reactor (LFR) model based on the conceptual design of the China Initiative Accelerator Driven System. Their results showed that such detectors can effectively reflect local power variations and fission-rate changes, but their main application remains reaction-rate- and power-related monitoring rather than high-resolution neutron spectrum reconstruction (Pu et al., 2024). The EAST fission-chamber system, which combines pulse-counting and Campbell modes to achieve wide-range real-time neutron yield measurement (Yang et al., 2023), also indicates, from the application perspective, that the main advantage of this type of detector lies in wide-dynamic-range monitoring rather than event-by-event spectroscopic analysis. Therefore, if finer spectral information or the relative contributions of D-D and D-T reactions are to be obtained using the nuclear fission method, it is usually necessary to combine multiple fissile targets, response-function modeling, or cross-calibration with other spectroscopic diagnostics, rather than relying on a single fission chamber alone.

#### 1.3.5 Neutron Activation Method

The neutron activation method infers neutron fluence and neutron yield based on the activation phenomenon, in which stable nuclides absorb neutrons and are converted into radioactive isotopes. In this method, activation materials are placed in a neutron radiation field for irradiation, and after irradiation, the radioactivity of the activation products is measured by gamma-ray spectrometry. The neutron fluence or total neutron yield is then inferred from the measured activity. Because different activation materials have different threshold energies and reaction cross sections, an appropriate combination of activation materials can be used not only for total yield measurement, but also, to some extent, for distinguishing D-D and D-T neutron components. Therefore, neutron activation has long been an important technical route for time-integrated neutron yield measurement in fusion devices (Fonnesu et al., 2025). Taking the activated cooling water measurement in JET DTE3 as an example, after neutron-induced activation of the coolant medium, the characteristic gamma rays emitted by the activation products can be measured using NaI and BGO scintillation detectors arranged near the cooling water pipelines. The experimental setup is shown in Fig. 12.

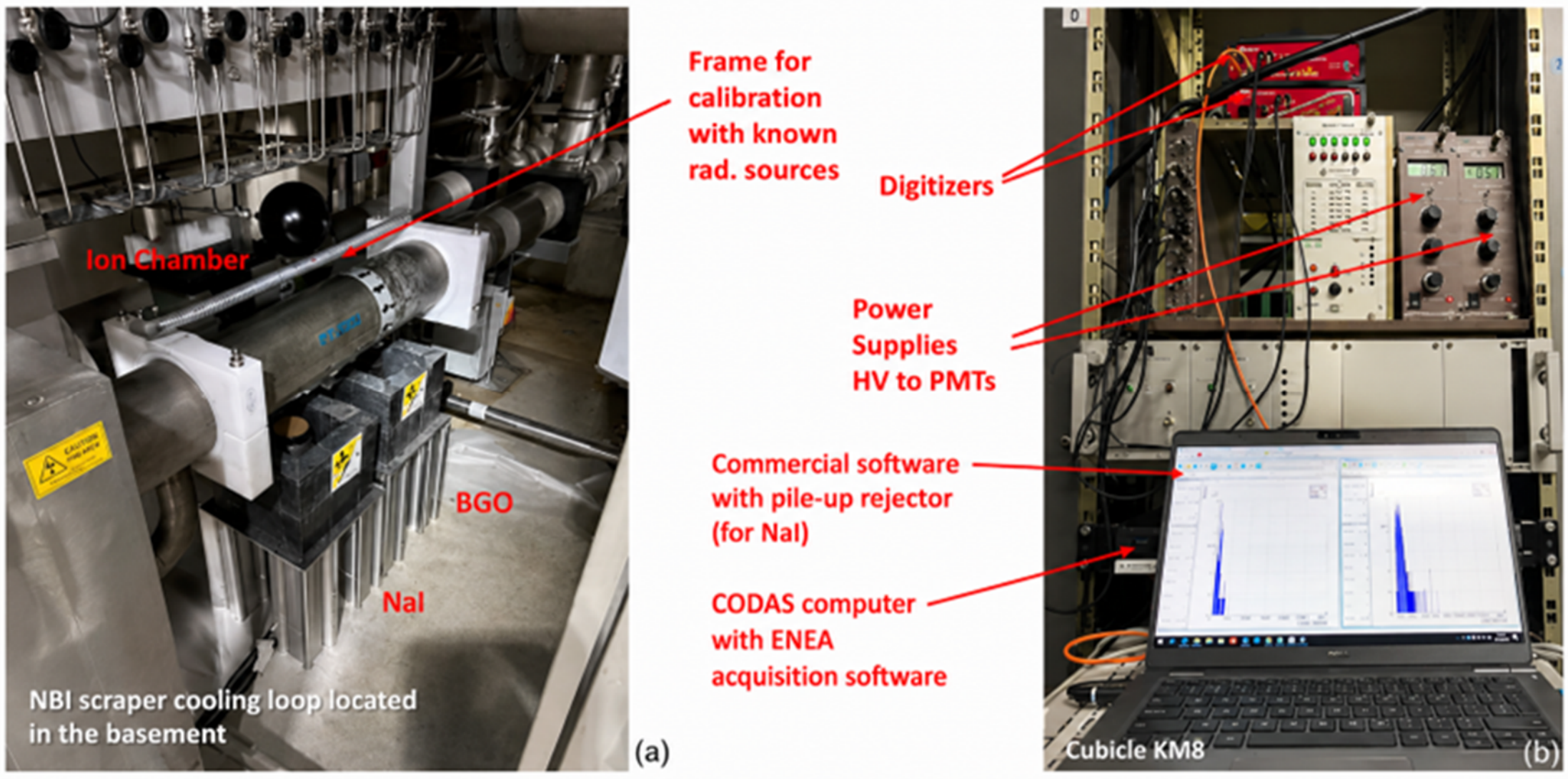

Fig. 12. Schematic diagram of the JET activated cooling water measurement system and data acquisition setup (Fonnesu et al., 2025)

As shown in Fig. 12, the neutron activation method can be applied not only to measurements of solid samples such as activation foils, but also extended to online or quasi-online measurements of activation products in fusion device coolants and other media. By detecting and analyzing the characteristic gamma rays emitted from activated media, important information related to neutron irradiation can be obtained. Therefore, this method has significant application value in fusion neutronics experiments, radiation safety assessment, and activation effect studies.

Current studies show that the practical application of neutron activation in fusion remains highly active. Device-design studies for DTT indicate that the neutron activation system is still an important component of its neutron/gamma diagnostic system, and that activation foils will be used to measure the neutron yield of the device (Marocco et al., 2024). This demonstrates that the neutron activation method still has clear engineering application value in future fusion facilities. For JET, which has completed D-T operation, recent studies further indicate that activation measurements remain an important means of obtaining absolute time-integrated neutron yield and supporting neutronics validation. Related neutronics experiments conducted during JET DTE3 and subsequent summaries of nuclear operation have emphasized the calibration and validation value of activation measurements in real fusion environments (Kappatou et al., 2025). However, the inherent limitations of neutron activation are also clear. It is essentially a time-integrated method and cannot provide truly real-time information on spectral evolution. Therefore, it is not suitable as the sole diagnostic approach for fast-ion transport, transient spectral-shape changes, or real-time power feedback. Rajput et al. performed activation-foil neutronics simulations for D-D plasma operation in the ST40 spherical tokamak and used neutron-induced activity to measure neutron yield, indicating that neutron activation is more suitable for integral yield evaluation and radiological analysis than for millisecond-scale real-time dynamic diagnostics (Rajput et al., 2025). Lahmann et al. further developed a real-time nuclear activation detector at NIF, with the aim of overcoming the limitations of conventional activation measurements that require post-irradiation readout and have limited time response. This also indirectly shows that conventional neutron activation itself does not possess true real-time capability for measuring spectral evolution (Lahmann et al., 2025).

Among the various methods for fusion neutron spectrum measurement, TOF and the recoil proton method are the most suitable for high-resolution fast-neutron spectral diagnostics. The former can directly determine the peak position and spectral broadening through the time-of-flight relationship, whereas the latter enables detailed characterization of fast-neutron spectra through recoil-proton kinematics combined with magnetic analysis or response-function reconstruction. Both methods have been applied to diagnostics related to ion temperature, fuel ratio, and fusion power. In contrast, the Bonner sphere spectrometer relies on multiple response functions and unfolding algorithms, making it more suitable for broad-energy-range spectral-field characterization, shielding validation, and auxiliary calibration, rather than serving as the primary real-time spectrometer for core plasma diagnostics (Santos Oliveira et al., 2023). The nuclear fission method has clear advantages under conditions of high count rates, strong gamma backgrounds, and wide dynamic ranges, and is therefore more suitable for neutron yield, flux, and power monitoring than for high-resolution spectrum reconstruction (Moreno-Pérez et al., 2025). The neutron activation method evaluates total neutron yield and the overall D-D and D-T components through time-integrated measurements of neutron-induced activity, but it is difficult to provide real-time information on spectral evolution.

## 2. Applications of Neutron/Gamma Discrimination Technologies in Fusion

### 2.1 Plasma Diagnostics

In fusion research, neutrons are not confined by magnetic fields and can directly escape from the reaction region. Therefore, neutron yield, spectral shape, and temporal evolution have long been regarded as key diagnostic information for characterizing fusion power, ion temperature, fuel composition, and fast-ion behavior. Sangaroon et al. investigated neutral-beam-heated deuterium plasmas in LHD and showed that the energy shift, broadening, and non-Maxwellian structures of D-D neutron spectra can directly reflect the velocity-space distribution and confinement behavior of fast ions. This indicates that high-fidelity neutron signals are essential for reliable subsequent physics inference (Sangaroon et al., 2026). Under D-T operating conditions, JET has recently developed a 14 MeV neutron counting and spectroscopic diagnostic system based on single-crystal diamond detectors, achieving separation of 2.5 MeV and 14 MeV neutrons as well as identification of thermal and non-thermal neutron components. This further demonstrates that neutron/gamma discrimination and high-resolution spectral measurement have become important components of burning-plasma diagnostic systems (Rigamonti et al., 2024). Fig. 13 shows the 14 MeV neutron spectrum measured by the JET single-crystal diamond detector and its multi-component fitting results.

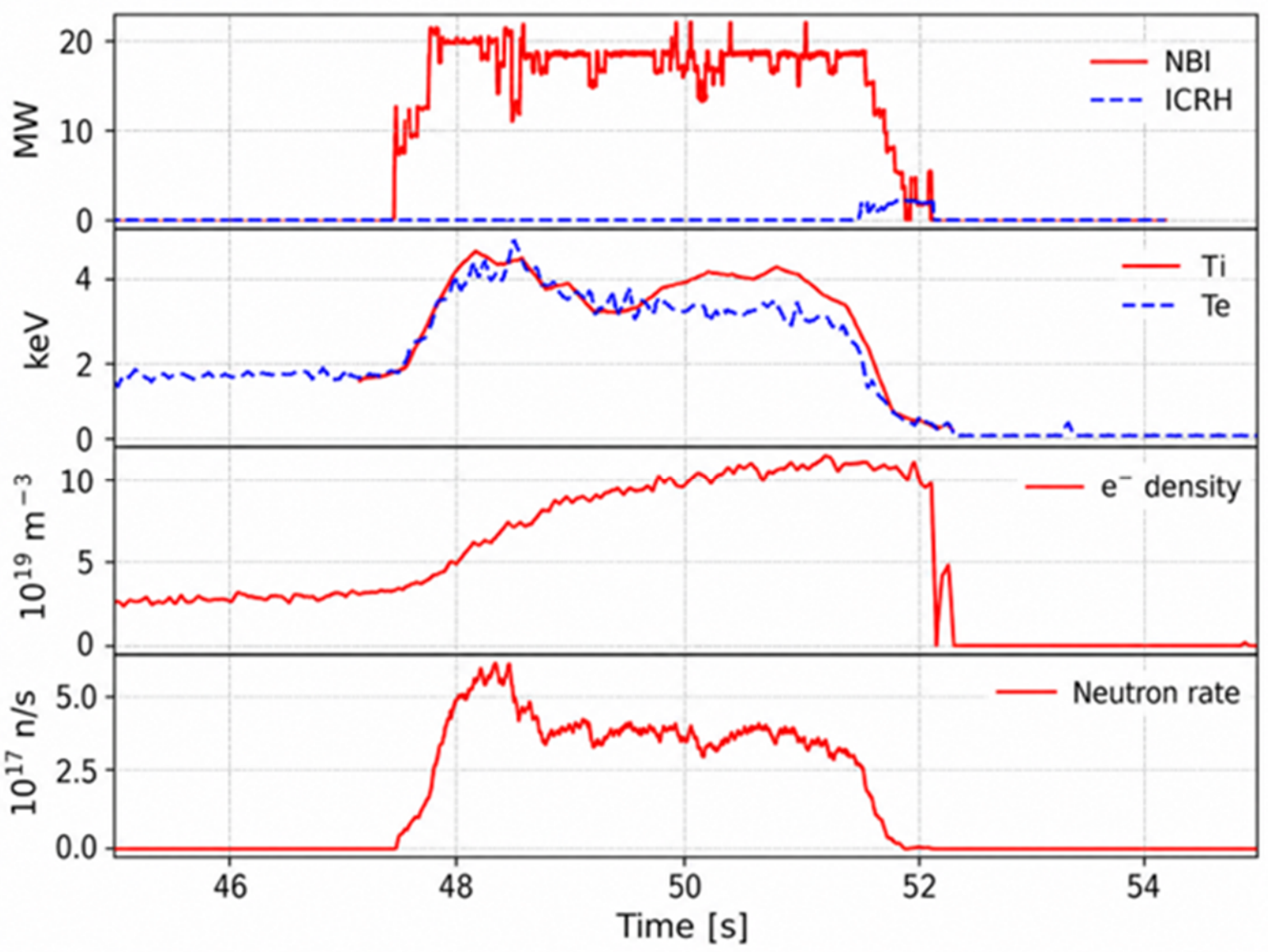

Fig. 13. 14 MeV neutron spectrum measured by the JET single-crystal diamond detector and fitted neutron components (Rigamonti et al., 2024)

The 14 MeV neutron spectral peak measured by the single-crystal diamond detector can be further decomposed into thermal plasma components, beam-thermal components, beam-beam components, and scattered neutron components. Through fitting analysis of these different components, information such as ion temperature, fuel ion distribution, and non-thermal reaction contributions induced by external heating can be extracted. This demonstrates that high-resolution neutron spectrometry is capable not only of neutron counting but also of providing more comprehensive physical information for D-T plasma state diagnostics.

In addition to spectrum measurement, neutron/gamma discrimination also plays a key role in time-resolved diagnostics. Ogawa et al. developed a scintillating-fiber detector on the Korea Superconducting Tokamak Advanced Research (KSTAR) device for high-time-resolution measurement of secondary D-T neutrons. Their results showed that time-evolution information for analyzing tritium burnup and fast-ion slowing-down processes can only be obtained when neutron pulses are stably extracted under strong gamma-background conditions (Kunihiro Ogawa et al., 2024). On LHD, Ogawa et al. further observed upward and downward shifts of D-D neutron energies relative to 2.45 MeV using neutron diagnostics, and related these shifts to fast-ion anisotropy under tangential neutral beam injection. This indicates that the role of neutron/gamma discrimination is not merely to improve counting purity, but also to ensure that subtle structures in neutron spectra can be reliably identified, thereby supporting studies of fast-ion transport and velocity-space distributions (K. Ogawa et al., 2024).

For practical fusion devices, neutron/gamma discrimination is also directly related to the miniaturization, engineering implementation, and online operation of diagnostic systems. Iliasova et al. systematically characterized diamond detectors and organic scintillation detectors according to the requirements of fusion neutron diagnostics under continuous plasma operation. Their results showed that these detector schemes not only possess neutron/gamma discrimination capability, but can also be calibrated and operated stably for continuous-operation scenarios. This indicates that compact detectors can also undertake fusion neutron measurements with clear physics diagnostic value (Iliasova et al., 2024). Zhang et al. developed real-time digital pulse acquisition and processing algorithms for a compact neutron spectrometer on EAST, enabling online signal processing under relatively high count-rate conditions. This demonstrates that neutron/gamma discrimination in fusion devices depends not only on the detector material itself, but also strongly on the capability of back-end electronics and digital algorithms to separate neutron pulses from gamma backgrounds in real time (Zhang et al., 2024). Overall, in plasma diagnostics, the core value of neutron/gamma discrimination lies in ensuring the authenticity of neutron counting, the accuracy of spectrum unfolding, and the resolvability of dynamic processes, thereby providing a reliable basis for fusion power evaluation, ion temperature analysis, fuel-ratio determination, and fast-ion physics studies.

To more intuitively present the application scenarios, implementation routes, and roles of neutron/gamma discrimination technologies in representative fusion devices, Table 2 summarizes relevant applications in typical international fusion facilities.

Table 2 Applications of Neutron/Gamma Discrimination Technologies in Fusion Devices

| Device | Main application scenario | Discrimination/measurement object | Key technical route | Main role or result |
|---|---|---|---|---|
| ITER | Design of an integrated neutron/gamma diagnostic system | Fusion-power-related radiation signals, neutron-related backgrounds, and high-energy gamma channels | Radial gamma-ray spectrometers, radial neutron cameras, and optimized collimation, shielding, and nuclear-analysis design (Moro et al., 2024) | Fusion power measurement and system-level diagnostic optimization under high-radiation environments (Scioscioli et al., 2025) |
| JT-60SA | Enhanced deployment of neutron/gamma diagnostics | 2.5 MeV neutron spectra and gamma-ray channels | Coordinated design of compact chlorine-based scintillator neutron spectrometers and gamma-ray spectrometers (Sozzi et al., 2025) | Support for future fast-ion and plasma-parameter studies through neutron spectrum analysis in JT-60SA-relevant deuterium plasmas. (Matsuura et al., 2025) |
| EAST | Online neutron signal processing and spectrum | Neutron/gamma pulse flux and D-D neutron spectra | Real-time digital pulse acquisition and processing algorithms | Online acquisition, processing, and storage of |
| | measurement in long-pulse operation | | for neutron/gamma flux data (Zhang et al., 2024); $LaCl_3(Ce)$-based compact neutron spectrometer with PSD and D-D neutron-response characterization (Fridrikhsen et al., 2026) | neutron/gamma pulse data; demonstration of D-D neutron spectroscopy on EAST using a $LaCl_3(Ce)$-based detector (Pankratenko et al., 2025) |
| KSTAR | High-time-resolution operational monitoring | Secondary D-T neutrons under a gamma background and D-D fusion neutrons | Scintillating-fiber detector for high-time-resolution secondary D-T neutron measurement (Lee et al., 2025); single-crystal CVD diamond-based detector for D-D neutron spectroscopy (K Ogawa et al., 2025) | Analysis of time evolution related to triton burnup and D-D neutron flux/spectrum measurements on KSTAR |
| W7-X | Development of a diagnostic platform for high-performance steady-state operation | Radiation and diagnostic constraints under long-pulse high-performance operation | Scintillator-based fast-ion loss detector (sFILD) for fast-ion diagnostics (Jansen Van Vuuren et al., 2024) | Demonstration of W7-X long-pulse, high-performance operation, illustrating the representative diagnostic value of sFILD (Grulke et al., 2026) |
| ASDEX Upgrade | Compact device-level neutron spectrum measurement | 2.45 MeV D-D neutrons and gamma-ray response | COSMONAUT CLYC-7-based compact neutron spectrometer (Nocente et al., 2024) | Improved particle-discrimination capability and high-count-rate adaptability (Rigamonti et al., 2025) |

As shown in the table, neutron/gamma discrimination technology has gradually evolved from signal separation at the level of individual detectors into an important supporting technology that underpins neutron spectrum measurement, operational monitoring, fast-ion diagnostics, and reactor-level radiation diagnostic system design in fusion devices.

### 2.2 Monitoring the Safe Operation of Fusion Devices

For fusion devices, neutron/gamma-related radiation measurements increasingly serve online operational monitoring rather than merely post-processing correction of neutron data. As fusion devices move toward higher power, longer pulses, and higher radiation backgrounds, neutron- and gamma-based signals play increasingly important roles in fusion power characterization and discharge-state assessment. Therefore, measurement systems must be able to stably extract fusion-relevant radiation information under strong background and complex scattering-field conditions (Fugazza et al., 2026; Landsmeer et al., 2025). Rechena et al. investigated the collective Thomson scattering diagnostic system for ITER and pointed out that, under unprecedented neutron irradiation, magnetic-field, and fusion-power conditions, diagnostic systems serve not only physics experiments but also plasma control and machine protection functions. Therefore, their availability and maintenance constraints must be considered early in the design stage (Rechena et al., 2026). Mehrara et al. conducted nuclear-analysis optimization for the ITER radial gamma-ray spectrometer and showed that, for reactor-level radiation diagnostic systems operating under D-T conditions, measurement performance is jointly constrained by the signal-to-background ratio, collimation and shielding layout, and nuclear-analysis optimization. This indicates that online operational monitoring in future fusion devices is no longer a problem of a single detector alone, but a systems-engineering issue closely coupled with neutron/gamma background control, structural arrangement, and system-level engineering design (Mehrara et al., 2026).

DTT, which is currently under construction, provides a representative example of how neutron monitoring in future fusion devices must combine detector selection, calibration strategy, and system-level response correction. As discussed above, the DTT Neutron Yield Monitor system adopts multiple detector types to cover different neutron-yield ranges and neutron-energy components (Marocco et al., 2026). Anagnostopoulou et al. further investigated the preliminary calibration procedure for the Phase 1 neutron and gamma-ray diagnostics of DTT, showing that the Neutron Yield Monitors (NYM) and Neutron Activation System (NAS) require absolute calibration using $^{252}$Cf sources and compact D-T neutron generators, together with MCNP-based correction factors and staged D-D/D-T calibration schemes (Anagnostopoulou et al., 2026). This indicates that, for online monitoring in fusion devices, neutron/gamma discrimination should be considered together with detector calibration and system-level response correction, so that neutron-yield measurements remain reliable under complex mixed-radiation and device-geometry conditions.

Online safe-operation monitoring also relies on the integrated design of diagnostic systems and device structures. Cesaroni et al. pointed out in their study of the position monitoring system for the ITER radial neutron camera that positional deviations of detectors and lines of sight can directly affect the accuracy of neutron emissivity reconstruction. Therefore, online confirmation of the geometric status of diagnostic components is itself a prerequisite for ensuring the effectiveness of neutron monitoring (Cesaroni et al., 2024). Marzullo et al. further demonstrated in the mechanical design of the ITER Ex-Port radial neutron camera that the system must integrate collimation, shielding, and detection units within limited port space to ensure reliable measurement of uncollided 2.5 MeV and 14 MeV neutrons (Marzullo et al., 2024). Fig. 14 shows the ITER Ex-Port radial neutron camera and its line-of-sight arrangement.

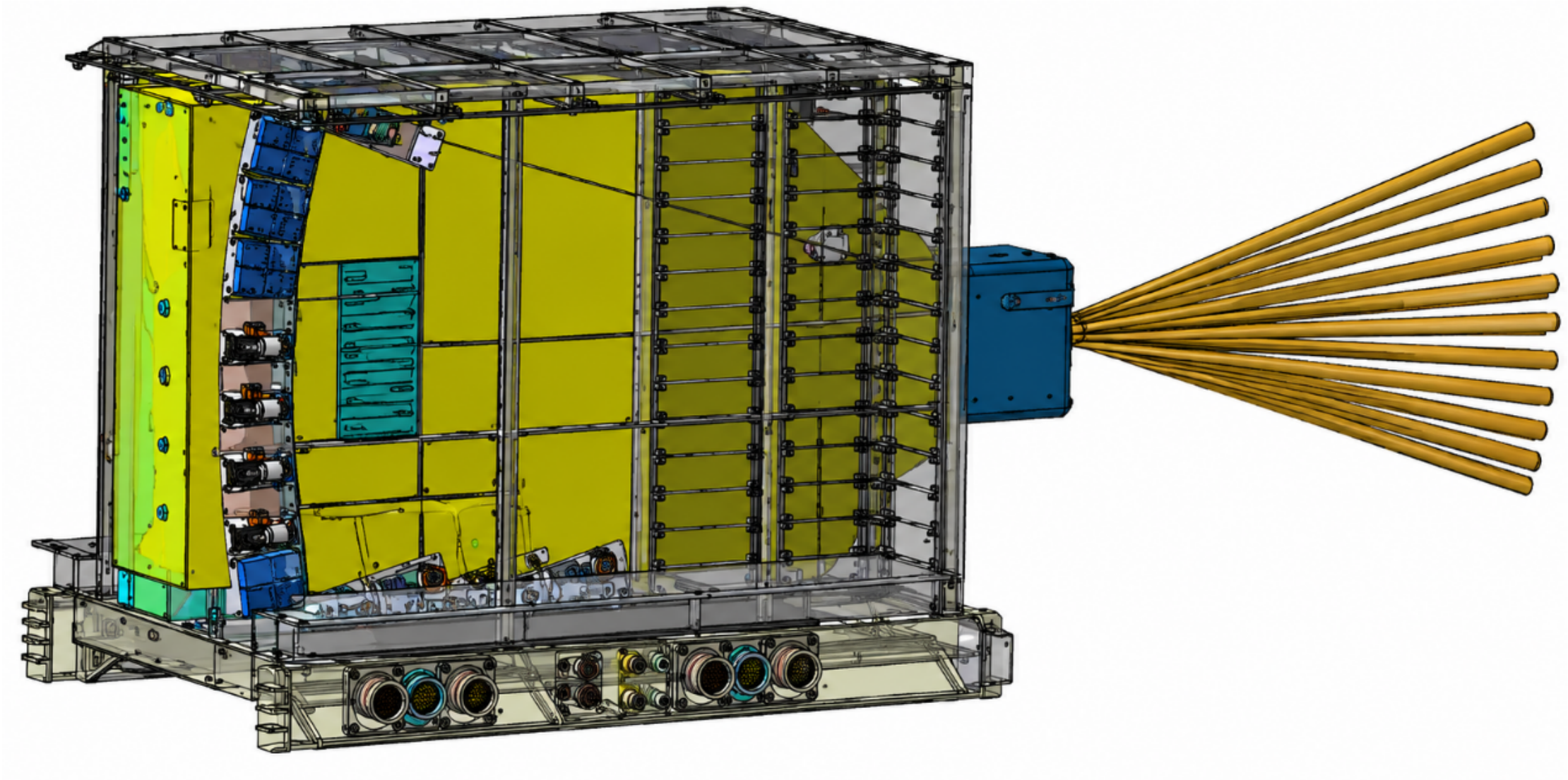

Fig. 14. Schematic diagram of the ITER Ex-Port radial neutron camera and its line-of-sight arrangement (Marzullo et al., 2024)

Reactor-grade neutron monitoring systems are required to simultaneously accomplish functions such as collimation, shielding, detector installation, and line-of-sight coverage within the limited port space available. Such structural designs demonstrate that online monitoring for the safe operation of fusion devices is not merely a single-detector problem, but rather a systematic engineering task jointly determined by viewing geometry, shielding structures, detector units, and data acquisition chains. Therefore, in the operational monitoring of fusion devices, neutron/gamma discrimination should not be narrowly understood as merely a pulse-separation algorithm. Instead, it should be regarded as an operational monitoring capability jointly supported by detector selection, structural arrangement, shielding design, acquisition chains, and calibration strategies. Online safe-operation monitoring of fusion devices is not independently accomplished by a single detector, but is realized through a complete monitoring chain composed of detectors, signal processing, data acquisition, and central control, as shown in Fig. 15.

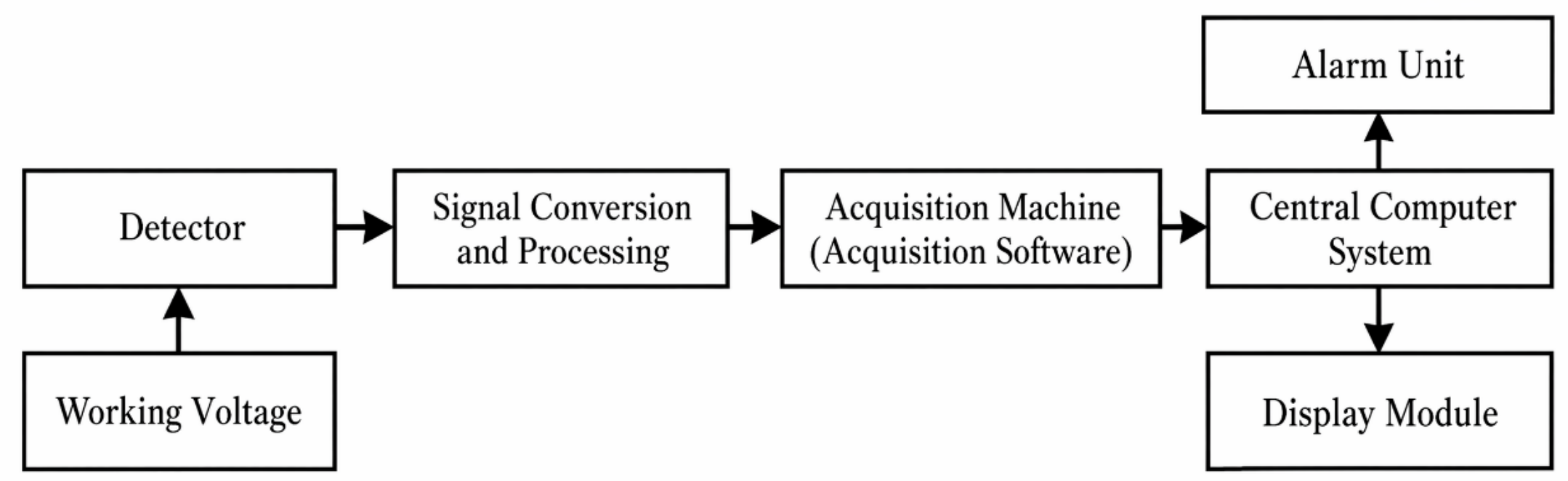

Fig. 15. Overall block diagram of the online radiation monitoring system for a fusion device (LI et al., 2023)

Such a system architecture indicates that the effectiveness of online operational monitoring depends not only on the performance of front-end detectors but also on back-end signal conversion, data processing, alarm interlocking, and centralized management capabilities. Therefore, from an engineering perspective, neutron/gamma discrimination is essentially part of the system-level capability of fusion devices.

### 2.3 Radiation Protection Monitoring

Compared with operational-state monitoring, radiation protection monitoring places greater emphasis on accurately identifying the sources, distribution, and evolution of mixed radiation fields around fusion devices (Martínez-Albertos et al., 2025). In fusion facilities, neutron and gamma fields are interrelated, but they also exhibit significant differences at different locations and on different time scales. Therefore, the key task of radiation protection monitoring is not simply to read out dose rates, but to identify dose sources and use this information to support area zoning, personnel access control, and maintenance planning (Guo et al., 2022). Full-model MCNP analyses of ITER developed by Juarez et al. showed that ITER safety analysis can be performed more robustly and radiation protection justification can be made more complete only when the tokamak body and building structures are incorporated into a unified radiation transport model (Juarez et al., 2024). ITER radiation safety analyses must simultaneously consider the tokamak itself, plant structures, and the major radiation source terms during different operational phases. The relationships among the radiation sources in the full-model configuration are illustrated in Fig. 16.

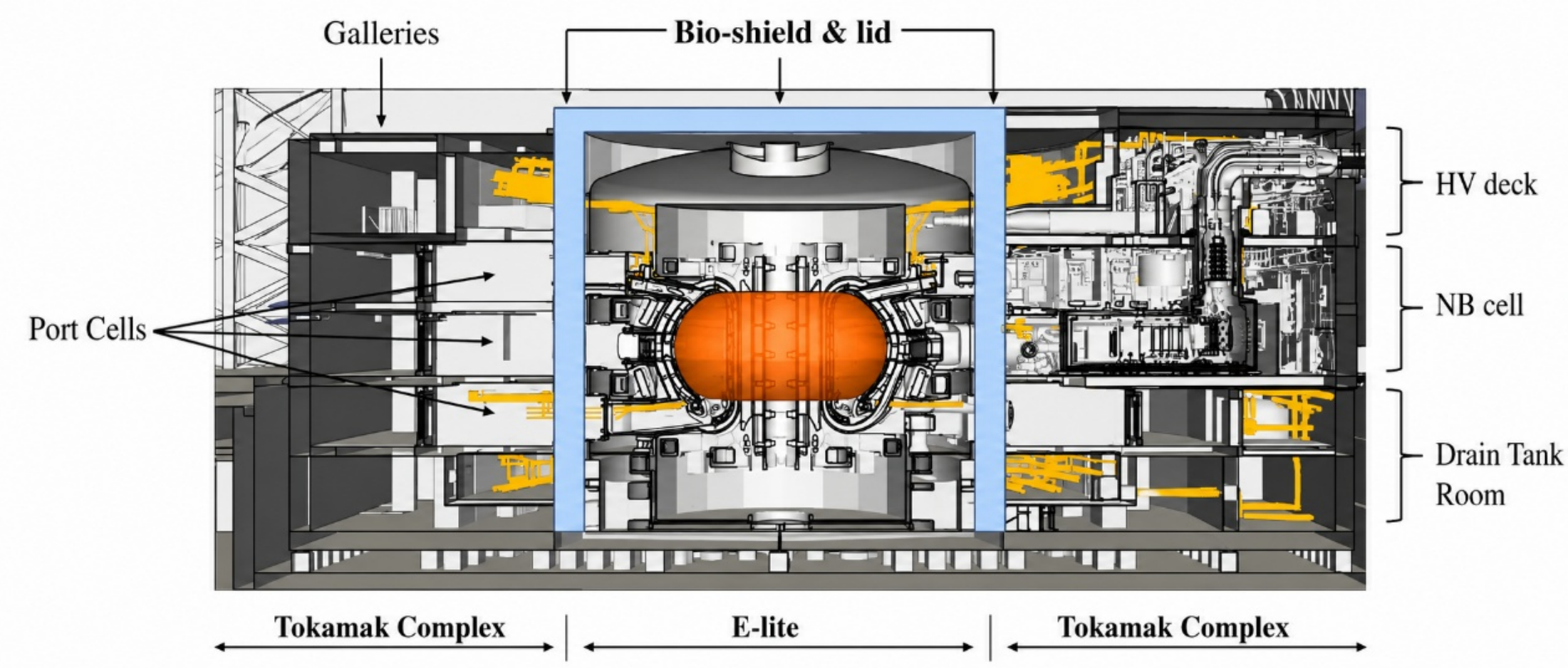

Fig. 16. Schematic diagram of the relationships between major radiation sources and facility structures in the ITER full-model configuration (Juarez et al., 2024)

ITER radiation protection monitoring involves multiple categories of source terms, including plasma neutron sources, activated cooling water, and activated structural components. Moreover, these source terms spatially extend across the tokamak itself, biological shielding, and plant structures. Therefore, radiation protection monitoring cannot rely solely on local dose measurements, but instead requires a comprehensive evaluation of dose origins and radiation field evolution in different regions by combining three-dimensional neutron/gamma transport models. Afanasenko et al. also investigated the radiation environment of ITER equatorial port 11 and pointed out that radiation-field distributions in the port plug, gap region, and port cell exhibit significant spatial differences. This indicates that radiation protection monitoring in fusion devices must be based on joint modeling and zoned assessment of neutron fields and secondary gamma fields (Afanasenko et al., 2024).

In practical maintenance scenarios, neutron/gamma discrimination also plays an important role in identifying the residual radiation environment after shutdown. Occhiuto et al. pointed out in their study of the maintenance scheme for the ITER radial neutron camera that, even under post-shutdown conditions where personnel access is permitted, the radiation field in the interspace region still requires further optimization of manual operation procedures according to the as low as reasonably achievable (ALARA) principle. Virtual-reality tools are also needed to verify maintenance paths and replacement procedures so as to minimize occupational exposure (Occhiuto et al., 2025). Fonnesu et al. analyzed shutdown dose-rate experiments during JET DTE2 and showed that the shutdown gamma dose rate after real D-T operation exhibits clear temporal evolution. Dose variations from several hours to several weeks after shutdown must be jointly evaluated through experimental monitoring and uncertainty analysis to provide a reliable basis for subsequent maintenance and personnel protection (Fonnesu et al., 2024b). Therefore, neutron/gamma discrimination in radiation protection monitoring essentially aims to distinguish, in both space and time, the fusion neutron source during operation from the activated gamma source after shutdown, thereby avoiding confused judgments of radiation risks in fusion devices.

As future fusion devices move toward the fusion demonstration reactor (DEMO) level, radiation protection monitoring increasingly depends on standardized shutdown dose-rate analysis methods and high-accuracy source-term reconstruction capabilities. Peterson et al. implemented and verified a fully open-source Rigorous Two-Step shutdown dose-rate calculation capability in OpenMC, completing validation of the full chain from neutron transport, activation, and decay gamma-source generation to photon transport, and experimentally benchmarking the method using an ITER benchmark based on the Frascati Neutron Generator (Peterson et al., 2024). Afanasenko et al. further investigated shutdown dose rates inside the vacuum vessel of a fusion demonstration reactor and pointed out that maintenance accessibility and personnel access criteria must be based on three-dimensional, spatially resolved calculations of residual gamma fields, because differences in activation responses among structural materials and regions directly determine maintenance windows and dose levels (R. Afanasenko et al., 2026). In radiation protection monitoring for fusion applications, the ultimate significance of neutron/gamma discrimination lies in supporting radiation safety decision-making throughout the entire process from operation and shutdown to maintenance.

### 2.4 Fusion Reactor Design

In the fusion reactor design stage, the value of neutron/gamma discrimination is more prominently reflected in blanket experiments, reactor-level diagnostic system design, and the measurement of key nuclear parameters under intense mixed radiation fields. On the one hand, reactor-level diagnostic systems must fully optimize the signal-to-background ratio and suppress the interference of indirect and backscattered neutrons in line-of-sight measurements, so as to ensure the reliability of relevant neutron channels for diagnosing fusion power and radiation source distributions (Reviakin et al., 2026). On the other hand, studies on neutron measurements for DEMO blankets have shown that online neutron measurement inside blankets is often accompanied by strongly coupled gamma-flux backgrounds. Therefore, the measurement of key nuclear parameters in blanket and shielding structures must rely on detectors with gamma-resistance capability and corresponding discrimination schemes, so that true neutron-reaction signals can be extracted from intense gamma fields (Filliatre, 2024).

This point was also demonstrated in the online tritium production measurement of the helium-cooled pebble bed test blanket module (HCPB TBM) mock-up during JET DTE2. Related studies verified the feasibility of characterizing tritium production through neutron detection in harsh TBM-like radiation environments by deploying diamond detectors (Fonnesu et al., 2024a). In the HCPB TBM mock-up, the diamond detector employs a $^{6}$LiF converter layer to achieve thermal neutron sensitivity, while fast neutron measurements can also be realized through neutron interactions within the diamond crystal itself. Its structure and detection mechanism are illustrated in Fig. 17.

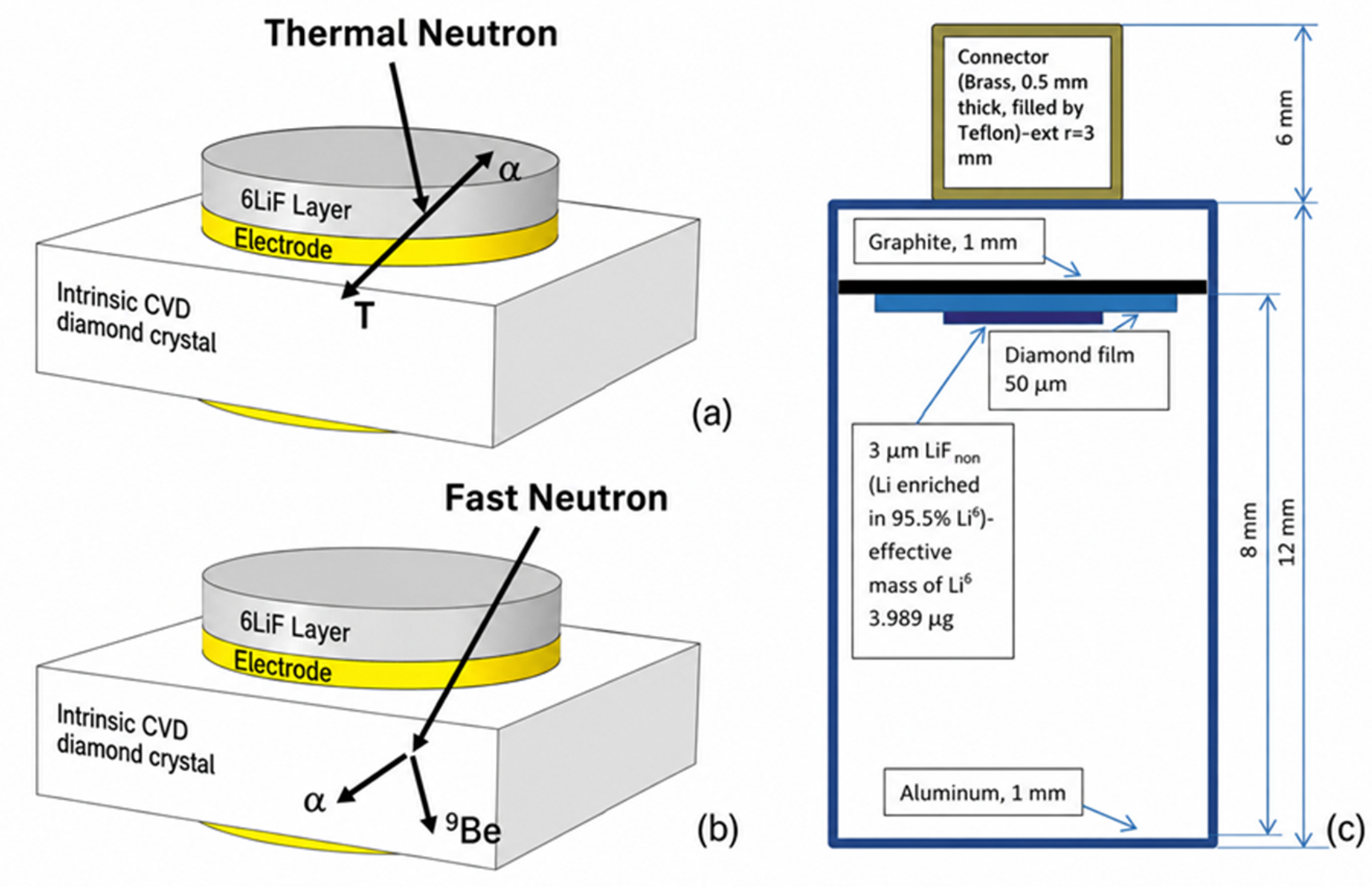

Fig. 17. Schematic diagram of the TBM diamond detector structure and the thermal/fast neutron detection mechanisms (Fonnesu et al., 2024a)

This structure demonstrates that single-crystal diamond detectors can expand their neutron response channels by combining intrinsic material reactions with external converter layers, thereby adapting to the complex radiation environments in blanket experiments where thermal neutrons, fast neutrons, and gamma rays coexist. Moro et al. pointed out in the nuclear design study of the electronics shielding cabinet for the ITER radial neutron camera that only through the coordinated optimization of detection units, electronics layout, and shielding structures can such diagnostic systems stably support measurements of fusion power and radiation source distributions (Moro et al., 2023). Kobayashi et al. developed a measurement method based on a single-crystal CVD diamond detector and LiF foils, enabling simultaneous fast-neutron spectrometry and tritium production measurement in mixed radiation fields. This demonstrates that compact detection systems with discrimination capability can directly support blanket performance validation (Kobayashi et al., 2024). Li et al. developed a compact back-to-back $^{6}Li/^{7}Li$ lithium-glass detector, in which two scintillators with similar responses were used to subtract the gamma background. The detector was applied to online tritium production measurement under strong gamma environments, providing a more engineering-adaptable solution for in-situ monitoring in reactor blanket experiments (Li et al., 2024).

At the level of blanket neutronics design and nuclear data validation, neutron/gamma discrimination further extends to the capability requirements of measurable reaction rates, separable heating components, and verifiable calculations. In another study, Kobayashi et al. used a single-crystal diamond detector and thin LiF layers to measure the $^{6}Li$ burnup reaction rate, and distinguished $^{6}Li$ reaction-product signals from fast-neutron and gamma backgrounds through pulse-shape discrimination. This indicates that local reaction-rate validation experiments themselves are highly dependent on neutron/gamma discrimination (Kobayashi et al., 2023). Ebiwonjumi et al. showed that the validation of fusion neutronics codes involves not only neutron spectra and dosimetric reaction rates, but also coupled quantities such as photon spectra, gamma heating rates, and tritium production. Therefore, experimentally distinguishable neutron- and photon-related responses are important for imposing meaningful constraints on fusion neutronics codes and nuclear data (Ebiwonjumi et al., 2025). Subsequent studies in the same research direction performed uncertainty propagation analyses for tritium production and gamma heating rates in fusion neutronics clean benchmarks, further indicating that nuclear data uncertainties can affect the prediction of key reactor-design quantities, including tritium breeding and energy deposition, through neutron transport results. This suggests that future reactor design requires not only benchmark measurements themselves, but also uncertainty-informed constraints on neutron- and gamma-related response quantities (Ebiwonjumi and Peterson, 2024).

Fusion reactor design also requires neutron/gamma discrimination results to be reliably transferred into the engineering analysis chain. Afanasenko et al. investigated the synchronization of ITER neutronics models between OpenMC and MCNP, showing that model conversion and consistency verification between open-source tools and conventional engineering codes have become important components of design iteration and nuclear-analysis verification for ITER and subsequent fusion reactors (E. S. Afanasenko et al., 2026). Therefore, neutron/gamma discrimination in fusion reactor design is no longer merely a detector-level signal-processing issue, but a fundamental supporting capability that extends throughout the entire process of diagnostic system design, blanket nuclear-parameter measurement, nuclear data validation, and engineering-model verification.

## 3. Future Development Potential of Neutron/Gamma Discrimination Technologies in Fusion Reactors

### 3.1 Development of High-Performance Detectors

For future fusion reactors, the development potential of neutron/gamma discrimination technologies is reflected not only in algorithm-level signal separation but also in their driving role in the evolution of high-performance detector materials and system architectures. As fusion devices evolve toward high-power, long-pulse, intense-irradiation, and high-heat-load operating conditions, detectors must not only maintain high neutron-identification purity in complex mixed radiation fields, but also satisfy the requirements of intense-flux measurement capability, high-temperature stability, and radiation hardness (Kushoro et al., 2024). Therefore, high-performance detectors with real application prospects in future fusion reactors should not merely be devices with higher detection efficiency, but integrated diagnostic units that combine intrinsic neutron/gamma discrimination capability, radiation hardness, temperature compatibility, and in-situ calibratability.

From the perspective of detector materials, wide-bandgap semiconductors represented by CVD diamond are becoming an important development direction for high-performance radiation detectors in fusion reactors, providing material and device-level support for fast-response, high-reliability, and radiation-hard neutron/gamma diagnostic systems, as shown in Fig. 18. Weiss and Griesmayer summarized the application of CVD diamond detectors in fusion neutron diagnostics and showed that single-crystal diamond can not only support response-function measurements and fast-neutron spectroscopic analysis for D-D and D-T fusion neutrons, but also exhibits clear advantages in lifetime, energy resolution, and high-temperature compatibility. These characteristics make it highly suitable for advanced neutron diagnostic tasks in ITER and subsequent fusion reactors (Weiss and Griesmayer, 2024). Potiron et al. further pointed out that solid-state detectors under high-temperature 14 MeV neutron measurement conditions are already approaching the environmental requirements for online flux monitoring near the blanket, indicating that future neutron/gamma discrimination technologies will increasingly rely on solid-state platforms capable of stable operation in high-temperature and intense-irradiation environments (Potiron et al., 2026). Meanwhile, Pérez et al. investigated SiC P-N detectors and found that such devices exhibit linear response, no obvious saturation, and good environmental adaptability under both thermal-neutron and fast-neutron irradiation. This provides a new device basis for discrimination-capable detectors that combine engineering feasibility with irradiation stability in future fusion-relevant harsh-radiation monitoring scenarios (Pérez et al., 2025).

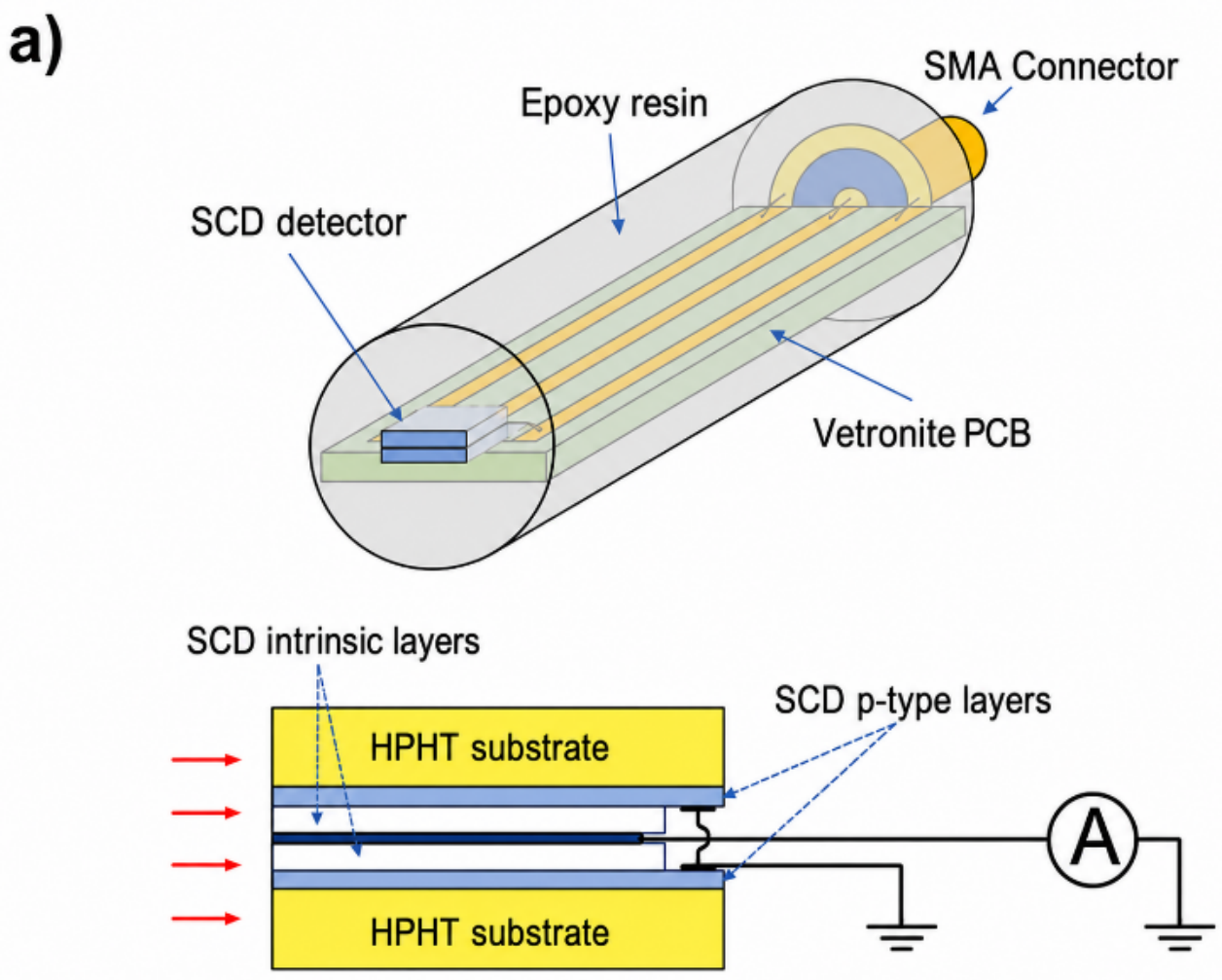

Fig. 18. Schematic layout and simplified electrical diagram of a lateral-irradiation CVD diamond detector prototype for fusion plasma diagnostics. Reproduced from Ref (Cesaroni et al., 2023)

In addition to new detector materials themselves, the development of future high-performance detectors will increasingly depend on the deep integration of neutron/gamma discrimination technologies with digital electronics and artificial intelligence algorithms. Panda et al. investigated EJ-276D plastic scintillators and showed that machine-learning methods can further improve neutron/gamma discrimination capability in low-energy regions where conventional pulse-shape discrimination performs relatively weakly. This suggests that future discrimination technologies will not only improve classification accuracy but may also substantially extend the effective operating threshold and usable energy range of detectors (Panda et al., 2026). Choi et al. further optimized pulse-shape discrimination for liquid scintillators using machine-learning methods, demonstrating that approaches such as autoencoders and principal component analysis can extract more robust discriminative features from high-dimensional waveforms, thereby improving the integrated performance of waveform compression, feature extraction, and classification (Choi et al., 2025). These results indicate that high-performance detectors in future fusion reactors are likely to evolve beyond single detection elements into integrated systems composed of advanced detector materials, high-speed digital acquisition, and intelligent discrimination algorithms. In such systems, neutron/gamma discrimination will become a key link driving substantial improvements in detector performance.

The development of high-performance detectors for future fusion reactors is essentially a multi-parameter optimization process centered on neutron/gamma discrimination capability. On the one hand, radiation-resistant and high-temperature-stable materials such as diamond and SiC are needed to improve detector usability under extreme environments. On the other hand, scintillators with intrinsic PSD potential and intelligent algorithms are required to improve identification accuracy under strong gamma backgrounds and high count-rate conditions. In this sense, neutron/gamma discrimination technologies will not be marginalized in future fusion reactors; rather, they will become an essential enabling technology that defines the performance boundaries of next-generation fusion diagnostic detectors.

### 3.2 Material Damage Assessment

In future fusion reactors, the potential of neutron/gamma discrimination technologies in material damage assessment is first reflected in their ability to provide more reliable neutron source-term inputs for irradiation damage analysis. Recent studies on the coupling between damage simulations and nuclear processes have shown that material damage prediction increasingly depends on nuclear data and particle source descriptions that are consistent with neutron transport and nuclear inventory calculations (Gilbert, 2025). Studies on the International Fusion Materials Irradiation Facility-DEMO-Oriented Neutron Source (IFMIF-DONES) have indicated that, to obtain irradiation effects closer to the blanket conditions of a demonstration fusion reactor (DEMO), the neutron energy spectrum needs to be tailored, and the uniformity of the neutron field and irradiation effects in the sample region should be ensured as much as possible (Mota et al., 2026). Relevant studies have achieved optimization and homogenization of the spatial neutron flux distribution by introducing neutron spectrum shifters (neutron spectrum shifter, NSS) with non-uniform thickness. Different NSS configurations and their corresponding neutron field layout schemes are shown in Fig. 19.

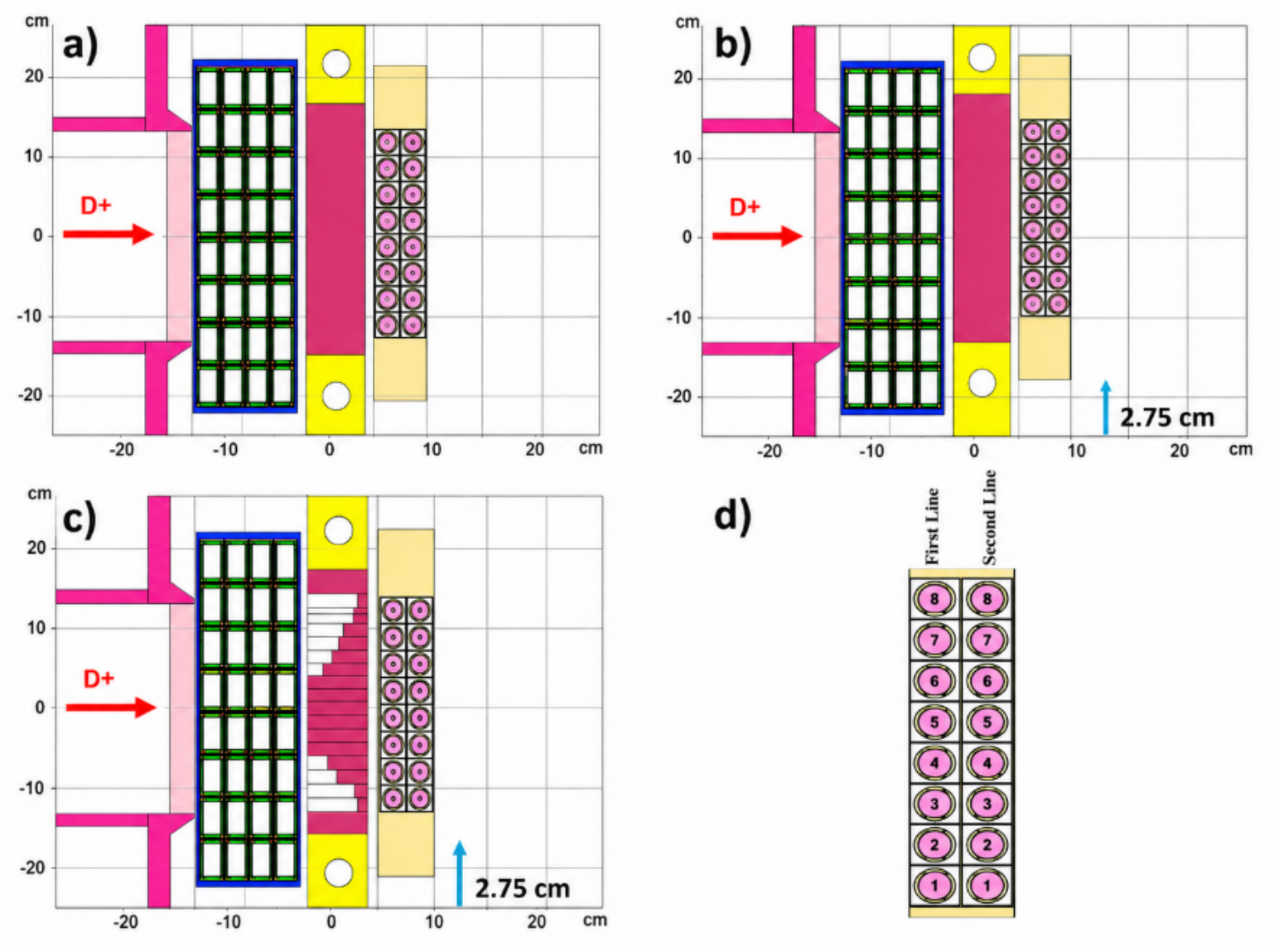

Fig. 19. Different neutron spectrum shifter configurations and neutron field optimization schemes in IFMIF-DONES (Mota et al., 2026)

Taking tungsten as an example, recent experimental validation has shown that the accumulation of neutron-irradiation-induced transmutation products, as well as helium and hydrogen production, can significantly alter the material composition and further affect its physical and mechanical properties (Chatzikos et al., 2024). Therefore, if excessive gamma background, scattered components, or non-target responses are mixed into neutron measurements, the credibility of subsequent evaluations of displacements per atom (dpa), gas production rates, activation levels, and lifetime prediction may be systematically affected.

This indicates that the key issue in future material damage assessment is not merely whether material data are available, but whether the neutron-field inputs on which these data depend are sufficiently accurate. Marian et al. showed in their multiscale assessment of materials for stripping neutron sources that fusion material damage analysis has evolved from simple fluence conversion into a coupled problem involving neutronics, primary damage, defect evolution, chemical inventory, and thermodynamic analysis (Marian et al., 2025). Álvarez et al. further evaluated EUROFER97, tungsten, and CuCrZr under IFMIF-DONES irradiation conditions, demonstrating that different candidate materials exhibit significant differences in damage and transmutation responses to fusion-spectrum neutrons. Therefore, material damage assessment must be based on a more refined characterization of neutron fields (Álvarez et al., 2026). In this process, the value of neutron/gamma discrimination lies in improving the purity of neutron fluence, energy spectrum, and spatial distribution measurements, thereby providing material damage models with input parameters that are closer to real fusion environments.

The true significance of neutron/gamma discrimination technologies for material damage assessment lies in promoting a closed loop between experimental measurements and computational predictions. Progress in IFMIF/EVEDA indicates that dedicated neutron irradiation infrastructures for fusion materials research are gradually becoming mature, providing a practical pathway for establishing irradiation damage databases under fusion-spectrum conditions and for conducting experimental validation (Carin et al., 2024). Pettinari et al. evaluated structural materials for compact fusion devices and pointed out that material selection increasingly relies on comprehensive comparisons of neutron response, activation behavior, and structural compatibility using neutronics tools such as OpenMC. This means that future assessments will depend on higher-quality inputs of neutron spectra, fluence rates, and spatial distributions (Pettinari et al., 2025). The credibility of these inputs ultimately depends directly on whether neutron/gamma discrimination can stably distinguish true neutron signals from gamma backgrounds in complex mixed radiation fields.

Therefore, for future fusion reactors, the role of discrimination technologies in material damage assessment will be reflected in at least the following aspects: improving the accuracy of neutron fluence and spectrum measurements at key locations and reducing the interference of gamma backgrounds in dpa evaluation; supporting cross-validation among neutron activation, online flux monitoring, and numerical simulations to establish a unified source-term basis for material lifetime assessment; and serving material qualification procedures in IFMIF-DONES, DEMO, and subsequent fusion reactors. Neutron/gamma discrimination technologies are not only a diagnostic issue, but will also become a fundamental supporting capability for material design, lifetime assessment, and operational safety justification in future fusion reactors.

### 3.3 Development Potential of Intelligent and Real-Time Neutron/Gamma Discrimination Systems

The demand for neutron/gamma discrimination technologies in future fusion reactors will no longer be limited to offline waveform analysis or post-experimental data processing, but will increasingly shift toward intelligent and real-time discrimination systems for operational monitoring and feedback control. Morales Argueta et al. developed an embedded online neutron/gamma discrimination system based on a CLYC detector with SiPM readout, which can operate on a commercial FPGA and perform real-time event-by-event classification using machine learning (Morales et al., 2024b). Garnett and Byun proposed the NeutralNet framework and systematically evaluated the applicability of neural-network architectures to pulse-shape discrimination in liquid scintillators, showing that machine-learning methods can improve the classification performance of conventional PSD methods in complex mixed radiation fields (Garnett and Byun, 2024). Garnett et al. subsequently extended NeutralNet to discrimination tasks under different neutron-source conditions, indicating that future neutron/gamma discrimination systems have the potential to evolve toward cross-source and cross-condition generalization (Garnett et al., 2025). These studies suggest that future discrimination systems in fusion reactors will increasingly rely on integrated architectures rather than simple comparisons based on a single threshold or a single waveform parameter.

In addition to algorithmic advances, the development potential of future neutron/gamma discrimination systems is also reflected in the parallel upgrading of hardware implementation and evaluation frameworks. Liu et al. proposed the STFT-DFF model, which combines time-frequency feature extraction with dynamic feature fusion, and further validated its implementation on FPGA platforms, indicating that next-generation discrimination systems are already moving toward online deployability (B. Liu et al., 2026). Similarly, Morales et al. proposed a frequency-domain classification method based on an SiPM-coupled CLYC detector. Without relying on complex preprocessing, this method achieved neutron/gamma classification over a relatively wide energy range and demonstrated potential for deployment on FPGA or DSP platforms, as shown in Fig. 20. These results indicate that the improvement of future intelligent real-time discrimination systems will depend not only on detector materials themselves, but increasingly also on the integrated design of readout electronics and feature-extraction algorithms. After systematically classifying and benchmarking a large number of pulse-shape discrimination algorithms, Liu et al. pointed out that future evaluation frameworks should not remain limited to the traditional figure-of-merit (FOM), but should also consider robustness and reproducibility under different thresholds, noise levels, and energy ranges (H. Liu et al., 2026).

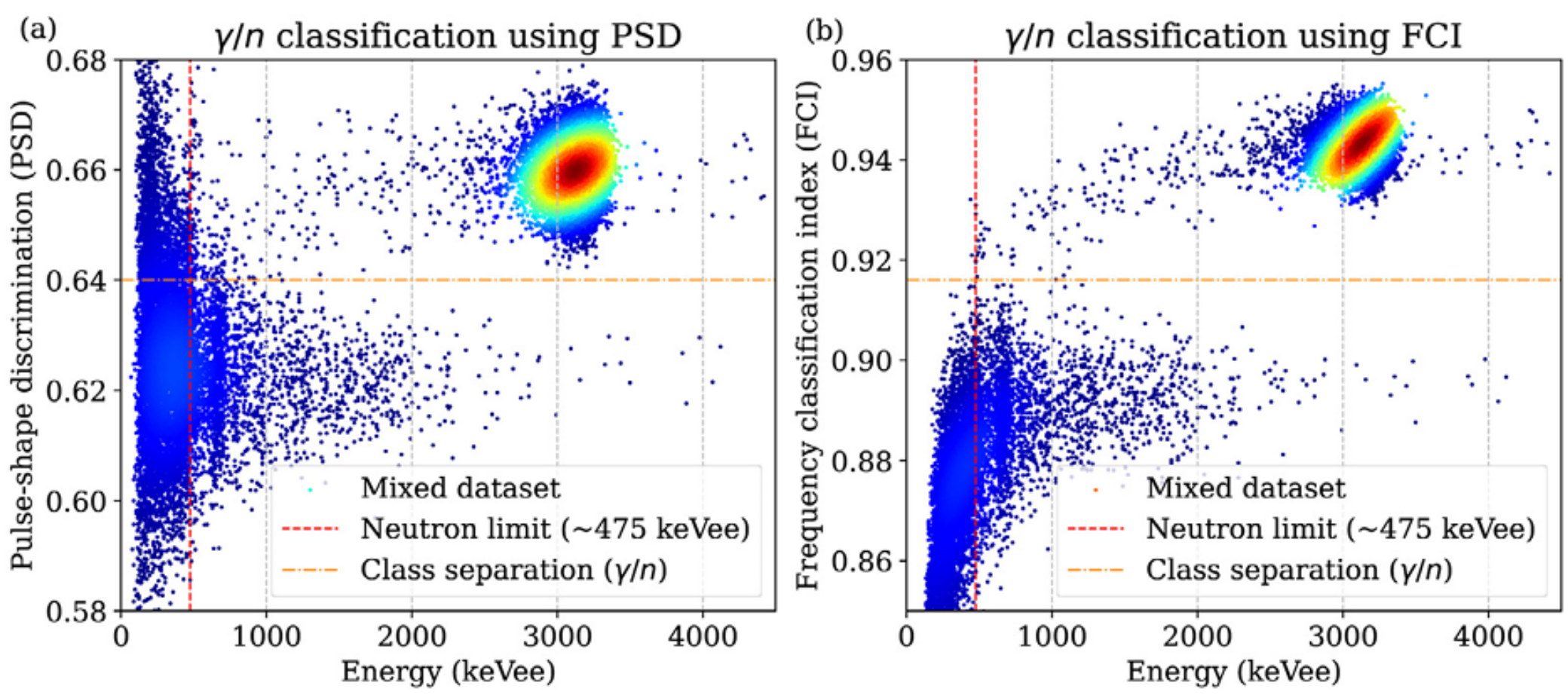

Fig. 20. Comparison of neutron/gamma classification results for an SiPM-coupled CLYC detector (Morales et al., 2024a)

Perelli Cippo et al. further proposed a simplified lifetime prediction method for single-crystal diamond neutron spectrometers in steady-state fusion reactors, indicating that future discrimination systems should not only improve identification accuracy, but also possess operational-state awareness and detector health-management capability (Perelli Cippo et al., 2026). Therefore, the future development of neutron/gamma discrimination technologies will not be limited to improving classification accuracy alone, but will promote the evolution of discrimination systems toward real-time, intelligent, and autonomous operation.

### 3.4 Multi-Detector Coordination and System-Level Diagnostic Integration

In future fusion reactors, neutron/gamma discrimination technologies will also exhibit a clear trend toward system-level development. Their role will gradually evolve from localized discrimination by individual detectors to multi-detector coordination, cross-channel information fusion, and system-level diagnostic integration. Battye and Perinpanayagam pointed out in their review that digital twin technology is integrating data analysis, real-time control, and simulation-based prediction, and is regarded as an important tool for full-process perception and optimization in future fusion systems (Battye and Perinpanayagam, 2025). Schissel et al. further summarized discussions from a digital twin workshop and indicated that future fusion research will increasingly depend on the coordinated development of physical models, data assimilation, artificial intelligence, and unified interfaces (Schissel et al., 2025). This means that the role of neutron/gamma discrimination technologies in future fusion reactors should be understood as a core link for providing high-confidence radiation information inputs to multi-source diagnostic data systems.

From the perspective of device-level application prospects, multi-detector coordination has become an important development direction for diagnostic systems in future fusion reactors. Jardin et al. pointed out that energy-resolved X-ray and neutron diagnostics have clear complementarity in tokamaks and can jointly support plasma-parameter inference and fusion-performance assessment (Jardin et al., 2024). The gamma-ray diagnostic system developed by Ogawa et al. on LHD shows that gamma-ray channels can provide complementary measurements to neutron diagnostics while also supporting plasma physics studies and activation-related analyses (Ogawa et al., 2026b). This indicates that neutron/gamma discrimination technologies in future fusion reactors will be increasingly embedded in system-level diagnostic architectures involving multiple spectrometers, multiple channels, and multiple time scales.

The measurement innovation needs report released by the U.S. Fusion Energy Sciences Advisory Committee (FESAC) has identified radiation-hardened sensing, in-situ calibration, AI-enhanced data interpretation, integrated data analysis, and digital twins as important directions for future fusion devices (Fusion Energy Sciences Advisory Committee and U.S. Department of Energy, Office of Fusion Energy Sciences, 2024). This indicates that the future development potential of neutron/gamma discrimination technologies lies not only in detectors and algorithms themselves, but also in whether they can be integrated into reactor-level diagnostic, control, and maintenance systems as standardized sensing modules. Therefore, as diagnostic tasks in future fusion reactors shift from single-point measurements toward system-level integration, neutron/gamma discrimination technologies will also evolve into a key component of system-level information fusion and operational support capabilities.

## 4. Conclusions

This paper reviews the current status and prospects of neutron detection and neutron/gamma discrimination technologies in fusion applications. It systematically summarizes the main sources, characteristics, and measurement methods of fusion neutrons, analyzes the operating principles, applicable ranges, and application characteristics of different types of neutron detectors in fusion scenarios, and further discusses the role of neutron/gamma discrimination technologies in plasma diagnostics, safe operation monitoring of fusion devices, radiation protection monitoring, and fusion reactor design. On this basis, the future development potential of neutron/gamma discrimination technologies in fusion reactors is discussed from the perspectives of high-performance detector development, material damage assessment, intelligent real-time discrimination systems, and multi-detector coordination with system-level diagnostic integration.

The comprehensive review presented in this paper shows that fusion neutrons are not only core information carriers for characterizing the state of fusion reactions, but also important environmental factors affecting material service performance, structural safety, and engineering realization in fusion reactors. The 14.1 MeV neutrons produced by the D-T reaction constitute the most important high-energy neutron source for future fusion reactors, whereas the 2.45 MeV neutrons produced by the D-D reaction are the most commonly encountered neutron source in most current experimental devices. In addition, neutrons generated by secondary reactions further increase the complexity of real fusion environments. Therefore, neutron measurement is not merely a process of obtaining counting results, but requires the accurate characterization of neutron flux, energy spectrum, spatial distribution, and temporal evolution in complex mixed radiation fields.

In terms of detection technologies, different neutron measurement methods and detector types have their own emphases, which also determine their different roles in neutron/gamma discrimination. The nuclear reaction and nuclear recoil methods are more suitable for real-time diagnostics and fast-neutron spectrum measurement, whereas the nuclear fission and neutron activation methods show clear advantages in high-flux monitoring, calibration, and integral measurements. Liquid scintillators and some organic crystals remain the most practically applicable routes for fusion fast-neutron measurement and neutron/gamma discrimination at present. Inorganic Li-containing crystals are more suitable for thermal-neutron-enhanced detection and mixed-field discrimination. Gas detectors have advantages in thermal-neutron monitoring and low-gamma-background scenarios, while semiconductor detectors show significant potential in compact design, radiation tolerance, and fast-neutron spectrometry. Therefore, there is no single optimal solution for neutron detection in fusion applications; instead, it depends more on task-oriented selection and coordinated design of multiple detector types.

In terms of applications, neutron/gamma discrimination technologies have evolved from simple detector-level signal separation into an important enabling capability supporting multiple key tasks in fusion devices. In plasma diagnostics, discrimination capability directly affects the authenticity of neutron counting, the accuracy of spectrum unfolding, and the determination of fast-ion behavior and fuel composition. In device operation monitoring, it is related to fusion power evaluation, abnormal-condition identification, and the reliability of online monitoring results. In radiation protection monitoring, it determines the ability to identify the sources and spatial distributions of mixed radiation fields, as well as the temporal evolution of residual gamma fields after shutdown. In fusion reactor design, discrimination capability further extends to blanket experiments and related reactor-level nuclear measurements. Therefore, neutron/gamma discrimination technology is no longer merely a local function attached to neutron detection, but has become a system-level supporting technology running through the entire process of fusion diagnostics, operation, maintenance, and engineering design.

For future fusion reactors, the development of neutron/gamma discrimination technologies will mainly be reflected in the following directions. First, high-performance detectors with discrimination capability will develop toward high count-rate capability, wide dynamic range, high-temperature tolerance, radiation resistance, and long service life. Wide-bandgap semiconductors, advanced scintillators, and compact integrated readout systems will become important technical routes. Second, neutron/gamma discrimination technologies will further support material damage assessment by improving the accuracy of neutron fluence, energy spectrum, and spatial distribution measurements, thereby providing more reliable source-term inputs for irradiation damage evaluation, activation behavior analysis, and lifetime prediction. With the development of digital electronics, artificial intelligence, and edge computing, neutron/gamma discrimination will gradually evolve from conventional offline analysis into intelligent, real-time, and autonomous systems, and is expected to be directly integrated into operational monitoring and control chains. Finally, neutron/gamma discrimination technologies in future fusion reactors will be increasingly embedded in frameworks of multi-detector coordination and system-level diagnostic integration, becoming important sensing modules that connect multi-source radiation measurements, digital twin modeling, and device operation decision-making.

Overall, the future development of neutron/gamma discrimination technologies in fusion will not remain limited to the single objective of improving the discrimination accuracy between neutrons and gamma rays. Instead, it will gradually evolve into an integrated technical system combining detector materials, signal processing, intelligent algorithms, system integration, and engineering applications. With the continuous advancement of ITER, DEMO, and subsequent fusion demonstration reactor studies, neutron detection and neutron/gamma discrimination technologies will play an increasingly critical role in improving the credibility of fusion diagnostics, ensuring safe device operation, optimizing material and structural design, and supporting the engineering realization of fusion reactors.

## Author contributions

**Zhuo Zuo:** Conceptualization, Methodology, Software, Writing – original draft. **BingqiLiu**: Conceptualization, Methodology, Software, Writing –review and editing. **Hao Feng**: Conceptualization, Visualization, Investigation. **Jie Zhang**: Software, Visualization. **Xianghe Liu**: Writing –review and editing. **Haoran Liu**: Supervision, Validation. **Peng Li**: Funding acquisition, Formal analysis. **Qibiao Wang**: Funding acquisition, Investigation. **Mingzhe Liu**: Funding acquisition, Investigation.

## References:

Abdelnour, M.R., Liu, J., Hossny, K., Wajid, A.M., Li, W., Liu, Z., 2025. Prompt gamma neutron activation analysis: A review of applications, design, analytics, challenges, and prospects. Radiation Physics and Chemistry 234, 112693. https://doi.org/10.1016/j.radphyschem.2025.112693

Abdou, M., Riva, M., Ying, A., Day, C., Loarte, A., Baylor, L.R., Humrickhouse, P., Fuerst, T.F., Cho, S., 2021. Physics and technology considerations for the deuterium–tritium fuel cycle and conditions for tritium fuel self sufficiency. Nucl. Fusion 61, 013001. https://doi.org/10.1088/1741-4326/abbf35

Abu-Shawareb, H., Acree, R., Adams, P., Adams, J., Addis, B., Aden, R., Adrian, P., Afeyan, B.B., Aggleton, M., Aghaian, L., Aguirre, A., Aikens, D., Akre, J., Albert, F., Albrecht, M., Albright, B.J., Albritton, J., Alcala, J., Alday, C., Alessi, D.A., Alexander, N., Alfonso, J., Alfonso, N., Alger, E., Ali, S.J., Ali, Z.A., Allen, A., Alley, W.E., Amala, P., Amendt, P.A., Amick, P., Ammula, S., Amorin, C., Ampleford, D.J., Anderson, R.W., Anklam, T., Antipa, N., Appelbe, B., Aracne-Ruddle, C., Araya, E., Archuleta, T.N., Arend, M., Arnold, P., Arnold, T., Arsenlis, A., Asay, J., Atherton, L.J., Atkinson, D., Atkinson, R., Auerbach, J.M., Austin, B., Auyang, L., Awwal, A.A.S., Aybar, N., Ayers, J., Ayers, S.,

Ayers, T., Azevedo, S., Bachmann, B., Back, C.A., Bae, J., Bailey, D.S., Bailey, J., Baisden, T., Baker, K.L., Baldis, H., Barber, D., Barberis, M., Barker, D., Barnes, A., Barnes, C.W., Barrios, M.A., Barty, C., Bass, I., Batha, S.H., Baxamusa, S.H., Bazan, G., Beagle, J.K., Beale, R., Beck, B.R., Beck, J.B., Bedzyk, M., Beeler, R.G., Beeler, R.G., Behrendt, W., Belk, L., Bell, P., Belyaev, M., Benage, J.F., Bennett, G., Benedetti, L.R., Benedict, L.X., Berger, R.L., Bernat, T., Bernstein, L.A., Berry, B., Bertolini, L., Besenbruch, G., Betcher, J., Bettenhausen, R., Betti, R., Bezzerides, B., Bhandarkar, S.D., Bickel, R., Biener, J., Biesiada, T., Bigelow, K., Bigelow-Granillo, J., Bigman, V., Bionta, R.M., Birge, N.W., Bitter, M., Black, A.C., Bleile, R., Bleuel, D.L., Bliss, E., Bliss, E., Blue, B., Boehly, T., Boehm, K., Boley, C.D., Bonanno, R., Bond, E.J., Bond, T., Bonino, M.J., Borden, M., Bourgade, J.-L., Bousquet, J., Bowers, J., Bowers, M., Boyd, R., Boyle, D., Bozek, A., Bradley, D.K., Bradley, K.S., Bradley, P.A., Bradley, L., Brannon, L., Brantley, P.S., Braun, D., Braun, T., Brienza-Larsen, K., Briggs, R., Briggs, T.M., Britten, J., Brooks, E.D., Browning, D., Bruhn, M.W., Brunner, T.A., Bruns, H., Brunton, G., Bryant, B., Buczek, T., Bude, J., Buitano, L., Burkhart, S., Burmark, J., Burnham, A., Burr, R., Busby, L.E., Butlin, B., Cabeltis, R., Cable, M., Cabot, W.H., Cagadas, B., Caggiano, J., Cahayag, R., Caldwell, S.E., Calkins, S., Callahan, D.A., Calleja-Aguirre, J., Camara, L., Camp, D., Campbell, E.M., Campbell, J.H., Carey, B., Carey, R., Carlisle, K., Carlson, L., Carman, L., Carmichael, J., Carpenter, A., Carr, C., Carrera, J.A., Casavant, D., Casey, A., Casey, D.T., Castillo, A., Castillo, E., Castor, J.I., Castro, C., Caughey, W., Cavitt, R., Celeste, J., Celliers, P.M., Cerjan, C., Chandler, G., Chang, B., Chang, C., Chang, J., Chang, L., Chapman, R., Chapman, T.D., Chase, L., Chen, H., Chen, H., Chen, K., Chen, L.-Y., Cheng, B., Chittenden, J., Choate, C., Chou, J., Chrien, R.E., Chrisp, M., Christensen, K., Christensen, M., Christiansen, N.S., Christopherson, A.R., Chung, M., Church, J.A., Clark, A., Clark, D.S., Clark, K., Clark, R., Claus, L., Cline, B., Cline, J.A., Cobble, J.A., Cochrane, K., Cohen, B., Cohen, S., Collette, M.R., Collins, G.W., Collins, L.A., Collins, T.J.B., Conder, A., Conrad, B., Conyers, M., Cook, A.W., Cook, D., Cook, R., Cooley, J.C., Cooper, G., Cope, T., Copeland, S.R., Coppari, F., Cortez, J., Cox, J., Crandall, D.H., Crane, J., Craxton, R.S., Cray, M., Crilly, A., Crippen, J.W., Cross, D., Cuneo, M., Cuotts, G., Czajka, C.E., Czechowicz, D., Daly, T., Danforth, P., Danly, C., Darbee, R., Darlington, B., Datte, P., Dauffy, L., Davalos, G., Davidovits, S., Davis, P., Davis, J., Dawson, S., Day, R.D., Day, T.H., Dayton, M., Deck, C., Decker, C., Deeney, C., DeFriend, K.A., Deis, G., Delamater, N.D., Delettrez, J.A., Demaret, R., Demos, S., Dempsey, S.M., Desjardin, R., Desjardins, T., Desjarlais, M.P., Dewald, E.L., DeYoreo, J., Diaz, S., Dimonte, G., Dittrich, T.R., Divol, L., Dixit, S.N., Dixon, J., Do, A., Dodd, E.S., Dolan, D., Donovan, A., Donovan, M., Döppner, T., Dorrer, C., Dorsano, N., Douglas, M.R., Dow, D., Downie, J., Downing, E., Dozieres, M., Draggoo, V., Drake, D., Drake, R.P., Drake, T., Dreifuerst, G., Drury, O., DuBois, D.F., DuBois, P.F., Dunham, G., Durocher, M., Dylla-Spears, R., Dymoke-Bradshaw, A.K.L., Dzenitis, B., Ebbers, C., Eckart, M., Eddinger, S., Eder, D., Edgell, D., Edwards, M.J., Efthimion, P., Eggert, J.H., Ehrlich, B., Ehrmann, P., Elhadj, S., Ellerbee, C., Elliott, N.S., Ellison, C.L., Elsner, F., Emerich, M., Engelhorn, K., England, T., English, E., Epperson, P., Epstein, R., Erbert, G., Erickson, M.A., Erskine, D.J., Erlandson, A., Espinosa, R.J., Estes, C., Estabrook, K.G., Evans, S., Fabyan, A., Fair, J., Fallejo, R., Farmer, N., Farmer,

W.A., Farrell, M., Fatherley, V.E., Fedorov, M., Feigenbaum, E., Fehrenbach, T., Feit, M., Felker, B., Ferguson, W., Fernandez, J.C., Fernandez-Panella, A., Fess, S., Field, J.E., Filip, C.V., Fincke, J.R., Finn, T., Finnegan, S.M., Finucane, R.G., Fischer, M., Fisher, A., Fisher, J., Fishler, B., Fittinghoff, D., Fitzsimmons, P., Flegel, M., Flippo, K.A., Florio, J., Folta, J., Folta, P., Foreman, L.R., Forrest, C., Forsman, A., Fooks, J., Foord, M., Fortner, R., Fournier, K., Fratanduono, D.E., Frazier, N., Frazier, T., Frederick, C., Freeman, M.S., Frenje, J., Frey, D., Frieders, G., Friedrich, S., Froula, D.H., Fry, J., Fuller, T., Gaffney, J., Gales, S., Le Galloudec, B., Le Galloudec, K.K., Gambhir, A., Gao, L., Garbett, W.J., Garcia, A., Gates, C., Gaut, E., Gauthier, P., Gavin, Z., Gaylord, J., Geddes, C.G.R., Geissel, M., Génin, F., Georgeson, J., Geppert-Kleinrath, H., Geppert-Kleinrath, V., Gharibyan, N., Gibson, J., Gibson, C., Giraldez, E., Glebov, V., Glendinning, S.G., Glenn, S., Glenzer, S.H., Goade, S., Gobby, P.L., Goldman, S.R., Golick, B., Gomez, M., Goncharov, V., Goodin, D., Grabowski, P., Grafil, E., Graham, P., Grandy, J., Grasz, E., Graziani, F.R., Greenman, G., Greenough, J.A., Greenwood, A., Gregori, G., Green, T., Griego, J.R., Grim, G.P., Grondalski, J., Gross, S., Guckian, J., Guler, N., Gunney, B., Guss, G., Haan, S., Hackbarth, J., Hackel, L., Hackel, R., Haefner, C., Hagmann, C., Hahn, K.D., Hahn, S., Haid, B.J., Haines, B.M., Hall, B.M., Hall, C., Hall, G.N., Hamamoto, M., Hamel, S., Hamilton, C.E., Hammel, B.A., Hammer, J.H., Hampton, G., Hamza, A., Handler, A., Hansen, S., Hanson, D., Haque, R., Harding, D., Harding, E., Hares, J.D., Harris, D.B., Harte, J.A., Hartouni, E.P., Hatarik, R., Hatchett, S., Hauer, A.A., Havre, M., Hawley, R., Hayes, J., Hayes, J., Hayes, S., Hayes-Sterbenz, A., Haynam, C.A., Haynes, D.A., Headley, D., Heal, A., Heebner, J.E., Heerey, S., Heestand, G.M., Heeter, R., Hein, N., Heinbockel, C., Hendricks, C., Henesian, M., Heninger, J., Henrikson, J., Henry, E.A., Herbold, E.B., Hermann, M.R., Hermes, G., Hernandez, J.E., Hernandez, V.J., Herrmann, M.C., Herrmann, H.W., Herrera, O.D., Hewett, D., Hibbard, R., Hicks, D.G., Higginson, D.P., Hill, D., Hill, K., Hilsabeck, T., Hinkel, D.E., Ho, D.D., Ho, V.K., Hoffer, J.K., Hoffman, N.M., Hohenberger, M., Hohensee, M., Hoke, W., Holdener, D., Holdener, F., Holder, J.P., Holko, B., Holunga, D., Holzrichter, J.F., Honig, J., Hoover, D., Hopkins, D., Berzak Hopkins, L.F., Hoppe, M., Hoppe, M.L., Horner, J., Hornung, R., Horsfield, C.J., Horvath, J., Hotaling, D., House, R., Howell, L., Hsing, W.W., Hu, S.X., Huang, H., Huckins, J., Hui, H., Humbird, K.D., Hund, J., Hunt, J., Hurricane, O.A., Hutton, M., Huynh, K.H.-K., Inandan, L., Iglesias, C., Igumenshchev, I.V., Ivanovich, I., Izumi, N., Jackson, M., Jackson, J., Jacobs, S.D., James, G., Jancaitis, K., Jarboe, J., Jarrott, L.C., Jasion, D., Jaquez, J., Jeet, J., Jenei, A.E., Jensen, J., Jimenez, J., Jimenez, R., Jobe, D., Johal, Z., Johns, H.M., Johnson, D., Johnson, M.A., Gatu Johnson, M., Johnson, R.J., Johnson, S., Johnson, S.A., Johnson, T., Jones, K., Jones, O., Jones, M., Jorge, R., Jorgenson, H.J., Julian, M., Jun, B.I., Jungquist, R., Kaae, J., Kabadi, N., Kaczala, D., Kalantar, D., Kangas, K., Karasiev, V.V., Karasik, M., Karpenko, V., Kasarky, A., Kasper, K., Kauffman, R., Kaufman, M.I., Keane, C., Keaty, L., Kegelmeyer, L., Keiter, P.A., Kellett, P.A., Kellogg, J., Kelly, J.H., Kemic, S., Kemp, A.J., Kemp, G.E., Kerbel, G.D., Kershaw, D., Kerr, S.M., Kessler, T.J., Key, M.H., Khan, S.F., Khater, H., Kiikka, C., Kilkenny, J., Kim, Y., Kim, Y.-J., Kimko, J., Kimmel, M., Kindel, J.M., King, J., Kirkwood, R.K., Klaus, L., Klem, D., Kline, J.L., Klingmann, J., Kluth, G., Knapp, P., Knauer, J., Knipping, J., Knudson, M., Kobs, D., Koch, J., Kohut, T., Kong, C., Koning,

J.M., Koning, P., Konior, S., Kornblum, H., Kot, L.B., Koziozemski, B., Kozlowski, M., Kozlowski, P.M., Krammen, J., Krasheninnikova, N.S., Krauland, C.M., Kraus, B., Krauser, W., Kress, J.D., Kritcher, A.L., Krieger, E., Kroll, J.J., Kruer, W.L., Kruse, M.K.G., Kucheyev, S., Kumbera, M., Kumpan, S., Kunimune, J., Kur, E., Kustowski, B., Kwan, T.J.T., Kyrala, G.A., Laffite, S., Lafon, M., LaFortune, K., Lagin, L., Lahmann, B., Lairson, B., Landen, O.L., Land, T., Lane, M., Laney, D., Langdon, A.B., Langenbrunner, J., Langer, S.H., Langro, A., Lanier, N.E., Lanier, T.E., Larson, D., Lasinski, B.F., Lassle, D., LaTray, D., Lau, G., Lau, N., Laumann, C., Laurence, A., Laurence, T.A., Lawson, J., Le, H.P., Leach, R.R., Leal, L., Leatherland, A., LeChien, K., Lechleiter, B., Lee, A., Lee, M., Lee, T., Leeper, R.J., Lefebvre, E., Leidinger, J.-P., LeMire, B., Lemke, R.W., Lemos, N.C., Le Pape, S., Lerche, R., Lerner, S., Letts, S., Levedahl, K., Lewis, T., Li, C.K., Li, H., Li, J., Liao, W., Liao, Z.M., Liedahl, D., Liebman, J., Lindford, G., Lindman, E.L., Lindl, J.D., Loey, H., London, R.A., Long, F., Loomis, E.N., Lopez, F.E., Lopez, H., Losbanos, E., Loucks, S., Lowe-Webb, R., Lundgren, E., Ludwigsen, A.P., Luo, R., Lusk, J., Lyons, R., Ma, T., Macallop, Y., MacDonald, M.J., MacGowan, B.J., Mack, J.M., Mackinnon, A.J., MacLaren, S.A., MacPhee, A.G., Magelssen, G.R., Magoon, J., Malone, R.M., Malsbury, T., Managan, R., Mancini, R., Manes, K., Maney, D., Manha, D., Mannion, O.M., Manuel, A.M., Manuel, M.J.-E., Mapoles, E., Mara, G., Marcotte, T., Marin, E., Marinak, M.M., Mariscal, D.A., Mariscal, E.F., Marley, E.V., Marozas, J.A., Marquez, R., Marshall, C.D., Marshall, F.J., Marshall, M., Marshall, S., Marticorena, J., Martinez, J.I., Martinez, D., Maslennikov, I., Mason, D., Mason, R.J., Masse, L., Massey, W., Masson-Laborde, P.-E., Masters, N.D., Mathisen, D., Mathison, E., Matone, J., Matthews, M.J., Mattoon, C., Mattsson, T.R., Matzen, K., Mauche, C.W., Mauldin, M., McAbee, T., McBurney, M., Mccarville, T., McCrory, R.L., McEvoy, A.M., McGuffey, C., Mcinnis, M., McKenty, P., McKinley, M.S., McLeod, J.B., McPherson, A., Mcquillan, B., Meamber, M., Meaney, K.D., Meezan, N.B., Meissner, R., Mehlhorn, T.A., Mehta, N.C., Menapace, J., Merrill, F.E., Merritt, B.T., Merritt, E.C., Meyerhofer, D.D., Mezyk, S., Mich, R.J., Michel, P.A., Milam, D., Miller, C., Miller, D., Miller, D.S., Miller, E., Miller, E.K., Miller, J., Miller, M., Miller, P.E., Miller, T., Miller, W., Miller-Kamm, V., Millot, M., Milovich, J.L., Minner, P., Miquel, J.-L., Mitchell, S., Molvig, K., Montesanti, R.C., Montgomery, D.S., Monticelli, M., Montoya, A., Moody, J.D., Moore, A.S., Moore, E., Moran, M., Moreno, J.C., Moreno, K., Morgan, B.E., Morrow, T., Morton, J.W., Moses, E., Moy, K., Muir, R., Murillo, M.S., Murray, J.E., Murray, J.R., Munro, D.H., Murphy, T.J., Munteanu, F.M., Nafziger, J., Nagayama, T., Nagel, S.R., Nast, R., Negres, R.A., Nelson, A., Nelson, D., Nelson, J., Nelson, S., Nemethy, S., Neumayer, P., Newman, K., Newton, M., Nguyen, H., Di Nicola, J.-M.G., Di Nicola, P., Niemann, C., Nikroo, A., Nilson, P.M., Nobile, A., Noorai, V., Nora, R.C., Norton, M., Nostrand, M., Note, V., Novell, S., Nowak, P.F., Nunez, A., Nyholm, R.A., O'Brien, M., Oceguera, A., Oertel, J.A., Oesterle, A.L., Okui, J., Olejniczak, B., Oliveira, J., Olsen, P., Olson, B., Olson, K., Olson, R.E., Opachich, Y.P., Orsi, N., Orth, C.D., Owen, M., Padalino, S., Padilla, E., Paguio, R., Paguio, S., Paisner, J., Pajoom, S., Pak, A., Palaniyappan, S., Palma, K., Pannell, T., Papp, F., Paras, D., Parham, T., Park, H.-S., Pasternak, A., Patankar, S., Patel, M.V., Patel, P.K., Patterson, R., Patterson, S., Paul, B., Paul, M., Pauli, E., Pearce, O.T., Pearcy, J., Pedretti, A., Pedrotti, B., Peer, A., Pelz, L.J., Penetrante, B.,

Penner, J., Perez, A., Perkins, L.J., Pernice, E., Perry, T.S., Person, S., Petersen, D., Petersen, T., Peterson, D.L., Peterson, E.B., Peterson, J.E., Peterson, J.L., Peterson, K., Peterson, R.R., Petrasso, R.D., Philippe, F., Phillion, D., Phipps, T.J., Piceno, E., Pickworth, L., Ping, Y., Pino, J., Piston, K., Plummer, R., Pollack, G.D., Pollaine, S.M., Pollock, B.B., Ponce, D., Ponce, J., Pontelandolfo, J., Porter, J.L., Post, J., Poujade, O., Powell, C., Powell, H., Power, G., Pozulp, M., Prantil, M., Prasad, M., Pratuch, S., Price, S., Primdahl, K., Prisbrey, S., Procassini, R., Pruyne, A., Pudliner, B., Qiu, S.R., Quan, K., Quinn, M., Quintenz, J., Radha, P.B., Rainer, F., Ralph, J.E., Raman, K.S., Raman, R., Rambo, P.W., Rana, S., Randewich, A., Rardin, D., Ratledge, M., Ravelo, N., Ravizza, F., Rayce, M., Raymond, A., Raymond, B., Reed, B., Reed, C., Regan, S., Reichelt, B., Reis, V., Reisdorf, S., Rekow, V., Remington, B.A., Rendon, A., Requieron, W., Rever, M., Reynolds, H., Reynolds, J., Rhodes, J., Rhodes, M., Richardson, M.C., Rice, B., Rice, N.G., Rieben, R., Rigatti, A., Riggs, S., Rinderknecht, H.G., Ring, K., Riordan, B., Riquier, R., Rivers, C., Roberts, D., Roberts, V., Robertson, G., Robey, H.F., Robles, J., Rocha, P., Rochau, G., Rodriguez, J., Rodriguez, S., Rosen, M.D., Rosenberg, M., Ross, G., Ross, J.S., Ross, P., Rouse, J., Rovang, D., Rubenchik, A.M., Rubery, M.S., Ruiz, C.L., Rushford, M., Russ, B., Rygg, J.R., Ryujin, B.S., Sacks, R.A., Sacks, R.F., Saito, K., Salmon, T., Salmonson, J.D., Sanchez, J., Samuelson, S., Sanchez, M., Sangster, C., Saroyan, A., Sater, J., Satsangi, A., Sauers, S., Saunders, R., Sauppe, J.P., Sawicki, R., Sayre, D., Scanlan, M., Schaffers, K., Schappert, G.T., Schiaffino, S., Schlossberg, D.J., Schmidt, D.W., Schmit, P.F., Smidt, J.M., Schneider, D.H.G., Schneider, M.B., Schneider, R., Schoff, M., Schollmeier, M., Schroeder, C.R., Schrauth, S.E., Scott, H.A., Scott, I., Scott, J.M., Scott, R.H.H., Scullard, C.R., Sedillo, T., Seguin, F.H., Seka, W., Senecal, J., Sepke, S.M., Seppala, L., Sequoia, K., Severyn, J., Sevier, J.M., Sewell, N., Seznec, S., Shah, R.C., Shamlian, J., Shaughnessy, D., Shaw, M., Shaw, R., Shearer, C., Shelton, R., Shen, N., Sherlock, M.W., Shestakov, A.I., Shi, E.L., Shin, S.J., Shingleton, N., Shmayda, W., Shor, M., Shoup, M., Shuldberg, C., Siegel, L., Silva, F.J., Simakov, A.N., Sims, B.T., Sinars, D., Singh, P., Sio, H., Skulina, K., Skupsky, S., Slutz, S., Sluyter, M., Smalyuk, V.A., Smauley, D., Smeltser, R.M., Smith, C., Smith, I., Smith, J., Smith, L., Smith, R., Smith, R., Schölmerich, M., Sohn, R., Sommer, S., Sorce, C., Sorem, M., Soures, J.M., Spaeth, M.L., Spears, B.K., Speas, S., Speck, D., Speck, R., Spears, J., Spinka, T., Springer, P.T., Stadermann, M., Stahl, B., Stahoviak, J., Stanley, J., Stanton, L.G., Steele, R., Steele, W., Steinman, D., Stemke, R., Stephens, R., Sterbenz, S., Sterne, P., Stevens, D., Stevers, J., Still, C.H., Stoeckl, C., Stoeffl, W., Stolken, J.S., Stolz, C., Storm, E., Stone, G., Stoupin, S., Stout, E., Stowers, I., Strauser, R., Streckart, H., Streit, J., Strozzi, D.J., Stutz, J., Summers, L., Suratwala, T., Sutcliffe, G., Suter, L.J., Sutton, S.B., Svidzinski, V., Swadling, G., Sweet, W., Szoke, A., Tabak, M., Takagi, M., Tambazidis, A., Tang, V., Taranowski, M., Taylor, L.A., Telford, S., Theobald, W., Thi, M., Thomas, A., Thomas, C.A., Thomas, I., Thomas, R., Thompson, I.J., Thongstisubskul, A., Thorsness, C.B., Tietbohl, G., Tipton, R.E., Tobin, M., Tomlin, N., Tommasini, R., Toreja, A.J., Torres, J., Town, R.P.J., Townsend, S., Trenholme, J., Trivelpiece, A., Trosseille, C., Truax, H., Trummer, D., Trummer, S., Truong, T., Tubbs, D., Tubman, E.R., Tunnell, T., Turnbull, D., Turner, R.E., Ulitsky, M., Upadhye, R., Vaher, J.L., VanArsdall, P., VanBlarcom, D., Vandenboomgaerde, M., VanQuinlan, R., Van Wonterghem, B.M.,

Varnum, W.S., Velikovich, A.L., Vella, A., Verdon, C.P., Vermillion, B., Vernon, S., Vesey, R., Vickers, J., Vignes, R.M., Visosky, M., Vocke, J., Volegov, P.L., Vonhof, S., Von Rotz, R., Vu, H.X., Vu, M., Wall, D., Wall, J., Wallace, R., Wallin, B., Walmer, D., Walsh, C.A., Walters, C.F., Waltz, C., Wan, A., Wang, A., Wang, Y., Wark, J.S., Warner, B.E., Watson, J., Watt, R.G., Watts, P., Weaver, J., Weaver, R.P., Weaver, S., Weber, C.R., Weber, P., Weber, S.V., Wegner, P., Welday, B., Welser-Sherrill, L., Weiss, K., Wharton, K.B., Wheeler, G.F., Whistler, W., White, R.K., Whitley, H.D., Whitman, P., Wickett, M.E., Widmann, K., Widmayer, C., Wiedwald, J., Wilcox, R., Wilcox, S., Wild, C., Wilde, B.H., Wilde, C.H., Wilhelmsen, K., Wilke, M.D., Wilkens, H., Wilkins, P., Wilks, S.C., Williams, E.A., Williams, G.J., Williams, W., Williams, W.H., Wilson, D.C., Wilson, B., Wilson, E., Wilson, R., Winters, S., Wisoff, P.J., Wittman, M., Wolfe, J., Wong, A., Wong, K.W., Wong, L., Wong, N., Wood, R., Woodhouse, D., Woodruff, J., Woods, D.T., Woods, S., Woodworth, B.N., Wooten, E., Wootton, A., Work, K., Workman, J.B., Wright, J., Wu, M., Wuest, C., Wysocki, F.J., Xu, H., Yamaguchi, M., Yang, B., Yang, S.T., Yatabe, J., Yeamans, C.B., Yee, B.C., Yi, S.A., Yin, L., Young, B., Young, C.S., Young, C.V., Young, P., Youngblood, K., Yu, J., Zacharias, R., Zagaris, G., Zaitseva, N., Zaka, F., Ze, F., Zeiger, B., Zika, M., Zimmerman, G.B., Zobrist, T., Zuegel, J.D., Zylstra, A.B., The Indirect Drive ICF Collaboration, 2024. Achievement of Target Gain Larger than Unity in an Inertial Fusion Experiment. Phys. Rev. Lett. 132, 065102. https://doi.org/10.1103/PhysRevLett.132.065102

Afanasenko, E.S., Portnov, D.V., Vysokikh, J.G., Kashchuk, Y.A., 2026. Synchronization of the OpenMC and MCNP Models for ITER Neutronics. Fusion Science and Technology 82, 805–813. https://doi.org/10.1080/15361055.2025.2476818

Afanasenko, R., Elbez-Uzan, J., Leichtle, D., Park, J.H., Pereslavtsev, P., 2026. DEMO Shutdown Dose Rate Assessment Inside the Vacuum Vessel. Applied Sciences 16, 1983. https://doi.org/10.3390/app16041983

Afanasenko, R., Vukolov, K., Andreenko, E., Orlovskiy, I., Rodionov, R., 2024. Radiation environment in ITER tokamak equatorial port 11. Fusion Engineering and Design 200, 114208. https://doi.org/10.1016/j.fusengdes.2024.114208

Ahnouz, I., Arahmane, H., Sebihi, R., 2024. A Review of Neutron–Gamma-Ray Discrimination Methods Using Organic Scintillators. Nuclear Science and Engineering 198, 2241–2273. https://doi.org/10.1080/00295639.2024.2316946

Akino, R., Sekiguchi, H., Ojima, C., Sonda, M., Hakamata, H., Suzuki, A., Tsubomatsu, M., Oda, Y., 2026. Development of a high-speed and precise automatic reading system of CR-39 plastic nuclear track detector for neutron dose measurements.

Ali, F., Surette, J., Atanackovic, J., 2024. Calculation of response parameters for a neutron long counter instrument. Applied Radiation and Isotopes 214, 111502. https://doi.org/10.1016/j.apradiso.2024.111502

Álvarez, I., Anguiano, M., Mota, F., Hernández, R., Serrano, M., Sosa, D., Palermo, I., Leon-Gutierrez, E., Noce, S., Moro, F., Arbeiter, F., Qiu, Y., Park, J.H., Ibarra, Á., 2026. Assessment of EUROFER97, tungsten and CuCrZr alloy transmutation effects under IFMIF-DONES and DEMO conditions. Journal of Nuclear Materials 618, 156228. https://doi.org/10.1016/j.jnucmat.2025.156228

Álvarez, I., De La Torre, J.A., Anguiano, M., Mota, F., Oliver, C., Klix, A., Qiu, Y., Leichtle, D.,

Arbeiter, F., Becerril, S., 2025. Assessment of SPND Signal in IFMIF-DONES using a Monte Carlo Modelling. J Fusion Energ 44, 69. https://doi.org/10.1007/s10894-025-00542-y

An, B., Deng, Y., Jin, Z., Sun, S., 2025. Scintillators for Neutron Detection and Imaging: Advances and Prospects. Adv Funct Materials 35, 2422522. https://doi.org/10.1002/adfm.202422522

Anagnostopoulou, V., Gandolfo, G., Panza, F., Colangeli, A., Flammini, D., Angelone, M., Pillon, M., Gelfusa, M., Marocco, D., 2026. Preliminary calibration procedure for the Phase1 neutron and gamma-ray diagnostics of the Divertor Tokamak Test (DTT) facility. Fusion Engineering and Design 222, 115537. https://doi.org/10.1016/j.fusengdes.2025.115537

Bachmann, C., Siccinio, M., Acampora, E., Aiello, G., Bajari, J., Boscary, J., Bruschi, A., Claps, V., Cufar, A., Elbez-Uzan, J., Federici, G., Franke, T., Germano, G., Giannini, L., Gliss, C., Härtl, T., Hauer, V., Hopf, C., Kannamüller, M., Leichtle, D., Lombroni, R., Luongo, C., Maisonnier, D., Marek, P., Maione, I., Marzullo, D., Maviglia, F., Mollicone, P., Moscato, I., Mozzillo, R., Muscat, M., Pagani, I., Park, J.H., Pautasso, G., Pereslavtsev, P., Quartararo, A., Renard, S., Schreck, S., Späh, P., Steinbacher, T., Tarallo, A., Valentine, A., Vinoni, P., Vallone, E., Vigano, F., Wiesen, S., Wu, C., Zammuto, I., 2025. Engineering concept of the VNS - a beam-driven tokamak for component testing. Fusion Engineering and Design 211, 114796. https://doi.org/10.1016/j.fusengdes.2024.114796

Baselga, S., Montbarbon, E., 2024. Neutron and Gamma Pulse Shape Discrimination by Robust Determination of the Decay Shape. Applied Sciences 14, 5532. https://doi.org/10.3390/app14135532

Battye, M.I., Perinpanayagam, S., 2025. Digital Twins in Fusion Energy Research: Current State and Future Directions. IEEE Access 13, 75787–75821. https://doi.org/10.1109/ACCESS.2025.3561920

Bencivenni, G., Balossino, I., Cibinetto, G., De Oliveira, R., Farinelli, R., Felici, G., Garzia, I., Gatta, M., Giovannetti, M., Gramigna, S., Lavezzi, L., Mezzadri, G., Morello, G., Papalino, G., Poli Lener, M., Scodeggio, M., 2023. Thermal neutron detection based on resistive gaseous devices. EPJ Web Conf. 288, 06012. https://doi.org/10.1051/epjconf/202328806012

Bielecki, J., Kurowski, A., 2019. Neutron Diagnostics for Tokamak Plasma: From a Plasma Diagnostician Perspective. J Fusion Energ 38, 386–393. https://doi.org/10.1007/s10894-018-0195-9

Brown, J.A., Goldblum, B.L., Gordon, J.M., Laplace, T.A., Nagel, T.S., Venkatraman, A., 2024. Proton discrimination in CLYC for fast neutron spectroscopy. Nuclear Instruments and Methods in Physics Research Section A: Accelerators, Spectrometers, Detectors and Associated Equipment 1069, 169859. https://doi.org/10.1016/j.nima.2024.169859

Cancelli, S., Caruggi, F., Perelli Cippo, E., Putignano, O., Celora, A., Gorini, G., Krzystyniak, M., Muraro, A., Romanelli, G., Pinna, R.S., Tardocchi, M., Croci, G., 2024. Development of a multi-layer high-efficiency GEM-based neutron detector for spallation sources. Sci Rep 14, 23954. https://doi.org/10.1038/s41598-024-74958-5

Carin, Y., Sugimoto, M., Dzitko, H., Hasegawa, K., Gex, D., Kondo, K., Chel, S., Facco, A., Jimenez-Rey, D., Kasugai, A., Cara, P.H., Micciche, G., Pisent, A., Radloff, D., Terentyev, D., 2024. IFMIF/EVEDA achievements overview. Fusion Engineering and Design 201,

114258. https://doi.org/10.1016/j.fusengdes.2024.114258

Cesaroni, S., Bombarda, F., Bollanti, S., Cianfarani, C., Claps, G., Cordella, F., Flora, F., Marinelli, M., Mezi, L., Milani, E., Murra, D., Pacella, D., Palomba, S., Verona, C., Verona-Rinati, G., 2023. Conceptual design of CVD diamond tomography systems for fusion devices. Fusion Engineering and Design 197, 114037. https://doi.org/10.1016/j.fusengdes.2023.114037

Cesaroni, S., Marocco, D., Anagnostopoulou, V., Belli, F., Colangeli, A., Fonnesu, N., Loreti, S., Moro, F., Pagano, G., Pirovano, E., Pompili, F., Pontesilli, M., Zimbal, A., Esposito, B., 2025. Characterization of a new 4He scintillator detector prototype for the ITER Radial Neutron Camera. Fusion Engineering and Design 215, 114948. https://doi.org/10.1016/j.fusengdes.2025.114948

Cesaroni, S., Marocco, D., Marzullo, D., Moro, F., Belli, F., Brolatti, G., Centioli, C., Occhiuto, E., Riva, M., Rocchi, G., Esposito, B., 2024. Design of a position monitoring system for the ITER radial neutron camera. Fusion Engineering and Design 203, 114439. https://doi.org/10.1016/j.fusengdes.2024.114439

Chatzikos, V., Savva, M.I., Vasilopoulou, T., Stamatelatos, I.E., Terentyev, D., Stankovskiy, A., Patronis, N., Mergia, K., 2024. Experimental Validation of Transmutation Products Calculations in Neutron Irradiated Tungsten. https://doi.org/10.2139/ssrn.5056825

Chen, S., 2022. New evaluation of neutron-induced displacement damage cross section for EUROFER97. Nuclear Materials and Energy 33, 101284. https://doi.org/10.1016/j.nme.2022.101284

Choi, J.Y., Joo, K.K., Park, J., Cheoun, M.-K., 2025. Machine-Learning-Assisted Optimization of Pulse Shape Discrimination in Gadolinium-Loaded Liquid Scintillators for the RENE Experiment. Progress of Theoretical and Experimental Physics 2025, 093C02. https://doi.org/10.1093/ptep/ptaf106

Cohen-Tanugi, D., Stapelberg, M.G., Short, M.P., Ferry, S.E., Whyte, D.G., Hartwig, Z.S., Buonassisi, T., 2024. Long-term research and design strategies for fusion energy materials. Matter 7, 4148–4160. https://doi.org/10.1016/j.matt.2024.08.017

D’Amico, N., Puri, S., Jones, I., Gillespie, A., Lin, C., Zhao, B., Duncan, R.V., 2026. Boron nitride coatings for the enhanced detection of neutrons in CR-39. Nuclear Engineering and Technology 58, 104027. https://doi.org/10.1016/j.net.2025.104027

Dankowski, J., Bielecki, J., Błądek, J., Conroy, S., Coriton, B., Croci, G., Dworaka, D., Ericsson, G., Eriksson, J., Wójcik-Gargula, A., Hjalmarsson, A., Jardin, A., Kantor, R., Kovalev, A., Król, K., Kulińska, A., Kurowski, A., Mariano, G., Mehrara, R., Morawski, D., Rebai, M., Scholz, M., Scioscioli, F., Tardocchi, M., Tracz, G., Turzański, M., Wiącek, U., 2025. Development and performance of the thin-foil proton recoil spectrometer for ITER plasma diagnostics. Fusion Engineering and Design 219, 115263. https://doi.org/10.1016/j.fusengdes.2025.115263

Deng, C., Hu, Q., Li, P., Wang, Q., Xie, B., Yang, J., Tuo, X., 2024. Research on a Neutron Detector with a Boron-Lined Multilayer Converter. Applied Sciences 14, 4269. https://doi.org/10.3390/app14104269

Di Chicco, A., Horst, F., Boscolo, D., Schuy, C., Weber, U., Zboril, M., 2024. Bonner sphere measurements of high-energy neutron spectra from a 1 GeV/u 56Fe ion beam on an aluminum target and comparison to spectra obtained by Monte Carlo simulations. Front.

Phys. 12, 1456472. https://doi.org/10.3389/fphy.2024.1456472

Diawara, Y. (Ed.), 2023. Neutron Detectors for Scattering Applications, Particle Acceleration and Detection. Springer Nature Switzerland, Cham. https://doi.org/10.1007/978-3-031-36546-1

Du, Q., Wu, F., Zhang, J., 2025. Comprehensive comparisons of different fusion fuels by transfer learning. Physics of Plasmas 32, 022706. https://doi.org/10.1063/5.0246387

Du, X., Zhang, J., Sheng, L., Qiu, M., Tang, C., 2022. Neutron spectrum unfolding of the magnetic proton recoil spectrometer using the GRAVEL and MLEM algorithms.

Duan, H., Zhang, J., Zhao, C., Zhang, Yunsheng, Zhang, Yipo, Wang, Z., 2025. Design of compact neutron detector for tokamak. Nuclear Instruments and Methods in Physics Research Section A: Accelerators, Spectrometers, Detectors and Associated Equipment 1075, 170456. https://doi.org/10.1016/j.nima.2025.170456

Duteil, M.P., Izarra, G.D., Jammes, C., 2024. Measurement and modelling signals from an optical fission chambers during reactor irradiation. Nuclear Instruments and Methods in Physics Research Section A: Accelerators, Spectrometers, Detectors and Associated Equipment 1061, 169141. https://doi.org/10.1016/j.nima.2024.169141

Ebiwonjumi, B., Peterson, E., 2024. Uncertainty Analyses of Tritium Production and Gamma Heating Rates in the Fns Clean Benchmark Experiments. https://doi.org/10.2139/ssrn.4949078

Ebiwonjumi, B., Segantin, S., Peterson, E., 2025. OpenMC Interpretation of FNS SINBAD Shielding Benchmark Experiments. Fusion Science and Technology 81, 18–31. https://doi.org/10.1080/15361055.2024.2323747

Ericsson, G., 2019. Advanced Neutron Spectroscopy in Fusion Research. J Fusion Energ 38, 330–355. https://doi.org/10.1007/s10894-019-00213-9

Esposito, B., Marocco, D., Gandolfo, G., Belli, F., Bertalot, L., Blocki, J., Bocian, D., Brolatti, G., Cecconello, M., Centioli, C., Pereira, R.C., Conroy, S., Crescenzi, F., Cruz, N., De Bilbao, L., Domenicone, A., Ducasse, Q., Di Mambro, G., Dongiovanni, D., Eletxigerra, I., Etxeita, B., Fernandez, A., Ficker, O., Gallina, P., Giacomin, T., Ginoulhiac, G., Godlewski, J., Hjalmarsson, A., Imrisek, M., Kantor, R., Kasprzak, K., Kotula, J., Krasilnikov, V., Lewandowska, M., Maffucci, A., Marotta, U., Marzullo, D., Mazzitelli, G., Mazzone, G., Miklaszewski, R., Mikszuta-Michalik, K., Maciocha, W., Magagnino, S., Misano, M., Mlynar, J., Monti, C., Moro, F., Ortwein, R., Passeri, M., Pinna, T., Pirovano, E., Pisciotta, V., Pompili, F., Podda, S., Riva, M., Santos, B., Sousa, J., Swierblewski, J., Szklarz, P., Tatí, A., Ventre, S., Villone, F., Virgili, N., Zimbal, A., 2022. Progress of Design and Development for the ITER Radial Neutron Camera. J Fusion Energ 41, 22. https://doi.org/10.1007/s10894-022-00333-9

Favalli, A., Wiggins, B.W., Iliev, M., Richards, C.G., Ogren, K., McLean, T.D., Ianakiev, K.D., Hehlen, M.P., 2025. Next-generation neutron detection using a 6Li glass scintillator composite. Commun Phys 8, 10. https://doi.org/10.1038/s42005-024-01903-3

Feng, T., Ju, W., Jia, X., Chong, X., Wei, H., Zhao, Z., Li, J., Wang, J., 2026. Research status of inorganic scintillators based on lithium for neutron detection. Nuclear Engineering and Technology 58, 103971. https://doi.org/10.1016/j.net.2025.103971

Filliatre, P., 2024. Applicability to DEMO breeding blankets of neutron measurements techniques from fission reactors. Fusion Engineering and Design 202, 114332.

https://doi.org/10.1016/j.fusengdes.2024.114332
Fonnesu, N., Angelone, M., Loreti, S., Pillon, M., Villari, R., Batistoni, P., Colangeli, A., Flammini, D., Lungaroni, M., Moro, F., Noce, S., Previti, A., Litaudon, X., JET Contributors, 2024a. Measurement of tritium production in the helium cooled pebble bed test blanket module mock-up at JET during DTE2. Eur. Phys. J. Plus 139, 893. https://doi.org/10.1140/epjp/s13360-024-05670-6
Fonnesu, N., Beaumont, P., Berry, T., Colangeli, A., Dacquait, F., Damiano, M., Flammini, D., Grove, C.L., Litaudon, X., Loreti, S., Lungaroni, M., Mianowski, S., Moro, F., Noce, S., Peric, J., Previti, A., Radulović, V., Villari, R., Zito, P., 2025. ITER-relevant experimental neutronic activities at JET during DTE3 and at the Frascati neutron generator. Fusion Engineering and Design 219, 115297. https://doi.org/10.1016/j.fusengdes.2025.115297
Fonnesu, N., Loreti, S., Villari, R., Flammini, D., Mariano, G., Batistoni, P., Colangeli, A., Moro, F., Previti, A., Klix, A., JET Contributors, 2024b. Shutdown dose rate experiment at JET during DTE2. Eur. Phys. J. Plus 139, 432. https://doi.org/10.1140/epjp/s13360-024-05208-w
Fridrikhsen, D.S., Obudovsky, S.Yu., Kormilitsyn, T.M., Pankratenko, A.V., Kashchuk, Yu.A., Moiseev, N.N., 2026. Characterisation of LaCl3(Ce)-based detectors for D-D fusion neutron diagnostic. Radiation Measurements 191, 107600. https://doi.org/10.1016/j.radmeas.2025.107600
Fugazza, S.L., Marcer, G., Nocente, M., Ciurlino, A., Croci, G., Rosa, M.D., Dal Molin, A., Gallo, E., Gorini, G., Parisi, M., Raj, P., Rebai, M., Reinke, M., Rigamonti, D., Scioscioli, F., Tardocchi, M., 2026. Feasibility study of gamma-ray spectroscopy for the determination of the fusion power at the SPARC tokamak. Fusion Engineering and Design 222, 115403. https://doi.org/10.1016/j.fusengdes.2025.115403
Fusion Energy Sciences Advisory Committee, U.S. Department of Energy, Office of Fusion Energy Sciences, 2024. Basic Research Needs for Measurement Innovation. Fusion Energy Sciences Advisory Committee.
Garcia, J., JET Contributors, 2025. Importance of the second D–T campaign at JET for future fusion tokamak devices. Rev. Mod. Plasma Phys. 9, 10. https://doi.org/10.1007/s41614-025-00182-x
Garnett, R.L., Amsellem, A., Persaud, A., Miller, A.L., Smith, M.B., Byun, S.H., 2025. NeutralNet: an application of deep neural networks to pulse shape discrimination for use with accelerator-based neutron sources. Applied Radiation and Isotopes 224, 111891. https://doi.org/10.1016/j.apradiso.2025.111891
Garnett, R.L., Byun, S.H., 2024. NeutralNet: Development and testing of a machine learning solution for pulse shape discrimination. Applied Radiation and Isotopes 211, 111384. https://doi.org/10.1016/j.apradiso.2024.111384
Gatu Johnson, M., Schlossberg, D., Appelbe, B., Ball, J., Bitter, M., Casey, D.T., Celora, A., Ceurvorst, L., Chen, H., Conroy, S., Crilly, A., Croci, G., Dal Molin, A., Delgado-Aparicio, L., Efthimion, P., Eriksson, B., Eriksson, J., Forrest, C., Fry, C., Frenje, J., Gao, L., Geppert-Kleinrath, H., Geppert-Kleinrath, V., Gilson, E., Heuer, P.V., Hill, K., Khater, H., Kraus, F., Laggner, F., Lawrence, Y., Mackie, S., Meaney, K., Milder, A., Moore, A., Nocente, M., Pablant, N., Panontin, E., Rebai, M., Reichelt, B., Reinke, M., Rigamonti, D., Ross, J.S., Rubery, M., Russell, L., Tardocchi, M., Tinguely, R.A., Wink, C., 2024.

Learning from each other: Cross-cutting diagnostic development activities between magnetic and inertial confinement fusion (invited). Review of Scientific Instruments 95, 093533. https://doi.org/10.1063/5.0218498

Gerenton, V., Jardin, A., Wiącek, U., Drozdowicz, K., Kulinska, A., Kurowski, A., Scholz, M., Woźnicka, U., Dąbrowski, W., Łach, B., Mazon, D., 2024. AI-supported Modelling of a Simple TPR System for Fusion Neutron Measurement. J Fusion Energ 43, 10. https://doi.org/10.1007/s10894-024-00403-0

Giancarli, L.M., Ahn, M.-Y., Cho, S., Kawamura, Y., Leal-Pereira, A., Merola, M., Poitevin, Y., Ricapito, I., Sheng, Q., Tanigawa, Hiroyasu, Tanigawa, Hisashi, Van Der Laan, J.G., Wang, X., 2024. Status of the ITER TBM Program and overview of its technical objectives. Fusion Engineering and Design 203, 114424. https://doi.org/10.1016/j.fusengdes.2024.114424

Gilbert, M.R., 2025. Coupling Nuclear Predictions into Damage Simulations with SPECTRA-PKA. Nuclear Science and Engineering 199, 48. https://doi.org/10.1080/00295639.2024.2342506

Grulke, O., Acton, G., Adamek, J., Aggelis, D., Alamo-Calderon, R.-M., Albert, C., Aleynikov, P., Aleynikova, K., Alonso, A., Amanekwe, G.C., Anda, G., Andreeva, T., Andrew, E., Arkuszewski, A., Arnold, S., Arranz, M., Arvanitou, M., Ascasibar, E., Astrain Etxezarreta, M., Asztalos, O., Avramidis, K., Aymerich, E., Baciero, A., Bähner, J.-P., Baek, S.-G., Balden, M., Baldzuhn, J., Ballinger, S., Banduch, M., Bannmann, S., Bañon Navarro, A., Baylor, L., Benndorf, A., Beidler, C.D., Beiersdorf, D., De Beij, M., Van Berkel, M., Bertelli, N., Biedermann, C., Bieg, B., Biewer, T.M., Birkenmeier, G., Björk, L., Blackwell, B., Blank, H., Bluhm, T., Böckenhoff, D., Boeyaert, D., Bold, D., Bonciarelli, A., Bongiovi, G., Borchardt, M., Borodin, D., Bosman, T., Boumendjel, Y., Bouvain, H., Bozhenkov, S., Bräuer, T., Brandt, C., Brezinsek, S., Brunner, K.J., Buhler, A., Buller, S., Burton, L., Büschel, C., Bussiahn, R., Buttenschön, B., Buzás, A., Bykov, V., Cai, J., Calvo, I., Cappa, A., Carls, A., Carovani, F., Carr, M., Carralero, D., Carroll, T., De Carvalho, B.B., Casas, J.R., Castano-Bardawil, D., Cavalier, J., Cavazzana, R., Chaudhary, N., Chelis, I., Cipciar, D., Conway, G., Cordella, F., Corre, Y., Costello, P., Crombé, K., Cseh, G., Csillag, B., Cu Castillo, H.I., Czymek, G., Damm, H., Davies, R.J., Degenkolbe, S., Dekeyser, W., Delgado-Aparicio, L., Demby, A., Desgranges, C., Dhard, C.-P., Dinklage, A., Dittmar, T., Dittrich, L., Dräger, S., Dreval, M., Drevlak, M., Droste, J., Duligal, R.K., Dumortier, P., Dunai, D., Edlund, E., Edmondson, A., Van Eeten, P., Ehrke, G., Endler, M., Ennis, D.A., Escoto, F.J., Estrada, T., Faber, B., Federici, F., Fellinger, J., Feng, Y., Fernando, D.L., Fischer, S., Ford, O.P., Fornal, T., Frank, J., Frerichs, H., Fuchert, G., Fujii, K., Fukuyama, M., Galdón Quiroga, J., Gallego Llorente, J., Gao, Y., Garcia, K., Garcia, O.E., Garcia-Munoz, M., García Regaña, J.M., Geiger, B., Geiger, J., Geißler, P., Gerard, M., Giudicotti, L., Gonda, T., Gonzalez, J.C., González Ganzábal, A., Goodman, A., Goriaev, A., Gradic, D., Grahl, M., Grasser, M., Grekov, D., Grelier, E., Grenfel, G., Griener, M., Groth, M., Gruca, M., Gudicotti, F., Guerrero Arnaiz, J.F., Haak, V., De Haas, M., Haeussler, A., Hakola, A., Van Ham, L., Hammond, K.C., Hamstra, B., Han, X., Hansen, S.K., Harris, A., Harris, J.H., Harting, D., Hartmann, D., Hathiramani, D., Hausten, E.V., Heinrich, S., Helander, P., Held, G., Henderson, P., Henke, F., Henneberg, S.A., Henschke, L., Herold, F., Hillebrecht, H., Hinson, E., Hirsch,

M., Hoffmeister, A., Holtz, A., Hörmann, S.J., Höschen, D., Houry, M., Hromadka, J., Hua, J., Huang, X., Hunger, K., Hwangbo, D., Ida, K., Igitkhanov, Y., Iglesias Fernandez, S., Igochine, V., Ioannidis, Z., d'Isa, F.A., Jablonski, S., Jabłoński, B., Jakubowski, M., Jenko, F., Johansson, A., Johnson, C., Van Kaathoven, T.C.W., Kaczmarczyk, J., Kajita, S., Kallmeyer, J.-P., Kasahara, H., Kasparek, W., Kawan, C., Kazakov, Ye.O., Keller, S.A., Kenmochi, N., Kernbichler, W., Kharwandikar, A.K., Khokhlov, M., Killer, C., Kirschner, A., Kleiber, R., Klepper, C.C., Klinger, T., Knauer, J., Knaup, M., Knieps, A., Kobayashi, M., Kocsis, G., Koelbl, M., Kolesnichenko, Y., Könies, A., Kontula, J., Kornejew, P., Kovtun, Y., Kozulia, M., Krämer-Flecken, A., Krause, M., Kremeyer, T., Krier, L., Kriete, D.M., Krings, T., Krychowiak, M., Ksiazek, I., Kubkowska, M., Kulla, D., Kulyk, Y., Kumar, A., Kunkel, F., Kurki-Suonio, T., Kuzmych, I., Kwak, S., Laguardia, L., Langenberg, A., Laqua, H., Laqua, H.P., Leche, K., Lee, B., Lee, W., Leyh, H., Liang, Y., Liao, L., Licchelli, M., Lin, Z., Lisaj, M., Litnovsky, A., Litovoli, F., Loizu, J., Lomanowski, B., Lopez Cansino, R., Lopez Miranda, B., Lopez-Rodriguez, D., Lore, J., Lorenz, A., Louwe, J., De La Luna, E., Lunsford, R., Luo, Y., Lutsenko, V., Maaziz, N., Machielsen, M., Mackenbach, R., Madeira, M., Makowski, D., Manz, P., Maragkoudakis, E., Marchuk, O., Marinoni, A., Markl, M., Marsen, S., Martinez Fernandez, J., Martseniuk, Y., Marushchenko, N., Masuzaki, S., Matsheza, S., Maurer, D., Mayer, M., Mazur, D., McCarthy, K., McCormack, O., McNeely, P., Medina Roque, D., Meineke, J., Meitner, S., Menzel-Barbara, A., Van Milligen, B., Misdanitis, S., Mishchenko, A., Mitteau, R., Moeyaert, E., Mohammed, A.I., Moiseenko, V.E., Möller, A., Möller, S., Molnar, B., Moncada, V., Morfin-Guerrero, D., Morren, M.C.L., Moseev, D., Motojima, G., Mulas, S., Mulholland, P., Murugesan, V., Nagel, M., Nagy, D., Nair, V., Narbutt, Y., Naujoks, D., Neilson, H.G., Nespoli, F., Neu, G., Neu, R., Neubauer, O., Neuner, U., Ngo, S.K., Nicolai, D., Nielsen, S.K., Nikulsin, N., Nishizawa, T., Nitzsche, L., Nührenberg, C., Ochoukov, R., Offermanns, G., Ogawa, K., Ongena, J., Oosterbeek, J.W., Otte, M., Overduin, E., Pablant, N., Pacios, L., Panadero, N., Pandey, A., Parks, K., Partesotti, G., Pasch, E.A., De Pascuale, S., Pavlichenko, R., Pavone, A., Pawelec, E., De La Pena Gomez, A., Pereira, A., Perseo, V., Peterson, B., Pisano, F., Pitcher, S., Plaum, B., Plunk, G., Podavini, L., Polei, N.S., Poloskei, P., Ponomarenko, S., Pons-Villalonga, P., Popov, A., Porkolab, M., Proll, J.H.E., Pueschel, M.J., Raak, A., Ragona, R., Rahbarnia, K., Rasiński, M., Rasmussen, J., Raths, O., Rattawongnara, E., Refy, D., Reimold, F., Richert, T., Richou, M., Ricken, J., Riemann, J.S., Riße, K., De La Riva Villen, J., Roberg-Clark, G., Rodriguez, E., Rogge, C., Rohde, V., Romazanov, J., Romba, T., Rondeshagen, D., Rong, P., Rud, M., Rummel, T., Runov, A., Rust, N., Ryc, L., Ryndyk, D., Sakai, H., Salewski, M., Sanchez, E., Sanchis Sanchez, L., Satake, S., Satheeswaran, G., Schacht, J., Scharff, E., Scharmer, F., Schlisio, G., Schmid, K., Schmidt, B.S., Schmidt, G.L., Schmitz, O., Schneider, M., Schröder, T., Schroeder, R., Schülke, M., Schweer, B., Sereda, S., Shanahan, B., Shin, J., Shiraiwa, S., Sias, G., Sichta, P., Siddiki, F.B.T., Simko, S., Singh, L., Sipilae, S., Slaby, C., Śleczka, M., Smith, B., Smith, D.R., Smith, H.M., Smoniewski, J., Spolaore, M., Spring, A., Stange, T., Von Stechow, A., Steinbrunner, P., Stepanov, I., Stephey, L., Stroth, U., Suzuki, C., Suzuki, Y., Swee, C., Syrocki, L., Szepesi, T., Szymanski, M., Takahashi, H., Tamura, N., Tanaka, K., Tantos, C., Thiede, S., Thienpondt, H., Thomsen, H., Thun, T., Togo, S., Tork, T., Torkler, M.,

Trimiño Mora, H., Tsikouras, A., Valougeorgis, D., Varoutis, S., Vavrik, M., Vaz Mendes, S., Vecsei, M., Velasco, J.L., Versemann, L., Verstraeten, M., Vervier, M., Vianello, N., Viezzer, E., Villalobos-Granados, E.-M., Wagner, J., Wang, E., Wappl, M., Warmer, F., Wegner, Th., Wei, W., Weir, G., Wendler, N., White, A., Willensdorfer, M., Windisch, T., Winter, A., Winters, V., Wischmeier, M., De Wolf, R., Wolf, R.C., Wright, J., Wurden, G., Xanthopoulos, P., Xu, S., Yamada, H., Yan, X., Yang, J., Yang, Y., Ye, K., Yokoyama, M., Yoshinuma, M., Zamorski, B., Zanini, M., Zarnstorff, M., Zhang, D., Zhu, C., Zilker, M., Zimmermann, J., Zocco, A., Zohm, H., Zoletnik, S., 2026. Overview of Wendelstein 7-X high-performance operation. Nucl. Fusion 66, 116003. https://doi.org/10.1088/1741-4326/ae5f32

Guo, D., Chen, C., Wang, Z., Lin, J., Zhang, B., Ge, D., Chen, Z., 2022. Occupational Radiation Exposure Estimated for Fusion Demonstration Reactors and Beyond. Fusion Science and Technology 78, 103–110. https://doi.org/10.1080/15361055.2021.1960089

He, J., Zhong, G., Hong, X., Zhu, Q., Xu, M., Wang, W., Yang, Y., Li, Z., 2026. Improvement of the neutron flux detector structural design for EAST. Fusion Engineering and Design 225, 115665. https://doi.org/10.1016/j.fusengdes.2026.115665

He, R., Niu, X.-Y., Wang, Y., Liang, H.-W., Liu, H.-B., Tian, Y., Zhang, H.-L., Zou, C.-J., Liu, Z.-Y., Zhang, Y.-L., Yang, H.-B., Huang, J., Wang, H.-K., Han, W.-J., Cao, B., Chen, G., Dai, C., Duan, L.-M., Fan, R.-R., Fu, F.-F., Guo, J.-H., Han, D., Jiang, W., Li, X.-Q., Li, X., Li, Z.-D., Liang, Y.-T., Liao, S., Lin, D.-X., Liu, C.-M., Liu, G.-R., Liu, J.-T., Long, Z., Niu, M.-C., Qiu, H., Ran, H., Sun, X.-M., Wang, B.-T., Wang, J., Wang, J.-X., Wang, Q.-L., Wang, Y.-S., Xia, X.-C., Xie, H.-Q., Yang, H.-R., Yin, H., Yuan, H., Zhang, C.-H., Zhao, R.-G., Zheng, R., Zhao, C.-X., 2023. Advances in nuclear detection and readout techniques. NUCL SCI TECH 34, 205. https://doi.org/10.1007/s41365-023-01359-0

Henzlova, D.C., Baker, M.P., Bartlett, K., Favalli, A., Iliev, M., Root, M.A., Sarnoski, S., Shin, T., Swinhoe, M. T., 2024. Neutron Detectors, in: Geist, W.H., Santi, P.A., Swinhoe, Martyn T. (Eds.), Nondestructive Assay of Nuclear Materials for Safeguards and Security. Springer Nature Switzerland, Cham, pp. 325–358. https://doi.org/10.1007/978-3-031-58277-6_15

Hong, X., Zhong, G., Yang, L., Li, Z., Xu, M., Wang, Y., He, J., 2026. Research on dual-mode cross-calibration method for neutron flux monitoring systems in fission chambers. J. Inst. 21, P01039. https://doi.org/10.1088/1748-0221/21/01/P01039

Hu, Z., Zhong, G., Ge, L., Du, T., Peng, X., Chen, Z., Xie, X., Yuan, X., Zhang, Y., Sun, J., Fan, T., Zhou, R., Xiao, M., Li, K., Hu, L., Chen, Jun, Zhang, H., Gorini, G., Nocente, M., Tardocchi, M., Li, X., Chen, Jinxiang, Zhang, G., 2018. Neutron field measurement at the Experimental Advanced Superconducting Tokamak using a Bonner sphere spectrometer. Nuclear Instruments and Methods in Physics Research Section A: Accelerators, Spectrometers, Detectors and Associated Equipment 895, 100–106. https://doi.org/10.1016/j.nima.2018.04.010

IAEA, 2024. Plasma Physics and Technology Aspects of the Deuterium-Tritium Fuel Cycle for Fusion Energy: Summary of a Technical Meeting. International Atomic Energy Agency, Vienna.

Iliasova, M.V., Mirfayzi, S.R., Rajput, M., Chandrasekhar, K., Fontana, M., Hoffman, D., O'Gorman, T., Kamal, G., McNamara, S.A.M., Sertoli, M., Sridhar, S., Varje, J., Wilson,

C., Zakhar, D., Naylor, G., 2024. Characterization of diamond and organic scintillation detectors utilizing radiation sources for continuous plasma operation. Review of Scientific Instruments 95, 083556. https://doi.org/10.1063/5.0218866

INTERNATIONAL ATOMIC ENERGY AGENCY, 2025. IAEA World Fusion Outlook 2025. INTERNATIONAL ATOMIC ENERGY AGENCY. https://doi.org/10.61092/iaea.kyle-8k70

Jansen Van Vuuren, A., Garcia-Dominguez, J., Hidalgo-Salaverri, J., Segado-Fernandez, J., LeViness, A., Ayllon-Guerola, J., Rueda-Rueda, J., Lazerson, S.A., Galdon-Quiroga, J., Garcia-Munoz, M., Pablant, N., 2024. Development of a scintillator based fast-ion loss detector for the Wendelstein 7-X stellarator. Fusion Engineering and Design 204, 114520. https://doi.org/10.1016/j.fusengdes.2024.114520

Jardin, A., Bielecki, J., Dąbrowski, W., Drozdowicz, K., Dworak, D., Gerenton, V., Guibert, D., Kantor, R., Król, K., Kulińska, A., Kurowski, A., Łach, B., Mazon, D., Savoye-Peysson, Y., Scholz, M., Walkowiak, J., Wiącek, U., Woźnicka, U., WEST team, 2024. Energy-resolved x-ray and neutron diagnostics in tokamaks: Prospect for plasma parameters determination. Physics of Plasmas 31, 082514. https://doi.org/10.1063/5.0213721

Jeet, J., Appelbe, B.D., Crilly, A.J., Divol, L., Eckart, M., Hahn, K.D., Hartouni, E.P., Hayes, A., Kerr, S., Kim, Y., Mariscal, E., Moore, A.S., Ramirez, A., Rusev, G., Schlossberg, D.J., 2024. Diagnosing up-scattered deuterium–tritium fusion neutrons produced in burning plasmas at the National Ignition Facility (invited). Review of Scientific Instruments 95, 093521. https://doi.org/10.1063/5.0219671

Juarez, R., Belotti, M., Kolsek, A., López, V., Alguacil, J., Pedroche, G., López-Revelles, A.J., Martínez-Albertos, P., De Pietri, M., Guijosa, P., Le Tonqueze, Y., Loughlin, M.J., Polunovskiy, E., Pampin, R., Fabbri, M., Sanz, J., 2024. ITER full model in MCNP for radiation safety demonstration. Nat Commun 15, 8563. https://doi.org/10.1038/s41467-024-52667-x

Kappatou, A., Baruzzo, M., Hakola, A., Joffrin, E., Keeling, D., Labit, B., Tsitrone, E., Vianello, N., Wischmeier, M., Balboa, I., Bernardo, J., Bernert, M., Bosman, T., Brezinsek, S., Brida, D., Carvalho, I.S., Carvalho, P., Ceelen, L., Challis, C.D., Coffey, I., Dittmar, T., Dunne, M., Faitsch, M., Field, A.R., Frassinetti, L., Garzotti, L., Ghani, Z., Giroud, C., Henderson, S., Henriques, R.B., Hobirk, J., Jacquet, P., Jepu, I., Kazakov, Y.O., King, D.B., Kirov, K.K., Kos, D., Krieger, K., Lennholm, M., Lerche, E., Litaudon, X., Litherland-Smith, E., Lomas, P., Lowry, C., Mailloux, J., Mantsinen, M.J., Maslov, M., Matveev, D., Meigs, A., Menmuir, S., Olde, C., Perez Von Thun, C., Piron, L., Pucella, G., Reimerdes, H., Rimini, F., Sauter, O., Schneider, P.A., Sieglin, B., Silburn, S., Solano, E.R., Sun, H., Valcarcel, D.F., Van Eester, D., Villari, R., Widdowson, A., Wiesen, S., Zlobinski, M., Zotta, V.K., Contributors, T.J., Tokamak Exploitation Team, T.Euro., 2025. Overview of the third JET deuterium-tritium campaign. Plasma Phys. Control. Fusion 67, 045039. https://doi.org/10.1088/1361-6587/adbd75

Karmakar, A., Pal, A., Anil Kumar, G., Bhavika, Vivek, Tyagi, M., 2025. Neutron-gamma pulse shape discrimination for organic scintillation detector using 2D CNN based image classification. Applied Radiation and Isotopes 217, 111653. https://doi.org/10.1016/j.apradiso.2024.111653

Kobayashi, M.I., Yoshihashi, S., Ogawa, K., Isobe, M., Aso, T., Hara, M., Sangaroon, S., Kusaka,

S., Tamaki, S., Murata, I., Toyama, S., Miwa, M., Osakabe, M., 2025. Application of a single crystal CVD diamond detector for fast neutron measurement in high dose and mixed radiation fields.
Kobayashi, M.I., Yoshihashi, S., Ogawa, K., Isobe, M., Aso, T., Hara, M., Sangaroon, S., Tamaki, S., Murata, I., Toyama, S., Miwa, M., Matsuyama, S., Osakabe, M., 2024. Simultaneous measurements for fast neutron flux and tritium production rate using pulse shape discrimination and single crystal CVD diamond detector. Nucl. Fusion 64, 066026. https://doi.org/10.1088/1741-4326/ad3f2e
Kobayashi, M.I., Yoshihashi, S., Ogawa, K., Isobe, M., Miwa, M., Toyama, S., Matsuyama, S., Osakabe, M., 2023. Measurement of 6Li burn-up reaction rate using a single crystal CVD diamond detector under fast neutron irradiation environment. Fusion Engineering and Design 193, 113799. https://doi.org/10.1016/j.fusengdes.2023.113799
Kono, S., Ishikawa, M., Sumida, S., Yamauchi, M., Ito, D., Yazawa, H., Kurosaki, M., Kobuchi, T., Ogawa, K., Isobe, M., Nunoya, Y., 2024. Estimation and Measurement of Alpha Decay Pulses in Fission Detectors and Their Practical Application for Verifying Detector Health. Plasma and Fusion Research 19, 1405015–1405015. https://doi.org/10.1585/pfr.19.1405015
Koshimizu, M., 2025. Fundamental processes and recent development of organic scintillators. Journal of Luminescence 278, 121008. https://doi.org/10.1016/j.jlumin.2024.121008
Kovalev, A.O., Rodionov, R.N., Portnov, D.V., Vorobiev, V.A., Vysokih, Yu.G., Obudovsky, S.Yu., Kashchuk, Yu.A., 2022. Analysis of Radiation Conditions of DNFM ITER Diagnostic Tool. Phys. Atom. Nuclei 85, 1271–1277. https://doi.org/10.1134/S106377882207016X
Kovalev, A.O., Rodionov, R.N., Vorobiev, V.A., Portnov, D.V., Kormilitsyn, T.M., Vysokih, Yu.G., Obudovsky, S.Yu., Kashchuk, Yu.A., 2023. Measurements of ITER Fusion Power by Neutron Flux Monitors. Phys. Atom. Nuclei 86, S187–S197. https://doi.org/10.1134/S1063778823140077
Kreusch Filho, D., Caetano, L.L., De Azevedo, A.M., Nunes, W.V., 2025. Progress in the Design of New Gas-Based Neutron Detectors: A Critical Review. Braz. J. Radiat. Sci. 13, e2893. https://doi.org/10.15392/2319-0612.2025.2893
Kushoro, M.H., Angelone, M., Bozzi, D., Cancelli, S., Dal Molin, A., Gallo, E., Gorini, G., La Via, F., Parisi, M., Perelli Cippo, E., Putignano, O., Tardocchi, M., Rebai, M., 2024. Operation of a 250µm-thick SiC detector with DT neutrons at high temperatures. Fusion Engineering and Design 204, 114486. https://doi.org/10.1016/j.fusengdes.2024.114486
Kushoro, M.H., Colombi, S., Croci, G., Dal Molin, A., Gallo, E., Nocente, M., Osipenko, M., Putignano, O., Rebai, M., Rigamonti, D., Tardocchi, M., Gorini, G., 2025. SiC detector response to tokamak neutron spectra mock-up validated on experimental results. Fusion Engineering and Design 214, 114875. https://doi.org/10.1016/j.fusengdes.2025.114875
Lahmann, B., Hahn, K.D., Henry, E.A., Munteanu, F., Schlossberg, D.J., Bionta, R.M., 2025. Using real-time nuclear activation detectors for measuring neutron yields from D(D, T)n reactions on the national ignition facility (NIF). Review of Scientific Instruments 96, 033506. https://doi.org/10.1063/5.0213464
Landsmeer, C., Marcer, G., Dal Molin, A., Rebai, M., Rigamonti, D., Coriton, B., Gorini, G., Guerini Rocco, M., Kovalev, A., Muraro, A., Nocente, M., Perelli Cippo, E., Polevoi, A., Putignano, O., Scioscioli, F., Croci, G., Tardocchi, M., 2025. A machine learning case study in nuclear fusion: Assessment of the absolute deuterium-tritium fusion power of ITER with gamma-ray spectroscopy. Energy and AI 21, 100526. https://doi.org/10.1016/j.egyai.2025.100526

Lanza, J.A., Cao, L.R., 2026. Review of self-powered neutron detectors for reactor instrumentation. Progress in Nuclear Energy 193, 106230. https://doi.org/10.1016/j.pnucene.2025.106230

Lee, Y., Ko, J., Nam, Y.-U., Kim, H.-S., 2025. D-D Fusion Neutron Spectrometer in KSTAR Tokamak.

Lewis, E.R., Anderson, G., Martinez De Luca, D., Young, B.A., Davis, T.P., 2026. Qualification Pathways for Fusion Structural Materials. JNE 7, 23. https://doi.org/10.3390/jne7010023

Li, A., Yang, J., Xiao, J., HAN, Z., 2024. Compact back-to-back lithium glass detector for online tritium production rate measurement. High Power Laser and Particle Beams 36, 106001. https://doi.org/10.11884/HPLPB202436.240256 (in Chinese)

LI C., HUO Z., ZHONG G., Hu L., 2023. Survey of radiation monitoring system for magnetic confinement fusion device. Nuclear Techniques 46, 020001. https://doi.org/10.11889/j.0253-3219.2023.hjs.46.020001 (in Chinese)

Li, K., Zhong, G., Hu, L., Zhou, R., Chen, W., Ogawa, K., 2025. Experimental measurements of triton burnup in EAST by neutron.

Li, Y., Li, T., Wang, Y., Hong, B., Wang, F., 2021. Determination method of high fluence rate for D-T neutron source with long counter. Radiation Measurements 148, 106662. https://doi.org/10.1016/j.radmeas.2021.106662

Liao, L.Y., Ogawa, K., Sangaroon, S., Paenthong, W., Kusaka, S., Tamaki, S., Murata, I., Isobe, M., 2024. The initial measurement of a compact D–T neutron spectrometer based on a single-crystal chemical vapor deposition diamond stack for fusion plasma diagnostic. Review of Scientific Instruments 95, 073533. https://doi.org/10.1063/5.0219460

Lin, S.Y., Zhou, R.J., Artemev, K.K., Zhang, J.Z., Zhao, J.L., Zhong, G.Q., 2024. Development of diamond detector as fast neutron spectroscopy in the EAST tokamak. J. Inst. 19, P07017. https://doi.org/10.1088/1748-0221/19/07/P07017

Liu, B., Tang, Y., Liu, X., Liu, H., Zhan, Y., Li, P., Zuo, Z., Li, S., Tang, L., Wang, Q., Liu, M., 2026. Neutron-gamma discrimination based on STFT-DFF model and FPGA implementation. Applied Radiation and Isotopes 232, 112567. https://doi.org/10.1016/j.apradiso.2026.112567

Liu, H., Zhan, Y., Liu, M., Liu, Y., Li, P., Zuo, Z., Liu, B., Liu, R., 2026. Pulse shape discrimination algorithms: Survey and benchmark. Radiation Measurements 193, 107653. https://doi.org/10.1016/j.radmeas.2026.107653

Mackie, S., Wink, C.W., Dalla Rosa, M., Berg, G.P.A., Ball, J.L., Wang, X., Carmichael, J., Tinguely, R.A., Rigamonti, D., Tardocchi, M., Raj, P., Frenje, J., Rice, J., 2024. Ion optical design of the magnetic proton recoil neutron spectrometer for the SPARC tokamak. Review of Scientific Instruments 95, 103502. https://doi.org/10.1063/5.0219551

Maggi, C.F., Abate, D., Abid, N., Abreu, P., Adabonyan, O., Afzal, M., Ahmad, I., Akhtar, M., Albanese, R., Aleiferis, S., Alessi, E., Aleynikov, P., Aleynikov, P., Alguacil, J., Alhage, J., Ali, M., Allen, H., Allinson, M., Alonzo, M., Alves, E., Ambrosino, R., Andersson Sundén, E., Andrew, P., Angelone, M., Angioni, C., Antoniou, I., Appel, L., Appelbee, C., Aramunde, C., Ariola, M., Arnoux, G., Artaserse, G., Artaud, J.-F., Arter, W., Artigues, V.,

Artola, F.J., Ash, A., Asztalos, O., Auld, D., Auriemma, F., Austin, Y., Avotina, L., Ayllón, J., Aymerich, E., Baciero, A., Bähner, L., Bairaktaris, F., Balboa, I., Balden, M., Balshaw, N., Bandaru, V.K., Banks, J., Banon Navarro, A., Barcellona, C., Bardsley, O., Barnes, M., Barnsley, R., Baruzzo, M., Bassan, M., Batista, A., Batistoni, P., Baumane, L., Bauvir, B., Baylor, L., Bearcroft, C., Beaumont, P., Beckett, D., Begolli, A., Beidler, M., Bekris, N., Beldishevski, M., Belli, E., Belli, F., Benkadda, S., Bentley, J., Bernard, E., Bernardo, J., Bernert, M., Berry, M., Bertalot, L., Betar, H., Beurskens, M., Bhat, P.G., Bickerton, S., Bielecki, J., Biewer, T., Bilato, R., Bílková, P., Birkenmeier, G., Bisson, R., Bizarro, J.P.S., Blatchford, P., Bleasdale, A., Bobkov, V., Boboc, A., Bock, A., Bodnar, G., Bohm, P., Bonalumi, L., Bonanomi, N., Bonfiglio, D., Bonnin, X., Bonofiglo, P., Booth, J., Borba, D., Borba, D., Borodin, D., Borodkina, I., Bosman, T.O.S.J., Bourdelle, C., Bowden, M., Božičević Mihalić, I., Bradnam, S.C., Breizman, B., Brezinsek, S., Brida, D., Brix, M., Brown, P., Brunetti, D., Buckley, M., Buermans, J., Bufferand, H., Buratti, P., Burckhart, A., Burgess, A., Buscarino, A., Busse, A., Butcher, D., Calabrò, G., Calacci, L., Calado, R., Canavan, R., Cannas, B., Cannon, M., Cappelli, M., Carcangiu, S., Card, P., Cardinali, A., Carli, S., Carman, P., Carnevale, D., Carvalho, B., Carvalho, I.S., Carvalho, P., Casiraghi, I., Casson, F.J., Castaldo, C., Catalan, J.P., Catarino, N., Causa, F., Cavedon, M., Cecconello, M., Ceelen, L., Challis, C.D., Chamberlain, B., Chandra, R., Chang, C.S., Chankin, A., Chapman, B., Chauhan, P., Chernyshova, M., Chiariello, A., Chira, G.-C., Chmielewski, P., Chomiczewska, A., Chone, L., Cieslik, J., Ciraolo, G., Ciric, D., Citrin, J., Ciupinski, Ł., Clarkson, R., Cleverly, M., Coates, P., Coccorese, V., Coelho, R., Coenen, J.W., Coffey, I.H., Colangeli, A., Colas, L., Collins, J., Conroy, S., Contré, C., Conway, N.J., Coombs, D., Cooper, P., Cooper, S., Cordaro, L., Corradino, C., Corre, Y., Corrigan, G., Coster, D., Craciunescu, T., Cramp, S., Craven, D., Craven, R., Croci, G., Croft, D., Crombé, K., Cronin, T., Cruz, N., Cufar, A., Cullen, A., Dal Molin, A., Dalley, S., David, P., Davies, A., Davies, J., Davies, S., Davis, G., Dawson, K., Dawson, S., Day, I., De Tommasi, G., Deane, J., Dearing, M., De Bock, M., Decker, J., Dejarnac, R., Delabie, E., De La Cal, E., De La Luna, E., Del Sarto, D., Dempsey, A., Deng, W., Dennett, A., Derks, G.L., De Temmerman, G., Devasagayam, F., De Vries, P., Devynck, P., Di Siena, A., Dickinson, D., Dickson, T., Diez, M., Dinca, P., Dittmar, T., Dittrich, L., Dobrashian, J., Dochnal, T., Donné, A.J.H., Dorland, W., Dorling, S., Dormido-Canto, S., Dotse, R., Douai, D., Dowson, S., Doyle, R., Dreval, M., Drews, P., Drummond, G., Duckworth, Ph., Dudding, H.G., Dumont, R., Dumortier, P., Dunai, D., Dunatov, T., Dunne, M., Ďuran, I., Durodié, F., Dux, R., Eade, T., Eardley, E., Edwards, J., Eich, T., Eksaeva, A., El-Haroun, H., Ellis, R.D., Ellwood, G., Elsmore, C., Emery, S., Ericsson, G., Eriksson, B., Eriksson, F., Eriksson, J., Eriksson, L.G., Eriksson, L.G., Ertmer, S., Evans, G., Evans, S., Fable, E., Fagan, D., Faitsch, M., Fajardo Jimenez, D., Falessi, M., Fanni, A., Farmer, T., Farquhar, I., Faugeras, B., Fazinić, S., Fedorczak, N., Felker, K., Felton, R., Fernandes, H., Ferreira, D.R., Ferreira, J., Ferrò, G., Fessey, J., Février, O., Ficker, O., Field, A.R., Figueiredo, A., Figueiredo, J., Fil, A., Fil, N., Finburg, P., Fischer, U., Fishpool, G., Fittill, L., Fitzgerald, M., Flammini, D., Flanagan, J., Foley, S., Fonnesu, N., Fontana, M., Fontdecaba, J.M., Fortuna, L., Fortuna-Zalesna, E., Fortune, M., Fowler, C., Fox, P., Franklin, O., Fransson, E., Frassinetti, L., Fresa, R., Frigione, D., Fülöp, T., Furseman, M., Gabriellini, S., Gadariya, D., Gadgil, S., Gál, K.,

Galeani, S., Galkowski, A., Gallart, D., Gambrioli, M., Gans, T., Garcia, J., García-Muñoz, M., Garzotti, L., Gaspar, J., Gatto, R., Gaudio, P., Gear, D., Gebhart, T., Gee, S., Gelfusa, M., George, R., Gerasimov, S.N., Gerru, R., Gervasini, G., Gethins, M., Ghani, Z., Gherendi, M., Gherghina, P.-I., Ghezzi, F., Giacomelli, L., Gibson, C., Gil, L., Gilbert, M.R., Gillgren, A., Giovannozzi, E., Giroud, C., Giruzzi, G., Goff, J., Goloborodko, V., Gomes, R., Gomez, J.-F., Gonçalves, B., Goniche, M., Gonzalez-Martin, J., Goodyear, A., Gore, S., Gorini, G., Görler, T., Gotts, N., Gow, E., Graves, J.P., Green, J., Greuner, H., Grigore, E., Griph, F., Gromelski, W., Groth, M., Grove, C., Grove, R., Gupta, N., Hacquin, S., Hägg, L., Hakola, A., Halitovs, M., Hall, J., Ham, C.J., Hamed, M., Hardman, M.R., Haresawa, Y., Harrer, G., Harrison, J.R., Harting, D., Hatch, D.R., Haupt, T., Hawes, J., Hawkes, N.C., Hawkins, J., Hazael, S., Hearmon, J., Heesterman, P., Heinrich, P., Held, M., Helou, W., Hemming, O., Henderson, S.S., Henriques, R., Henriques, R.B., Hepple, D., Herfindal, J., Hermon, G., Hillesheim, J.C., Hizanidis, K., Hjalmarsson, A., Ho, A., Hobirk, J., Hoenen, O., Hogben, C., Hollingsworth, A., Hollis, S., Hollmann, E., Hölzl, M., Hook, M., Hoppe, M., Horáček, J., Horsten, N., Horsten, N., Horton, A., Horton, L.D., Horvath, L., Hotchin, S., Hu, Z., Huang, Z., Hubenov, E., Huber, A., Huber, V., Huddleston, T., Huijsmans, G.T.A., Husain, Y., Huynh, P., Hynes, A., Iglesias, D., Iliasova, M.V., Imríšek, M., Ingleby, J., Innocente, P., Ioannou-Sougleridis, V., Isernia, N., Ivanova-Stanik, I., Ivings, E., Jachmich, S., Jackson, T., Jacobsen, A.S., Jacquet, P., Järleblad, H., Järvinen, A., Jaulmes, F., Jayasekera, N., Jenko, F., Jepu, I., Joffrin, E., Johnson, T., Johnston, J., Jones, C., Jones, E., Jones, G., Jones, L., Jones, T.T.C., Joyce, A., Juvonen, M., Kallenbach, A., Kalnina, P., Kalupin, D., Kanth, P., Kantor, A., Kappatou, A., Kardaun, O., Karhunen, J., Karsakos, E., Kazakov, Ye.O., Kazantzidis, V., Keeling, D.L., Kelly, W., Kempenaars, M., Kennedy, D., Khan, K., Khilkevich, E., Kiefer, C., Kim, H.-T., Kim, J., Kim, S.H., King, D.B., Kinna, D.J., Kiptily, V.G., Kirjasuo, A., Kirov, K.K., Kirschner, A., Kiviniemi, T., Kizane, G., Klepper, C., Klix, A., Kneale, G., Knight, M., Knight, P., Knights, R., Knipe, S., Knoche, U., Knolker, M., Kocan, M., Köchl, F., Kocsis, G., Koenders, J.T.W., Kolesnichenko, Y., Kominis, Y., Kong, M., Kool, B., Korovin, V., Korsholm, S.B., Kos, B., Kos, D., Koubiti, M., Kovtun, Y., Kowalska-Strzęciwilk, E., Koziol, K., Krasikov, Y., Krasilnikov, A., Krasilnikov, V., Kresina, M., Kreter, A., Krieger, K., Krivska, A., Kruezi, U., Książek, I., Kumpulainen, H., Kurzan, B., Kwak, S., Kwon, O.J., Labit, B., Lacquaniti, M., Lagoyannis, A., Laguardia, L., Laing, A., Laksharam, V., Lam, N., Lambertz, H.T., Lane, B., Langley, M., Lascas Neto, E., Łaszyńska, E., Lawson, K.D., Lazaros, A., Lazzaro, E., Learoyd, G., Lee, C., Lee, K., Leerink, S., Leeson, T., Lefebvre, X., Leggate, H.J., Lehmann, J., Lehnen, M., Leichtle, D., Leipold, F., Lengar, I., Lennholm, M., Leon Gutierrez, E., Leppin, L.A., Lerche, E., Lescinskis, A., Lesnoj, S., Lewin, L., Lewis, J., Likonen, J., Linsmeier, Ch., Litaudon, X., Litherland-Smith, E., Liu, F., Loarer, T., Loarte, A., Lobel, R., Lomanowski, B., Lomas, P.J., Lombardo, J., Lorenzini, R., Loreti, S., Loschiavo, V.P., Loughlin, M., Lowe, T., Lowry, C., Luce, T., Lucock, R., Luda Di Cortemiglia, T., Lungaroni, M., Lungu, C.P., Lunt, T., Lutsenko, V., Lyons, B., Macdonald, J., Macusova, E., Mäenpää, R., Maggi, C.F., Maier, H., Mailloux, J., Makarov, S., Manas, P., Manning, A., Mantica, P., Mantsinen, M.J., Manyer, J., Manzanares, A., Maquet, Ph., Maraschek, M., Marceca, G., Marcer, G., Marchetto, C.,

Marchuk, O., Mariani, A., Mariano, G., Marin, M., Marin Roldan, A., Marinelli, M., Markovič, T., Marot, L., Marren, C., Marsden, S., Marsen, S., Marsh, J., Marshall, R., Martellucci, L., Martin, A.J., Martin, C., Martone, R., Maruyama, S., Maslov, M., Mattei, M., Matthews, G.F., Matveev, D., Matveeva, E., Mauriya, A., Maviglia, F., Mayer, M., Mayoral, M.-L., Mazzi, S., Mazzi, S., Mazzotta, C., McAdams, R., McCarthy, P.J., McCullen, P., McDermott, R., McDonald, D.C., McGuckin, D., McKay, V., McNamee, L., McShee, A., Mederick, D., Medland, M., Medley, S., Meghani, K., Meigs, A.G., Meitner, S., Menmuir, S., Mergia, K., Mianowski, S., Middleton, P., Mietelski, J., Mikszuta-Michalik, K., Milanesio, D., Milani, E., Militello-Asp, E., Militello, F., Milnes, J., Milocco, A., Minucci, S., Miron, I., Mitchell, J., Mlynář, J., Mlynář, J., Moiseenko, V., Monaghan, P., Monakhov, I., Montisci, A., Moon, S., Mooney, R., Moradi, S., Morales, R.B., Morgan, L., Moro, F., Morris, J., Mrowetz, T., Msero, L., Munot, S., Muñoz-Perez, A., Muraglia, M., Murari, A., Muraro, A., N'Konga, B., Na, Y.S., Nabais, F., Naish, R., Napoli, F., Nardon, E., Naulin, V., Nave, M.F.F., Neu, R., Ng, S., Nicassio, M., Nicolai, D., Nielsen, A.H., Nielsen, S.K., Nina, D., Noble, C., Nobs, C.R., Nocente, M., Nordman, H., Nowak, S., Nyström, H., O'Callaghan, J., O'Mullane, M., O'Neill, C., Olde, C., Oliver, H.J.C., Olney, R., Ongena, J., Orsitto, G.P., Osipov, A., Otin, R., Pace, N., Packer, L.W., Pajuste, E., Palade, D., Palgrave, J., Pan, O., Panadero, N., Pandya, T., Panontin, E., Papadopoulos, A., Papadopoulos, G., Papp, G., Parail, V.V., Parsloe, A., Paschalidis, K., Passeri, M., Patel, A., Pau, A., Pautasso, G., Pavlichenko, R., Pavone, A., Pawelec, E., Paz-Soldan, C., Peacock, A., Pearce, M., Pearson, I.J., Peluso, E., Penot, C., Pepperell, K., Perdas, A., Pereira, T., Perelli Cippo, E., Perez Von Thun, C., Perry, D., Petersson, P., Petravich, G., Petrella, N., Peyman, M., Pigatto, L., Pillon, M., Pinches, S., Pintsuk, G., Piron, C., Pironti, A., Pisano, F., Pitts, R., Planck, U., Platt, N., Plyusnin, V., Podesta, M., Pokol, G., Poli, F.M., Pompilian, O.G., Poradzinski, M., Porkolab, M., Porosnicu, C., Poulipoulis, G., Poulsen, A.S., Predebon, I., Previti, A., Primetzhofer, D., Provatas, G., Pucella, G., Puglia, P., Purahoo, K., Putignano, O., Pütterich, T., Quercia, A., Radulescu, G., Radulovic, V., Ragona, R., Rainford, M., Raj, P., Rasinski, M., Rasmussen, D., Rasmussen, J., Rasmussen, J.J., Raso, A., Rattá, G., Ratynskaia, S., Rayaprolu, R., Rebai, M., Redl, A., Rees, D., Réfy, D., Reichle, R., Reimerdes, H., Reman, B.C.G., Reux, C., Reynolds, S., Rigamonti, D., Righi, E., Rimini, F.G., Risner, J., Rivero-Rodriguez, J.F., Roach, C.M., Roberts, J., Robins, R., Robinson, S., Robson, D., Rode, S., Rodrigues, P., Rodriguez-Fernandez, P., Romanelli, S., Romazanov, J., Rose, E., Rose-Innes, C., Rossi, R., Rowe, S., Rowlands, D., Rowley, C., Rubel, M., Rubinacci, G., Rubino, G., Rud, M., Ruiz Ruiz, J., Ryter, F., Saarelma, S., Sahlberg, A., Salewski, M., Salmi, A., Salmon, R., Salzedas, F., Sanchez, F., Sanders, I., Sandiford, D., Sanni, F., Sauter, O., Sauvan, P., Schettini, G., Shevelev, A., Schekochihin, A.A., Schmid, K., Schmidt, B.S., Schmuck, S., Schneider, M., Schneider, P.A., Schoonheere, N., Schramm, R., Scoon, D., Scully, S., Segato, M., Seidl, J., Senni, L., Seo, J., Sergienko, G., Sertoli, M., Sharapov, S.E., Sharma, R., Shaw, A., Shaw, R., Sheikh, H., Sheikh, U., Shi, N., Shigin, P., Shiraki, D., Sias, G., Siccinio, M., Sieglin, B., Silburn, S.A., Silva, A., Silva, C., Silva, J., Silvagni, D., Simfukwe, D., Simpson, J., Sirén, P., Sirinelli, A., Sjöstrand, H., Skinner, N., Slater, J., Smart, T., Smirnov, R.D., Smith, N., Smith, P., Smith, T., Snell, J., Snoj, L., Solano, E.R., Solokha, V., Sommariva, C., Soni, K., Sos, M., Sousa, J., Sozzi, C., Spelzini, T.,

Spineanu, F., Spolladore, L., Spong, D., Srinivasan, C., Staebler, G., Stagni, A., Stamatelatos, I., Stamp, M.F., Štancar, Ž., Staniec, P.A., Stankūnas, G., Stead, M., Stein-Lubrano, B., Stephen, A., Stephens, J., Stevenson, P., Steventon, C., Stojanov, M., St-Onge, D.A., Strand, P., Strikwerda, S., Stuart, C.I., Sturgeon, S., Sun, H.J., Surendran, S., Suttrop, W., Svensson, J., Svoboda, J., Sweeney, R., Szepesi, G., Szoke, M., Tadić, T., Tal, B., Tala, T., Tamain, P., Tanaka, K., Tang, W., Tardini, G., Tardocchi, M., Taylor, D., Teimane, A.S., Telesca, G., Teplukhina, A., Terra, A., Terranova, D., Terranova, N., Testa, D., Thomas, B., Thompson, V.K., Thorman, A., Thrysoe, A.S., Tierens, W., Tinguely, R.A., Tipton, A., Todd, H., Tomeš, M., Tookey, A., Tsavalas, P., Tskhakaya, D., Turică, L.-P., Turner, A., Turner, I., Turner, M., Turner, M.M., Tvalashvili, G., Tykhyy, A., Tyrrell, S., Uccello, A., Udintsev, V., Vadgama, A., Valcarcel, D.F., Valentini, A., Valisa, M., Vallar, M., Valovic, M., Van Berkel, M., Van De Plassche, K.L., Van Rossem, M., Van Eester, D., Varela, J., Varje, J., Vasilopoulou, T., Vayakis, G., Vecsei, M., Vega, J., Veis, M., Veis, P., Ventre, S., Veranda, M., Verdoolaege, G., Verona, C., Verona Rinati, G., Veshchev, E., Vianello, N., Viezzer, E., Vignitchouk, L., Vila, R., Villari, R., Villone, F., Vincenzi, P., Vitins, A., Vizvary, Z., Vlad, M., Voldiner, I., Von Toussaint, U., Vondráček, P., Wakeling, B., Walker, M., Walker, R., Walsh, M., Walton, R., Wang, E., Warren, F., Warren, R., Waterhouse, J., Watts, C., Webster, T., Weiland, M., Weisen, H., Weiszflog, M., Wendler, N., West, A., Wheatley, M., Whetham, S., Whitehead, A., Whittaker, D., Widdowson, A., Wiesen, S., Willensdorfer, M., Williams, J., Wilson, I., Wilson, T., Wischmeier, M., Withycombe, A., Witts, D., Wojcik-Gargula, A., Wolfrum, E., Wood, R., Woodley, R., Worrall, R., Wyss, I., Xu, T., Yadykin, D., Yakovenko, Y., Yang, Y., Yanovskiy, V., Yi, R., Young, I., Young, R., Zaar, B., Zabolockis, R.J., Zakharov, L., Zanca, P., Zarins, A., Zarzoso Fernandez, D., Zastrow, K.-D., Zayachuk, Y., Zerbini, M., Zhang, W., Zimmermann, B., Zlobinski, M., Zocco, A., Zotta, V.K., Zuin, M., Zwingmann, W., Zychor, I., 2024. Overview of T and D–T results in JET with ITER-like wall. Nucl. Fusion 64, 112012. https://doi.org/10.1088/1741-4326/ad3e16

Marian, J., Setyawan, W., Yang, Y., Manzoor, A., Zhong, W., Trelewicz, J.R., Yu, J., Peterson, E., Katoh, Y., Snead, L., Wirth, B.D., 2025. Computational materials assessment of the D/Li-stripping neutron source as a prototypical facility for fusion materials testing. Current Opinion in Solid State and Materials Science 38, 101231. https://doi.org/10.1016/j.cossms.2025.101231

Marocco, D., Angelone, M., Belli, F., Caruggi, F., Croci, G., Esposito, B., Gandolfo, G., Gorini, G., Grosso, G., Nocente, M., Panza, F., Pillon, M., Pompili, F., Rigamonti, D., Rocchi, G., Scionti, J., Tardocchi, M., 2024. Design status of the neutron and gamma-ray diagnostics for the Divertor Tokamak Test facility. Fusion Engineering and Design 202, 114308. https://doi.org/10.1016/j.fusengdes.2024.114308

Marocco, D., Gandolfo, G., Panza, F., Anagnostopoulou, V., Centioli, C., Pompili, F., Pontesilli, M., 2026. Neutron yield monitors for the divertor tokamak test (DTT) facility: Detector set, expected performance and integration. Fusion Engineering and Design 222, 115538. https://doi.org/10.1016/j.fusengdes.2025.115538

Martínez-Albertos, P., Sauvan, P., Catalán, J.P., Belotti, M., Javier, F., Germa, J., Le Tonqueze, Y., Dammann, A., Juárez, R., 2025. Optimizing Radiation Shielding for Fusion Maintenance Facilities: Insights From a Comprehensive Analysis of ITER Hot Cell. International

Journal of Energy Research 2025, 9222155. https://doi.org/10.1155/er/9222155
Marzullo, D., Occhiuto, E., Brolatti, G., Falco, D., Laghi, D., Marocco, D., Esposito, B., 2024. Mechanical design of ITER radial neutron camera Ex-Port system. Fusion Engineering and Design 203, 114477. https://doi.org/10.1016/j.fusengdes.2024.114477
Matsumoto, T., Masuda, A., Harano, H., Manabe, S., 2024. Simulation study for design of long counter for standard neutron fields from 1 keV to 20 MeV at NMIJ/AIST. Radiation Protection Dosimetry 200, 1251–1257. https://doi.org/10.1093/rpd/ncae151
Matsuura, H., Wakisaka, S., Egashira, K., Nishitani, T., Sumida, S., Shinohara, K., Sugiyama, S., Ogawa, K., 2025. Neutron spectrum analysis considering nuclear elastic scattering in beam-injected JT-60SA relevant deuterium plasma. Plasma Phys. Control. Fusion 67, 115028. https://doi.org/10.1088/1361-6587/ae1c6e
Mauri, G., Messi, F., Kanaki, K., Hall-Wilton, R., Piscitelli, F., 2019. Fast neutron sensitivity for 3He detectors and comparison with Boron-10 based neutron detectors. EPJ Techn Instrum 6, 3. https://doi.org/10.1140/epjti/s40485-019-0052-x
Mayer, S., Harm, C., Harzmann, S., Hohmann, E., Kasprzak, M., Pedrazzi, L., Wouters, C., Yukihara, E.G., 2024. Challenges of radiation measurements in an institute with various large-scale research facilities. Radiation Measurements 176, 107224. https://doi.org/10.1016/j.radmeas.2024.107224
Mazon, D., Vayakis, G., Walsh, M., Yun, G., Hong, S.-H., Peterson, B., Aumeunier, M.H., Bultel, A., Klepper, C., Rasmussen, D., Choi, H., Grisolia, C., Kim, K., Oh, S., Sun, C., Scholz, M., Esposito, B., Marocco, D., Belli, F., Bertalot, L., Coriton, B., Ginv, V., Gin, D., Dankowski, J., Hjalmarsson, A., Krasilnikov, V., Ericsson, G., Tardochi, M., Rigamonti, D., Nocente, M., Garcia-Munoz, M., Ishikawa, M., Cheon, M., Jo, J., Zoletnik, S., Asztalos, O., Bandyopadhay, M., Bharathi, P., De Bock, M., Ford, O., Von Hellermann, M.G., Johnson, D.W., Ko, J., Menmuir, S., Mertens, Ph., Nielsen, A.H., Pokol, G.I., Serov, S.V., Singh, M.J., Tugarinov, S.N., Vyas, G.L., O'Mullane, M.G., Zhang, L., Barnsley, R., Tieulent, R., Colette, D., Neverov, V.S., Scannell, R., Liu, H., Mukhin, E., Yatsuka, E., Gorbunov, A., Giudicotti, L., Kurskiev, G., Chen, J., Van Zeeland, M.A., Finkenthal, D., Imazawa, R., Brower, D., Sirinelli, A., Akiyama, T., Carlstrom, T., Lesher, M., Watts, C., Bassan, M., Austin, M., Korsholm, S.B., Liu, Y., Danani, S., Muscatello, C., Rowan, W.L., Vershkov, V., Wang, G., Xie, J.L., Zerbini, M., Zhu, Y.L., Ďuran, I., Gusarov, A., Vukolov, K., Litnovsky, A., Moser, L., Babinov, N., Dmitriev, A., Kim, B., Marot, L., Razdobarin, A., Rogov, A., Samsonov, D., Seon, C., Soni, K., Yan, R., De Baar, M.R., Zabeo, L., Schneider, M., Blanken, Th., Bosman, Th., Ravensbergen, T., Van De Boorn, B., Orrico, C., Fischer, R., Bock, A., Denk, S.S., Medvedeva, A., Salewski, M., Stieglitz, D., the ASDEX Upgrade Team, the WEST Team, 2025. Diagnostics: Chapter 8 of the special issue: on the path to tokamak burning plasma operation. Nucl. Fusion 65, 113001. https://doi.org/10.1088/1741-4326/adfc7c
Mehrara, R., Mariano, G., Wiącek, U., Ciurlino, A., Croci, G., Dworak, D., Kovalev, A., Morawski, D., Rebai, M., Scioscioli, F., Tardocchi, M., Nocente, M., Tracz, G., Coriton, B., 2026. Nuclear Analysis Supporting the Design Optimization of the ITER Radial Gamma-Ray Spectrometer (RGRS). Fusion Engineering and Design 228, 115768. https://doi.org/10.1016/j.fusengdes.2026.115768
Melbinger, J., Axiotis, M., Chasapoglou, S., Diakaki, M., Fragner, T., Griesmayer, E., Kaperoni,

K., Kokkoris, M., Weiss, C., 2026. Comparative analysis of sCVD and SiC detectors for neutron measurements in nuclear fusion. Nuclear Instruments and Methods in Physics Research Section A: Accelerators, Spectrometers, Detectors and Associated Equipment 1083, 171099. https://doi.org/10.1016/j.nima.2025.171099

Melbinger, J., Weiss, C., Griesmayer, E., Jericha, E., Hainz, D., 2024. Diamond based neutron detector in high temperature environments of 200 °C. J. Inst. 19, P07015. https://doi.org/10.1088/1748-0221/19/07/P07015

Meng, P., Wang, X., Zhou, J., Zhu, L., Guo, S., Zhang, C., Zeng, L., Wang, Y., Lu, Y., Sun, Z., Chen, Y., 2025. A sealed neutron beam monitor based on boron-lined multi-wire proportion chamber. Radiat Detect Technol Methods 9, 375–381. https://doi.org/10.1007/s41605-024-00514-4

Meng, P., Wang, Y., Wang, X., Lu, Y., Zeng, L., Zhou, J., Sun, Z., 2026. High-Efficiency Thermal Neutron Detector Based on Boron-Lined Multi-Wire Proportional Chamber. Applied Sciences 16, 1444. https://doi.org/10.3390/app16031444

Meng, X.-D., Han, Y.-C., Ren, L., Zhang, L.-X., Peng, C., Wang, X.-Y., He, H.-J., Hou, X.-H., Bai, S.-Y., Feng, S., Li, T.-S., 2024. A novel 4H–SiC thermal neutron detector based on a metal-oxide-semiconductor structure. Nuclear Instruments and Methods in Physics Research Section A: Accelerators, Spectrometers, Detectors and Associated Equipment 1068, 169683. https://doi.org/10.1016/j.nima.2024.169683

Meschini, S., Laviano, F., Ledda, F., Pettinari, D., Testoni, R., Torsello, D., Panella, B., 2023. Review of commercial nuclear fusion projects. Front. Energy Res. 11, 1157394. https://doi.org/10.3389/fenrg.2023.1157394

Mohamed, M., Zakuan, N.D., Tengku Hassan, T.N.A., Lock, S.S.M., Mohd Shariff, A., 2024. Global Development and Readiness of Nuclear Fusion Technology as the Alternative Source for Clean Energy Supply. Sustainability 16, 4089. https://doi.org/10.3390/su16104089

Molin, A.D., Guiotto, F., Putignano, O., Rosa, M.D., Franz, P., Grosso, G., Monguzzi, A., Cippo, E.P., Pollice, L., Rigamonti, D., Tedoldi, L.G., Zuin, M., Tardocchi, M., 2024. Development of fast 2.5 MeV neutron detectors for high-intensity stray magnetic field environments. Review of Scientific Instruments 95, 083544. https://doi.org/10.1063/5.0219121

Moore, A.S., Schlossberg, D.J., Appelbe, B.D., Chandler, G.A., Crilly, A.J., Eckart, M.J., Forrest, C.J., Glebov, V.Y., Grim, G.P., Hartouni, E.P., Hatarik, R., Kerr, S.M., Kilkenny, J., Knauer, J.P., 2023. Neutron time of flight (nToF) detectors for inertial fusion experiments. Review of Scientific Instruments 94, 061102. https://doi.org/10.1063/5.0133655

Morales, I.R., Crespo, M.L., Bogovac, M., Cicuttin, A., Kanaki, K., Carrato, S., 2024a. Gamma/neutron classification with SiPM CLYC detectors using frequency-domain analysis for embedded real-time applications. Nuclear Engineering and Technology 56, 745–752. https://doi.org/10.1016/j.net.2023.11.013

Morales, I.R., Soledad Molina, R., Bogovac, M., Jovalekic, N., Liz Crespo, M., Kanaki, K., Ramponi, G., Carrato, S., 2024b. Gamma/Neutron Online Discrimination Based on Machine Learning With CLYC Detectors. IEEE Trans. Nucl. Sci. 71, 2602–2614. https://doi.org/10.1109/TNS.2024.3498321

Morandi, A., Pettinari, D., Zucchetti, M., 2026. Activation analysis of a compact Tokamak using

Deuterium–Helium3 fuel. Fusion Engineering and Design 222, 115491. https://doi.org/10.1016/j.fusengdes.2025.115491

Moreno-Pérez, J.A., Marchena, Á., Araya, P., López-Peñalver, J.J., De La Torre, J.A., Lallena, A.M., Becerril, S., Anguiano, M., Palma, A.J., Carvajal, M.A., 2025. Characterization of Different Types of Micro-Fission and Micro-Ionization Chambers Under X-Ray Beams. Sensors 25, 1862. https://doi.org/10.3390/s25061862

Moro, F., Marocco, D., Belli, F., Bocian, D., Brolatti, G., Centioli, C., Colangeli, A., Mambro, G.D., Flammini, D., Fonnesu, N., Gandolfo, G., Kantor, R., Kotula, J., Maffucci, A., Mariano, G., Marzullo, D., Ortwein, R., Podda, S., Pompili, F., Riva, M., Sancristóbal, D., Villari, R., Esposito, B., 2023. Nuclear design of a shielded cabinet for electronics: The ITER radial neutron camera case study. Fusion Engineering and Design 191, 113542. https://doi.org/10.1016/j.fusengdes.2023.113542

Moro, F., Marocco, D., Belli, F., Bocian, D., Brolatti, G., Cesaroni, S., Colangeli, A., Flammini, D., Fonnesu, N., Falco, D., Gandolfo, G., Kantor, R., Kotula, J., Marzullo, D., Occhiuto, E., Ortwein, R., Previti, A., Sancristóbal, D., Villari, R., Esposito, B., 2024. Nuclear analyses in support of ITER ex-port Radial Neutron Camera design. Fusion Engineering and Design 202, 114295. https://doi.org/10.1016/j.fusengdes.2024.114295

Mota, F., Ortiz, M.I., Arranz, F., Brañas, B., Rapisarda, D., 2026. Non uniform thickness neutron spectrum shifter to fit radiation BB effects on material in IFMIF-DONES. Nuclear Engineering and Technology 58, 103983. https://doi.org/10.1016/j.net.2025.103983

Mullen, A.D., Andrews, M.T., Rondini, L.I., Thompson, C.J., 2024. Simulating gas-filled neutron detector responses with DRiFT. Nuclear Instruments and Methods in Physics Research Section A: Accelerators, Spectrometers, Detectors and Associated Equipment 1061, 169164. https://doi.org/10.1016/j.nima.2024.169164

Nocente, M., Molin, A.D., Rigamonti, D., Rosa, M.D., Fernández, B., Fugazza, S., Guerrero, C., Gorini, G., Kazakov, Y., Tardini, G., Tardocchi, M., ASDEX Upgrade Team, Eurofusion Tokamak Exploitation Team, 2024. COSMONAUT: A COmpact spectrometer for measurements of neutrons at the ASDEX upgrade tokamak. Review of Scientific Instruments 95, 083501. https://doi.org/10.1063/5.0218178

Occhiuto, E., Marzullo, D., Dongiovanni, D.N., Bonavolontà, U., Malaroda, F., Marocco, D., Esposito, B., 2025. Maintenance plan of the ITER Radial Neutron Camera: Verification and validation by virtual reality simulation. Fusion Engineering and Design 215, 114980. https://doi.org/10.1016/j.fusengdes.2025.114980

Ogawa, K., Isobe, M., Osakabe, M., 2021. Progress on Integrated Neutron Diagnostics for Deuterium Plasma Experiments and Energetic Particle Confinement Studies in the Large Helical Device During the Campaigns from FY2017 to FY2019. Plasma and Fusion Research 16, 1102023–1102023. https://doi.org/10.1585/pfr.16.1102023

Ogawa, K., Isobe, M., Sangaroon, S., Liao, L.Y., Zhong, G.Q., Seki, R., Nuga, H., Osakabe, M., 2024. Observation of energetic ion anisotropy using neutron diagnostics in the Large Helical Device. Nucl. Fusion 64, 076010. https://doi.org/10.1088/1741-4326/ad4909

Ogawa, Kunihiro, Jo, J., Kim, J., Liao, L., Sangaroon, S., Takada, E., Isobe, M., 2024. A scintillating-fiber detector for making high-time-resolution secondary D–T neutron measurements in KSTAR. Review of Scientific Instruments 95, 073539. https://doi.org/10.1063/5.0213697

Ogawa, K, Kim, J.H., Jo, J., Sangaroon, S., Liao, L.Y., Takada, E., Isobe, M., 2025. Experimental study of MHD instability effect on MeV ion confinement in KSTAR.

Ogawa, Kunihiro, Liao, L., Kobuchi, T., Kurita, S., Isobe, M., Ochiai, K., Kumagai, K., Kwon, S., Kondo, K., Sangaroon, S., Takada, E., Miwa, M., Toyama, S., Matsuyama, S., Kusaka, S., Tamaki, S., Murata, I., 2025a. Progress in the development of radiation detectors for fusion neutron source A-FNS. Fusion Engineering and Design 218, 115214. https://doi.org/10.1016/j.fusengdes.2025.115214

Ogawa, K., Sangaroon, S., Kobuchi, T., Isobe, M., 2026a. Neutron Camera in Large Helical Device. J Fusion Energ 45, 7. https://doi.org/10.1007/s10894-026-00548-0

Ogawa, K., Sangaroon, S., Matsuura, H., Kobayashi, M.I., Yoshihashi, S., Isobe, M., 2026b. Gamma Ray Diagnostics in Large Helical Device. J Fusion Energ 45, 14. https://doi.org/10.1007/s10894-026-00557-z

Ogawa, Kunihiro, Zhong, G., Liao, L., Chen, W., Zhou, R., Li, K., Takada, E., Sangaroon, S., Hu, L., Isobe, M., 2025b. Time-resolved deuterium–deuterium fusion born 1 MeV triton confinement study in EAST deuterium plasma. Plasma Phys. Control. Fusion 67, 025001. https://doi.org/10.1088/1361-6587/ada1f7

Olatujoye, D.O., Chandran, D., Mokhtar, A.A., 2025. Advancements in breeding blanket: Key to realizing spherical tokamak-based nuclear fusion power. Results in Engineering 28, 108406. https://doi.org/10.1016/j.rineng.2025.108406

Panda, S., Netrakanti, P.K., Behera, S.P., Sahu, R.R., Kumar, K., Sehgal, R., Mishra, D.K., Jha, V., 2026. Discrimination of neutron-$γ$ in the low energy regime using machine learning for an EJ-276D plastic scintillator. Nuclear Instruments and Methods in Physics Research Section A: Accelerators, Spectrometers, Detectors and Associated Equipment 1083, 171170. https://doi.org/10.1016/j.nima.2025.171170

Pankratenko, A., Kormilitsyn, T., Semenov, T., Obudovsky, S., Kashchuk, Yu., Zhong, G., Zhang, Y., Xu, M., 2025. Neutron spectroscopy with LaCl3(Ce)-based detector on EAST tokamak. Fusion Engineering and Design 218, 115236. https://doi.org/10.1016/j.fusengdes.2025.115236

Pankratenko, A.V., Kormilitsyn, T.M., Kashuck, Yu.A., Obudovsky, S.Yu., Fridrikhsen, D.S., Shevelev, A.E., Khilkevitch, E.M., Iliasova, M.V., Bakharev, N.N., Skrekel, O.M., 2023. Analysis of the LaCl3(Ce) scintillator response function to fast neutrons. Nuclear Instruments and Methods in Physics Research Section A: Accelerators, Spectrometers, Detectors and Associated Equipment 1052, 168282. https://doi.org/10.1016/j.nima.2023.168282

Perelli Cippo, E., Scioscioli, F., Bozzi, D., Marcer, G., Cazzaniga, C., Caruggi, F., Rebai, M., Rigamonti, D., 2026. Development of a simple prognostic method for SDD in neutron spectroscopy for steady-state fusion reactors. Fusion Engineering and Design 222, 115536. https://doi.org/10.1016/j.fusengdes.2025.115536

Pérez, M., Blostein, J.J., Zamorano, F., Fleta, C., Marín, J., Pellegrini, G., Guardiola, C., 2025. Experimental evaluation of silicon carbide P-N detectors under thermal and fast neutron irradiation at the RA-6 nuclear research reactor. Sci Rep 16, 2322. https://doi.org/10.1038/s41598-025-32175-8

Pérez, M., Zamorano, F., Fleta, C., Fernández, B., Guerrero, C., Godignon, P., Pellegrini, G., Pérez-Maroto, P., Guardiola, C., 2024. Characterization of new silicon carbide neutron detectors with thermal and fast neutrons. Nuclear Instruments and Methods in Physics Research Section A: Accelerators, Spectrometers, Detectors and Associated Equipment 1069, 169968. https://doi.org/10.1016/j.nima.2024.169968

Peterson, E.E., Romano, P.K., Shriwise, P.C., Myers, P.A., 2024. Development and validation of fully open-source R2S shutdown dose rate capabilities in OpenMC. Nucl. Fusion 64, 056011. https://doi.org/10.1088/1741-4326/ad32dd

Pettinari, D., Meschini, S., Testoni, R., 2025. Assessment of structural materials in compact fusion reactor design. Front. Nucl. Eng. 4, 1683702. https://doi.org/10.3389/fnuen.2025.1683702

Pietropaolo, A., 2023. The physical mechanisms of neutron detection. Contemporary Physics 64, 194–223. https://doi.org/10.1080/00107514.2024.2314817

Pietropaolo, A., Angelone, M., Bedogni, R., Colonna, N., Hurd, A.J., Khaplanov, A., Murtas, F., Pillon, M., Piscitelli, F., Schooneveld, E.M., Zeitelhack, K., 2020. Neutron detection techniques from μ eV to GeV. Physics Reports 875, 1–65. https://doi.org/10.1016/j.physrep.2020.06.003

Potiron, Q., Destouches, C., Houry, M., Llido, O., Lyoussi, A., Reynard-Carette, C., Dubus, L., Malard, P., Legou, P., Cheymol, B., 2026. Evaluating solid-state neutron detectors for measuring 14 MeV neutrons at high temperatures. Fusion Engineering and Design 223, 115586. https://doi.org/10.1016/j.fusengdes.2025.115586

Pu, N., Zhang, X.-C., Cai, H.-J., Jia, H., He, Y., 2024. Simulation of in-core neutron monitoring with U-235 micro fission chamber for CiADS-like LFR. Annals of Nuclear Energy 203, 110490. https://doi.org/10.1016/j.anucene.2024.110490

Qiao, Y., Wang, L., Li, W., Han, Y., Lv, J., Xu, M., Zhang, S., Wang, J., Fang, K., 2025. Research on neutron spectrum reconstruction methods based on space neutron spectrometer. Nuclear Instruments and Methods in Physics Research Section B: Beam Interactions with Materials and Atoms 564, 165712. https://doi.org/10.1016/j.nimb.2025.165712

Raj, P., Ball, J.L., Carmichael, J., Frenje, J.A., Gocht, R., Gorini, G., Holmes, I., Johnson, M.G., Kennedy, R., Mackie, S., Nocente, M., Panontin, E., Petruzzo, M., Rebai, M., Reinke, M., Rice, J., Rigamonti, D., Rosa, M.D., Saltos, A.A., Tardocchi, M., Tinguely, R.A., Wang, X., 2024. Overview of the neutron diagnostic systems for the SPARC tokamak. Review of Scientific Instruments 95, 103507. https://doi.org/10.1063/5.0219538

Rajput, M., Kamal, G., Mirfayzi, S.R., Chandrasekhar, K., Iliasova, M.V., Naylor, G., Wilson, C., 2025. Neutronics simulations of ST40 tokamak operations for activation foils in DD plasma operation. J. Inst. 20, P10003. https://doi.org/10.1088/1748-0221/20/10/P10003

Rechena, D., Da Silva, J., Amado, C., Infante, V., Varela, P., Santos, J.M., Silva, A., Gonçalves, B., Yong, L., Udintsev, V., 2026. Availability requirements for diagnostics in nuclear fusion: the ITER Collective Thomson Scattering case-study. Fusion Engineering and Design 222, 115493. https://doi.org/10.1016/j.fusengdes.2025.115493

Reinke, M.L., Abramovic, I., Albert, A., Asai, K., Ball, J., Batko, J., Brettingen, J., Brunner, D., Cario, M., Carmichael, J., Chrobak, C., Creely, A., Cykman, D., Dalla Rosa, M., Dubas, E., Downey, C., Ferrera, A., Frenje, J., Fox-Widdows, E., Gocht, R., Gorini, G., Granetz, R., Greenwald, M., Grieve, A., Hanson, M., Hawke, J., Henderson, T., Hicks, S., Hillesheim, J., Hoffmann, A., Holmes, I., Howard, N., Hubbard, A., Hughes, J.W., Ilagan, J., Irby, J., Jean, M., Kaur, G., Kennedy, R., Kowalski, E., Kuang, A.Q., Kulchy, R., LaCapra, M., Lafleur, C., Lagieski, M., Li, R., Lin, Y., Looby, T., Zubieta Lupo, R.,

Mackie, S., Marmar, E., McKanas, S., Moncada, A., Mumgaard, R., Myers, C.E., Nikolaeva, V., Nocente, M., Normile, S., Novoa, C., Ouellet, S., Panontin, E., Paz-Soldan, C., Pentecost, J., Perks, C., Petruzzo, M., Quinn, M., Raimond, J., Raj, P., Rebai, M., Riccardo, V., Rigamonti, D., Rice, J.E., Rosenthal, A., Safabakhsh, M., Saltos, A., Shanahan, J., Silva Sa, M., Song, I., Souza, J., Stein-Lubrano, B., Stewart, I.G., Sweeney, R., Tardocchi, M., Tinguely, A., Vezinet, D., Wang, X., Witham, J., 2024. Overview of the early campaign diagnostics for the SPARC tokamak (invited). Review of Scientific Instruments 95, 103518. https://doi.org/10.1063/5.0218254

Reviakin, P., Nemtsev, G., Vysokikh, I., Zharov, A., Mariano, G., 2026. Signal-to-Background Ratio Calculations for ITER Upper Vertical Neutron Camera. Fusion Science and Technology 1–15. https://doi.org/10.1080/15361055.2026.2642537

Richards, C.G., Ogren, K.B., Wiggins, B.W., Iliev, M., Favalli, A., Hehlen, M.P., 2025. Gamma-insensitive measurements of high neutron fluxes using solid-state composite scintillators. Optical Materials 165, 117074. https://doi.org/10.1016/j.optmat.2025.117074

Rigamonti, D., Dal Molin, A., Muraro, A., Rebai, M., Giacomelli, L., Gorini, G., Nocente, M., Perelli Cippo, E., Conroy, S., Ericsson, G., Eriksson, J., Kiptily, V., Ghani, Z., Štancar, Ž., Tardocchi, M., JET Contributors, 2024. The single crystal diamond-based diagnostic suite of the JET tokamak for 14 MeV neutron counting and spectroscopy measurements in DT plasmas. Nucl. Fusion 64, 016016. https://doi.org/10.1088/1741-4326/ad0a49

Rigamonti, D., Guarino, G., Camera, F., Cazzaniga, C., Croci, G., Dal Molin, A., Gorini, G., Muraro, A., Nocente, M., Lutz, B., Perelli Cippo, E., Rebai, M., Tardocchi, M., 2025. A chlorine based detector ($LaCl_3$ (Ce)) for 2.5 MeV neutron spectroscopy in deuterium nuclear fusion plasmas with enhanced particle discrimination algorithm. Meas. Sci. Technol. 36, 015907. https://doi.org/10.1088/1361-6501/ad8f4e

Rosa, M.D., Mackie, S., Rigamonti, D., Tedoldi, L.G., Colombi, S., Molin, A.D., Marcer, G., Nocente, M., Gorini, G., Raj, P., Rebai, M., Carmichael, J., Reinke, M., Scioscioli, F., Tinguely, R.A., Tardocchi, M., 2024. Design solutions for the hodoscope of the magnetic proton recoil neutron spectrometer of the SPARC tokamak. Review of Scientific Instruments 95, 083508. https://doi.org/10.1063/5.0219463

Ruddy, F.H., Ottaviani, L., Lyoussi, A., Destouches, C., Palais, O., Reynard-Carette, C., 2022. Silicon Carbide Neutron Detectors for Harsh Nuclear Environments: A Review of the State of the Art. IEEE Trans. Nucl. Sci. 69, 792–803. https://doi.org/10.1109/TNS.2022.3144125

Saengkaew, P., Ploykrachang, K., 2026. Comprehensive review of neutron techniques, detection, and dosimetry in science and technology. Discov Appl Sci 8, 251. https://doi.org/10.1007/s42452-025-08024-8

Sangaroon, S., Ogawa, K., Isobe, M., 2024a. Neutron emission spectrometer in magnetic confinement fusion. AAPPS Bull. 34, 34. https://doi.org/10.1007/s43673-024-00139-1

Sangaroon, S., Ogawa, K., Isobe, M., Liao, L., Zhong, G., Wisitsorasak, A., Takada, E., Kobayashi, M.I., Poolyarat, N., Murakami, S., Seki, R., Nuga, H., Osakabe, M., 2024b. Neutron Spectroscopy in Perpendicular Neutral Beam Injection Deuterium Plasmas Using Newly Developed Compact Neutron Emission Spectrometers. IEEE Trans. Instrum. Meas. 73, 1–11. https://doi.org/10.1109/TIM.2024.3446631

Sangaroon, S., Ogawa, K., Liao, L., Zhong, G., Isobe, M., 2026. Neutron Spectroscopy of Deuterium-deuterium Reactions in Neutral Beam-heated Plasmas on the Large Helical Device. J Fusion Energ 45, 23. https://doi.org/10.1007/s10894-026-00564-0
Santos Oliveira, R., Batista Da Silva, J., Machado Mendes, L.M., De Sousa Lacerda, M.A., 2023. Responses of a 6LiI Bonner Sphere Spectrometer using the Monte Carlo codes PHITS and MCNPX. Applied Radiation and Isotopes 200, 110976. https://doi.org/10.1016/j.apradiso.2023.110976
Schissel, D.P., Nazikian, R.M., Gibbs, T., 2025. Summary report from the mini-conference on Digital Twins for Fusion Research. Physics of Plasmas 32, 050601. https://doi.org/10.1063/5.0273586
Scholz, M., Wiącek, U., Drozdowicz, K., Jardin, A., Woźnicka, U., Król, K., Kurowski, A., Kulińska, A., Dąbrowski, W., Łach, B., Mazon, D., 2026. Neutron spectrometer based on a gas electron multiplier (GEM) detector for fusion reactors. Fusion Engineering and Design 222, 115480. https://doi.org/10.1016/j.fusengdes.2025.115480
Schwemmlein, A.K., Forrest, C.J., Knauer, J.P., McClow, H., Stanley, B., Romanofsky, M.H., Rosenberg, M.J., Glebov, V.Y., 2022. A New Neutron Time-of-Flight Detector for Yield and Ion-Temperature Measurements in DT implosions on OMEGA.
Scioscioli, F., Marcer, G., Ciurlino, A., Colombi, S., Coriton, B., Dal Molin, A., Dankowski, J., Gorini, G., Kovalev, A., Muraro, A., Nocente, M., Rebai, M., Rigamonti, D., Tardocchi, M., Croci, G., 2025. Design and development status of the ITER Radial Gamma Ray Spectrometer. Fusion Engineering and Design 221, 115376. https://doi.org/10.1016/j.fusengdes.2025.115376
Shen, H.-Y., Zhang, J.-L., Zhang, J., Zhou, J.-H., 2024. FPGA implementation of 500-MHz high-count-rate high-time-resolution real-time digital neutron-gamma discrimination for fast liquid detectors. NUCL SCI TECH 35, 136. https://doi.org/10.1007/s41365-024-01441-1
Simoni, M., Baldassarre, L., Cazzaniga, C., Fazi, L., Gaboardi, M., Gemmiti, L., Kastriotou, M., Krzystyniak, M., Marsicano, A., Martellucci, M., Minniti, T., Prioriello, A., Senesi, R., Suteica, V., Romanelli, G., 2025. Use of CR-39 Dosimeters for the Imaging of Neutron Beam Profiles in the 100 keV–10 MeV Energy Range. Sensors 25, 5865. https://doi.org/10.3390/s25185865
Sisodiya, D.S., Patra, G.D., Singh, S.G., Singh, A.K., Pitale, S., Ghosh, M., Sen, S., 2025. 6LiI:Eu/LaBr3:Ce phoswich detector for detection and discrimination of thermal neutron and gamma radiation. Nuclear Instruments and Methods in Physics Research Section A: Accelerators, Spectrometers, Detectors and Associated Equipment 1075, 170428. https://doi.org/10.1016/j.nima.2025.170428
Solano, E.R., 2025. Fusion research in a Deuterium-Tritium tokamak. Fundamental Plasma Physics 15, 100096. https://doi.org/10.1016/j.fpp.2025.100096
Song, R., Han, J., Yan, X., Luo, X., Ren, F., Han, Z., Wen, C., Zhang, X., Zhang, Y., Chen, L., Yi, C., Qu, G., Liu, X., Lin, W., Leng, Q., Zhu, J., Qian, S., Wang, Z., Tong, Y., Tang, G., Qin, L., Wang, X., Liu, J., 2023. Fast neutron response of 6Li enriched CLYC and CLLB scintillators within 0.9–5.2 MeV. Nuclear Instruments and Methods in Physics Research Section A: Accelerators, Spectrometers, Detectors and Associated Equipment 1055, 168533. https://doi.org/10.1016/j.nima.2023.168533
Sozzi, C., Ayllon-Guerola, J., Belpane, A., Buzas, A., Cabrera, S., Cavinato, M., Carraro, L., Cecconello, M., Coda, S., Cseh, G., Davis, S., D'Isa, F., Estrada, T., Fassina, A.,

Figueiredo, J., Garcia-Munoz, M., Gonzalez, J., Giudicotti, L., Jokinen, A., Hajnal, N., Kocsis, G., Martinez, J., Nakano, T., Nocente, M., Noel, M., Moro, A., Pasqualotto, R., Refy, D.I., Rigamonti, D., Soare, S., Szepesi, T., Tanaka, K., Valisa, M., 2025. OVERVIEW OF THE EUROPEAN CONTRIBUTION TO THE DIAGNOSTIC EQUIPMENT OF JT-60SA FOR THE NEXT OPERATIONAL PHASES.

Stefanescu, I., Guérard, B., Hall-Wilton, R., Jackson, A., Khaplanov, A., Klein, M., Lai, C.-C., Piscitelli, F., Raspino, D., Schmidt, C.J., Schweika, W., Svensson, P.-O., 2024. $^{10}$ Boron-film-based gas detectors at ESS. Journal of Neutron Research 26, 83–109. https://doi.org/10.3233/JNR-230012

Su, X., Zhou, P., Li, J.L., Ma, J.F., Yuan, D.W., Zhang, Z., Zhang, G.Q., Li, K.N., Chen, P., Wu, F.Y., Yuan, X.H., Zhang, J., 2025. Calibration of a gated neutron time-of-flight spectrometer with low-afterglow. Review of Scientific Instruments 96, 093502. https://doi.org/10.1063/5.0242176

Sui, Z., Qian, S., Niu, L., Hu, P., Hua, Z., Zheng, X., Sun, X., Tang, G., Cai, H., Yang, D., Li, W., Zhang, M., Han, J., Ren, J., 2025. Glass scintillator: A window to future high-energy radiation detection. The Innovation 6, 100878. https://doi.org/10.1016/j.xinn.2025.100878

Sykora, G.J., Mann, S.E., Mauri, G., Schooneveld, E.M., Rhodes, N.J., 2024. Review of thermal neutron scintillators: Evaluation metrics and future prospects for demanding applications. Optical Materials: X 24, 100373. https://doi.org/10.1016/j.omx.2024.100373

Takada, M., Endo, S., Kajimoto, T., Horiguchi, T., Yamanishi, H., Yagi, N., Masuda, A., Matsumoto, T., Tanaka, H., Nunomiya, T., Aoyama, K., Narita, M., Nakamura, T., 2024. Thickness-dependent neutron detection efficiency of LiF-Si-based active neutron detector for boron neutron capture therapy. Nuclear Instruments and Methods in Physics Research Section A: Accelerators, Spectrometers, Detectors and Associated Equipment 1064, 169352. https://doi.org/10.1016/j.nima.2024.169352

Tinguely, R.A., Rosenthal, A., Simpson, R., Ballinger, S.B., Creely, A.J., Frank, S., Kuang, A.Q., Linehan, B.L., McCarthy, W., Milanese, L.M., Montes, K.J., Mouratidis, T., Picard, J.F., Rodriguez-Fernandez, P., Sandberg, A.J., Sciortino, F., Tolman, E.A., Zhou, M., Sorbom, B.N., Hartwig, Z.S., White, A.E., 2019. Neutron diagnostics for the physics of a high-field, compact, $Q\geq1$ tokamak. https://doi.org/10.48550/arXiv.1903.09479

Tudisco, S., Altana, C., Amaducci, S., Ciampi, C., Cosentino, G., De Luca, S., La Via, F., Lanzalone, G., Muoio, A., Pasquali, G., Trifirò, A., 2025. Silicon Carbide devices for radiation detection: A review of the main performances. Nuclear Instruments and Methods in Physics Research Section A: Accelerators, Spectrometers, Detectors and Associated Equipment 1072, 170112. https://doi.org/10.1016/j.nima.2024.170112

Tursinah, R., Permana, S., Su'ud, Z., Maulana, A., Sukmabuana, P., 2025. A Review on the Development of Single-Moderator-Based Neutron Spectrometers. Nuclear Technology 211, 635–644. https://doi.org/10.1080/00295450.2024.2361177

Vega-Carrillo, H.R., Baltazar-Raigosa, A., Hernandez-Davila, V.M., Gallego, E., Garcia-Fernandez, G., Bedenko, S.V., 2025. Monte Carlo design of a novel passive Regular Parallelepiped neutron spectrometer (TLD600/RPS). Radiation Physics and Chemistry 236, 112861. https://doi.org/10.1016/j.radphyschem.2025.112861

Villari, R., Litaudon, X., Mailloux, J., Dentan, M., Fonnesu, N., Ghani, Z., Packer, L.W., Rimini, F., Vila, R., Afanasenko, R., Alguacil, J., Batistoni, P., Beaumont, P., Bradnam, S.C.,

Carman, P., Catalan, J.P., Cecchetto, M., Colangeli, A., Croft, D., Cufe, J., De Pietri, M., Fabbri, M., Figueiredo, J., Flammini, D., Grove, C.L., Hjalmarsson, A., Ioannou-Sougleridis, V., Jones, L., Kierepko, R., Klosowski, M., Kolsek, A., Kos, B., Laszynska, E., Lerche, E., Tonqueze, Y.L., Leichtle, D., Leon-Gutierrez, E., Lengar, I., Loughlin, M., Lobel, R., Loreti, S., Mariano, G., Mianowski, S., Milnes, J., Moindjie, S., Moro, F., Mietelski, J., Naish, R., Noce, S., Peric, J., Pontier, J.B., Reynolds, S., Radulović, V., Savva, M.I., Sauvan, P., Shand, C.R., Snoj, L., Stamatelatos, I.E., Štancar, Z., Stokes, T., Terranova, N., Turner, A.N., Vasilopoulou, T., Wójcik-Gargula, A., 2025. Overview of deuterium-tritium nuclear operations at JET. Fusion Engineering and Design 217, 115133. https://doi.org/10.1016/j.fusengdes.2025.115133

Wang, C., Yu, R., Xia, W., Gong, J., 2025. Advances in High-Temperature Irradiation-Resistant Neutron Detectors. Sensors 25, 7554. https://doi.org/10.3390/s25247554

Wang, X., Gocht, R., Ball, J., Mackie, S., Panontin, E., Peterson, E., Raj, P., Holmes, I., Saltos, A.A., Johnson, A., Grieve, A., Tinguely, R.A., 2025. An OpenMC model of the SPARC tokamak for the diagnostic scoping studies. Fusion Engineering and Design 221, 115390. https://doi.org/10.1016/j.fusengdes.2025.115390

Wang, X., Gocht, R., Ball, J., Mackie, S., Panontin, E., Tinguely, R.A., Raj, P., Holmes, I., Saltos, A.A., Johnson, A., Grieve, A., 2024. Neutronics simulations for the design of neutron flux monitors in SPARC. Review of Scientific Instruments 95, 083560. https://doi.org/10.1063/5.0219508

Weiss, C., Griesmayer, E., 2024. Fusion neutron diagnostics with CVD diamond detectors. Fusion Engineering and Design 203, 114453. https://doi.org/10.1016/j.fusengdes.2024.114453

Wen, Z., Feng, L., Yuan, G., Zhao, W., Shi, J., Shen, F., 2026. Neutronic Modeling and Physical Analysis of the Neutron Yield Measurement System for the HL-3 Tokamak.

Xu, M., Zhong, G., Zhang, Yubo, Huang, L., Hu, L., Zhang, R., Yang, L., Zhang, Yongqiang, Chen, W., Li, Y., Sangaroon, S., Ogawa, K., Isobe, M., Liao, L., Fan, T., 2024. Investigation of Cs27LiYCl6:Ce scintillator energy response for D-D fusion neutron spectrometer. Fusion Engineering and Design 204, 114490. https://doi.org/10.1016/j.fusengdes.2024.114490

Yang, L., Hu, L.-Q., Zhong, G.-Q., Cao, H.-R., Zhao, J.-L., Zhang, Y.-Q., Chen, W.-K., 2023. Investigation of Campbell mode with automatic calibration operated in fission chamber-based neutron flux monitoring system on EAST. Annals of Nuclear Energy 192, 109986. https://doi.org/10.1016/j.anucene.2023.109986

Yücel, H., Karanfil, E.C., Karaman, M.C., Tüzel, O., Dağıstan, Z., Hanci, K.Ö., Özkan, M., 2025. Investigation of thermal neutron/gamma discrimination capability of a CLYC Scintillator with a SiPM in a mixed field of 241Am-Be neutron source: Pulse shape discrimination using constant fraction time over threshold (CF-TOT) and cumulative charge ratio (CCR). Radiation Measurements 189, 107540. https://doi.org/10.1016/j.radmeas.2025.107540

Zaitseva, N., 2026. Comparative analysis of new crystal and plastic scintillators for fast and thermal neutron detection.

Zeng, G., Ji, A., Wang, Y., 2026. Investigating deep level defects in neutron-irradiated large size 4H-SiC vertical Schottky barrier diodes. Nuclear Instruments and Methods in Physics Research Section A: Accelerators, Spectrometers, Detectors and Associated Equipment

1084, 171215. https://doi.org/10.1016/j.nima.2025.171215

Zhang H., Mao C., Liao Y., Wu Z., Liu G., Yan Q., 2026. Physical simulation design and optimization of micro structure semiconductor neutron detector. NT 49, 030504. https://doi.org/10.3724/j.0253-3219.2026.hjs.49.240434 (in Chinese)

Zhang, J., Chen, X.L., Shen, H.Y., Zhang, J.L., Zhang, Y.S., Han, Y.X., Yang, Q.L., Zhang, Y.P., Shi, Z.B., 2025. Time-Resolved Characterization of Fusion Neutron Spectra via a Liquid Scintillation Detection System in the HL-3 Tokamak.

Zhang, Y.-Q., Hu, L.-Q., Lu, W., Zhong, G.-Q., Cao, H.-R., Zhao, J.-L., Yang, L., Zhang, R.-X., Xu, M.-Y., Li, Q., 2024. Development of a real-time digital pulse acquisition and processing algorithm for compact neutron spectrometer on EAST. Fusion Engineering and Design 202, 114394. https://doi.org/10.1016/j.fusengdes.2024.114394

Zhang, Y.S., Zhang, Y.P., Zhang, J., Zhong, W.L., Ogawa, K., Yang, Q.L., Han, Y.X., Gao, Z., Zhu, Y.X., Guan, S., Zhang, C.X., Isobe, M., Shi, Z.B., 2026. Design and simulation study of scintillation fiber detector for deuterium-tritium fusion neutron measurement on the HL-3 tokamak. Fusion Engineering and Design 222, 115502. https://doi.org/10.1016/j.fusengdes.2025.115502